\documentclass[a4paper,fleqn,usenatbib,useAMS]{mnras}

\usepackage{graphicx}	
\usepackage{amsmath}	
\usepackage{amssymb}	
\usepackage{multicol}        
\usepackage{bm}		
\usepackage{pdflscape}	
\usepackage{rotating}
\usepackage{tabularx}
\usepackage{ragged2e}
\usepackage{lipsum}
\usepackage{booktabs} 

\newcommand{\kms}{\,km\,s$^{-1}$}

\newcommand{\msun}{M_\odot}
\newcommand{\mstar}{M_{*}}

\newcommand{\mcluster}{M_{\rm cluster}}

\newcommand{\nbody}{\texttt{NBODY6}}
\newcommand{\nbodyplus}{\texttt{NBODY6++}}
\newcommand{\nb}{\texttt{NBODY6++GPU}}
\newcommand{\rebound}{\texttt{REBOUND}}
\newcommand{\amuse}{\texttt{AMUSE}}
\newcommand{\chain}{\texttt{CHAIN}}
\newcommand{\mercury}{\texttt{MERCURY6}}
\newcommand{\mpi}{\texttt{MPI}}
\newcommand{\gpu}{\texttt{GPU}}
\newcommand{\hdf}{\texttt{HDF5}}
\newcommand{\caigpu}{\texttt{h5nb6xx}}

\newcommand{\lp}{\texttt{LonelyPlanets}}

\newcommand{\nstars}{N_{\rm s}}
\newcommand{\nhosts}{N_{\rm hosts}}

\newcommand{\rvir}{r_{\rm vir}}
\newcommand{\rtidal}{r_{\rm tidal}}
\newcommand{\halfmass}{r_{\rm hm}}

\newcommand{\crossingtime}{t_{\rm cr}}
\newcommand{\relaxationtime}{t_{\rm rh}}
\newcommand{\segregationtime}{t_{\rm ms}}

\usepackage[T1]{fontenc}
\usepackage{ae,aecompl}

 \title[Planets in Star Clusters: the Solar System]{Planetary systems in a star cluster I: the Solar system scenario}

\author[Flammini Dotti, Kouwenhoven, Cai \& Spurzem]{Francesco Flammini Dotti$^{1,2}$\thanks{Contact e-mail: \href{mailto:flammini.francesco@xjtlu.edu.cn}{flammini.francesco@xjtlu.edu.cn}},
M.B.N. Kouwenhoven$^{1}$\thanks{Contact e-mail: \href{mailto:t.kouwenhoven@xjtlu.edu.cn}{t.kouwenhoven@xjtlu.edu.cn}},
Maxwell Xu Cai$^{3}$,
\newauthor{and Rainer Spurzem$^{4,5,6}$\thanks{Research  Fellow  at  Frankfurt  Institute  for  Advanced  Studies (FIAS)}}
\\
$^{1}$Department of Mathematical Sciences, Xi{'}an Jiaotong-Liverpool University, 111 Ren{'}ai Rd., \\
Suzhou Dushu Lake Science and Education Innovation District, Suzhou Industrial Park, Suzhou 215123, P.R. China\\
$^{2}$Department of Mathematical Sciences, University of Liverpool, Liverpool L69 3BX, UK\\
$^3$Leiden Observatory, Leiden University, PO Box 9513, 2300 RA, Leiden, Netherlands\\
$^4$National Astronomical Observatories and Key Laboratory of Computational Astrophysics, \\
Chinese Academy of Sciences, 20A Datun Rd., Chaoyang District, 100101, Beijing, China\\
$^5$Kavli Institute for Astronomy and Astrophysics at Peking University, 5 Yiheyuan Rd., Haidian District, 100871, Beijing, China\\
$^6$Zentrum f\"{u}r Astronomie der Universit\"{a}t Heidelberg, Astronomisches Rechen-Institut, M\"{o}nchhofstr. 12-14, 69120 Heidelberg, Germany}

\date{Last updated \today}

\pubyear{2019}

\begin{document}
\label{firstpage}
\pagerange{\pageref{firstpage}--\pageref{lastpage}}
\maketitle

\bibliographystyle{mnras}

\begin{abstract}
Young stars are mostly found in dense stellar environments, and even our own Solar system may have formed in a star cluster. Here, we numerically explore the evolution of planetary systems similar to our own Solar system in star clusters.
We investigate the evolution of planetary systems in star clusters. Most stellar encounters are tidal, hyperbolic, and adiabatic. A small fraction of the planetary systems escape from the star cluster within 50~Myr; those with low escape speeds often remain intact during and after the escape process. While most planetary systems inside the star cluster remain intact, a subset is strongly perturbed during the first 50~Myr. Over the course of time, $0.3\%-5.3\%$ of the planets escape, sometimes up to tens of millions of years after a stellar encounter occurred.  Survival rates are highest for Jupiter, while  Uranus and Neptune have the highest escape rates. Unless directly affected by a stellar encounter itself, Jupiter frequently serves as a barrier that protects the terrestrial planets from perturbations in the outer planetary system. In low-density environments, Jupiter provides protection from perturbations in the outer planetary system, while in high-density environments, direct perturbations of Jupiter by neighbouring stars is disruptive to habitable-zone planets. The diversity amongst planetary systems that is present in the star clusters at 50~Myr, and amongst the escaping planetary systems, is high, which contributes to explaining the high diversity of observed exoplanet systems in star clusters and in the Galactic field.
\end{abstract}

\begin{keywords}
stars: solar-type -- stars: planetary systems -- planets: dynamical evolution and stability -- planets: terrestrial planets -- stars: statistics -- Galaxy: stellar content
\end{keywords}

\section{Introduction} \label{sec:intro}

It is commonly accepted that a large fraction of stars in the Galaxy hosts one or more planetary companions \citep[e.g.,][]{1a, Thompson2018}; and even binary star systems are known to host exoplanets \citep{1b}. As most stars form in clustered environments \citep[e.g.,][]{1c}, close stellar encounters with proto-planetary disks and young planetary systems may leave an imprint on the much older population in the Galactic neighbourhood. In recent decades, the possible relationship between planetary systems and star clusters was gradually recognised. As a result of the rapid advances in observational techniques, over 4100 exoplanets have now been identified in 3057 extra-solar planetary systems, among which 667 are multi-planetary systems\footnote{http://exoplanet.eu, accessed on 21 August 2019.}.
To fully understand the origin and dynamical evolution of planetary systems, it is necessary to carefully study the effect of their environments, i.e., that of the star-forming region in which they form and spend the first million years, and that of the Galactic field, the open cluster, or the globular cluster in which they may spend the billions of years that follow.

The current paradigm for the formation of stars suggests that stars are formed in groups in gaseous environments, which are similar to, but slightly less massive and concentrated, than the progenitors of the longer-lived open clusters. Most of these young groups of stars disperse within $10-50$~Myr, after which their member stars become part of the field star population, while others remain bound for hundreds of millions to billions of years \citep[e.g.,][]{grijs2008, 11a}. In such clustered environments, protoplanetary disks may be perturbed following encounters with neighbouring stars \citep[e.g.,][]{thies2005, olczak2012, 11b, kirsten2018}. In many dense star forming molecular clouds, however, the fraction of stars with proto-stellar disks depends more sensitively on the stellar age rather than on the dynamical properties of their host clusters \citep{1d}. Isotope analysis in meteorites has shown that the proto-planetary disk of our Sun was polluted by a nearby supernova \citep[e.g.,][]{11c,11d}. This suggests that even our own Solar System may have formed in a clustered stellar environment, in a star cluster of size $\rvir = \ 0.75 \pm 0.25$~pc, together with $2500 \pm 300$ other stars \citep{adams2010,1e}.

Modelling the evolution of planetary systems in star clusters remains a challenge, as the numerical noise in a star cluster simulation can be comparable to the precision required to accurately model the evolution of a planetary system: the relative energy error necessary for accurate integration of the planetary systems is over five orders of magnitude smaller than that for the stars \citep[see, e.g.,][]{1i}. Different approaches have been taken to overcome this challenge, including (i) modelling the evolution of single-planet systems using the existing binary regularisation methods in $N$-body codes; (ii) using scattering experiments to model the evolution of multi-planet systems; and recently (iii) using full $N$-body simulations of planetary systems in star clusters, under the assumption that the stellar dynamics is unaffected by the planetary bodies.

Stellar encounters lead to the disruption of planetary systems and the presence of free-floating planets in star clusters. With sufficiently high velocities, these free-floating planets can immediately escape from the star cluster \citep{wangkouwenhoven2015}. Alternatively, these free-floating planets gradually migrate to the outskirts of the cluster where they may be stripped off by the Galactic tidal field, or re-captured by other stars \citep[e.g.,][]{perets2012}. \cite{1g} provide a comprehensive analysis of the evolution of star-planet systems in star clusters. They compare numerical results obtained using two approaches, $\nbodyplus$ \citep{1l} and a hybrid Monte Carlo code \citep{spurzemgiersz1996, gierszspurzem2000, mac1}, and find that their outcomes are consistent with those of theoretical estimates. Their star clusters contain $19\,000$ equal-mass stars, of among which 1000 host a single planet. Differential cross sections for changes in the orbital elements of the star-planet binaries are obtained, and the regimes for the strengths of the stellar encounters are identified. The study of \cite{1g} provides a framework for studying planetary systems in star clusters. Important limitations include the lack of a stellar mass spectrum, the absence of stellar binaries, and the absence of multi-planet systems. 
\cite{11z} used $\nbody$ to study the evolution of single-planet systems in multi-mass open star clusters of different degrees of initial substructure, total initial masses, and initial virial ratios, and find analytical prescriptions for the retention rate of planetary companions and free-floating planets as a function of initial semi-major axis and cluster properties. 
\cite{fujii2019} followed a similar, but more comprehensive approach, and focus on planets in the Pleiades, Hyades and Praesepe clusters. These star clusters are thought to have formed from highly-substructured star forming regions \citep[e.g.,][]{fujii2012,sabbi2012,fujii2015}. The authors study single-planet systems in orbit around Solar-like stars, and find that planets with initial semi-major axes of $a_p=10-100$~AU have a relatively high probability of being ejected from their host systems. They find an escape probability of $p_{esc} \propto a_p^{-0.76}$, which is  consistent with the findings of \cite{11z}. Planets with $a_p>100$~AU are unlikely to have survived until the present day in these star clusters.

Further progress in modelling multi-planet systems in star clusters was made by \cite{shara2016}, who evolve two-planet systems in star containing $18\,000$ single stars and $2000$ binary systems. In each model, 100 stars in the mass range $0.5-1.1~\msun$ are assigned two planets with semi-major axes of 5.2~AU and 9.5~AU, respectively. They use \nbody{}, and evolved the planetary systems while making use of the three-body stability criteria algorithms of \cite{mardling2008} and \cite{gnbs}. They find that, over the course of billions of years, the innermost planet has a probability of 1.5\% of evolving into a hot Jupiter, i.e., that dynamical interactions with neighbour stars can provide a viable formation mechanism for at least a subset of the hot Jupiters discovered in star clusters.
\cite{puu} take an alternative approach to modelling two-planet systems in star clusters. They carry out simulations using \rebound{} \citep{1o} and compare these results to hybrid secular equations. Their approach contributes to our understanding of the origin of super-Earths and sub-Neptunes in general, and the Kepler-11 system in particular, and explains why systems with multiple transiting planets appear to be dynamically colder than those with a single transiting planet.

The two-body and three-body approaches described above allow accurate modelling of small planetary systems in star clusters, but in the case of more than two planets the classical star cluster simulation codes are not appropriate. One approach to overcome these limitations is to separate the evolution of the star cluster and the planetary systems, under the assumption that the presence of planets does not affect the dynamics of the stars. With this in mind,  \cite{1h} studied the evolution of multi-planet systems in star clusters using scattering experiments. The \chain{} package \citep{1m} was used to carry out the planetary system integrations during a stellar encounter, integrating multi-body star encounters, and \mercury{} \citep{1n} was used for the planetary integration during the intermittent periods. The stellar encounter properties were drawn from analytic distributions appropriate for each star cluster.  \cite{1h} modelled different realisations  of planetary systems containing four gas giants, and find that the lower-mass planets (Saturn, Uranus, and Neptune) are affected by both stellar encounters and planet-planet interactions, while the more massive planet Jupiter, when perturbed, is almost exclusively perturbed by stellar encounters. 
\cite{1f} and \cite{malmberg2011} model the evolution of multi-planet systems in star clusters through recording all close stellar encounters in a modified version of $\nbody$, and subsequently carrying out simulations of perturbed planetary systems using \mercury{}, they identify the effects of stellar fly-by's on planetary systems containing the Solar system's four gas giants, and determine survival rates and the timescales after which a close stellar encounter may trigger a planetary instability. 

Substantial progress was made by \cite{1i,caisignatures2018,cai2019}, who carry out full $N$-body simulations of multi-planet systems in star clusters.  \cite{1i} model open star clusters with planetary systems containing Solar-type stars with five equal-mass planets separated by 10 to 100 mutual Hill radii. They find that, although most planetary systems retain their planets, stellar encounters and planet-planet interactions can trigger substantial perturbations in the orbital eccentricities and inclinations, which can lead to decay of the planetary system. \cite{caisignatures2018} and \cite{cai2019} take a similar approach, and identify how the signatures of the parental star cluster may affect the observed characteristics of exoplanet systems in the Galactic field. 

In this paper we carry out simulations of multi-planet systems similar to our own Solar system in star clusters, in order to deepen our understanding of the evolution of perturbed unequal-mass planetary systems, and its implications for planets in the habitable zone. We use the \lp{} code \citep[e.g.,][]{1i, flamminidotti2018}, which combines the planetary systems $N$-body evolution code \rebound{} \citep{1o} with the \nb{} \citep{1p} star cluster evolution code in the \amuse{} multi-physics environment \citep{Portegies-Zwart2011, 1q, Pelupessy2013, spzbook}. 

This paper is organised as follows. In Section~2 we introduce our numerical method and the initial conditions. In Section~3 we describe our results, and place these into the context of star cluster evolution and close stellar encounters.  Finally, we present our conclusions and discussion in Section~4.

\section{Method and initial conditions}

\subsection{Initial conditions - star clusters}

\begin{table}
\caption{Initial conditions for the star clusters: the model ID (column 1, using the syntax C-$Q$-$\nstars$), the cluster initial number of stars (column 2), the initial total star cluster mass (column 3), the initial virial ratio (column 4), the  initial crossing time and the initial half-mass relaxation time (columns 5 and 6), and the number of planet-hosting stars realisation of the star cluster model (column 7).  
\label{tab:table}}
\begin{tabular}{lrccccc}
\hline
Model ID & $\nstars$ & $\mcluster$ & $Q$ & $\crossingtime$ & $\relaxationtime$ & $\nhosts$  \\
  &     & $\msun$ & & Myr & Myr &  \\
\hline
C045E2 & 500 &$ 2.78\times10^2$ & 0.4 & 0.86 & 10.19 & 25 \\
C055E2 & 500 &$ 2.45\times10^2$ & 0.5 & 0.81 & \ 9.57 & 25 \\
C065E2 & 500 & $2.61\times10^2$ & 0.6 & 0.84 & \ 9.87 & 25 \\
C045E3 & 5\,000 &$ 2.73\times10^3$ & 0.4 & 0.26 & 21.30 & 125  \\
C055E3 & 5\,000 & $2.87\times10^3$ & 0.5 & 0.25 & 20.77 & 125  \\
C065E3 & 5\,000 & $2.90\times10^3$ & 0.6 & 0.25 & 20.67 &  125  \\
C041E4 & 10\,000 & $5.87\times10^3$ & 0.4 & 0.18 & 26.59 & 200   \\
C051E4 & 10\,000 & $5.87\times10^3 $ & 0.5 & 0.18 & 26.59 & 200   \\
C061E4 & 10\,000 & $5.87\times10^3$ & 0.6 & 0.18 & 26.59 & 200   \\
\bottomrule
\end{tabular}
\label{table:initialconditions}
\end{table}

Our models represent open clusters containing $N=500-10\,000$ stars. The initial conditions for these models are summarised in Table~\ref{table:initialconditions}. We draw stellar positions and velocities from the  \cite{2a} model with an initial virial radius of $\rvir=1$~pc, corresponding to an initial half-mass radius of $\halfmass \approx 0.77$~pc. We study models with different initial virial ratios $Q = |T/U|$ where $T$ is the cluster total kinetic energy and $U$ is the total gravitational energy of the star cluster. Stellar masses are drawn from the Kroupa  initial mass distribution \citep{2b} in the mass range  $\mstar = 0.1-25~\msun$, and are assigned a Solar metallicity. We adopt an external tidal field corresponding to the Solar orbit in the Milky Way. We do not include primordial binaries, we assume that the star clusters are not rotating, and we ignore the presence of any gas remaining from the star-formation process. All models are evolved for $t=50$~Myr. 

We analyse how the evolution of planetary systems depends on time, on the initial number of cluster members, and the on the initial virial ratio of the star cluster. As all clusters initially have the same size, the initial number of cluster members, $N=500$, $5000$, and $10\,000$, respectively, determines the initial stellar density, $\rho_\star\propto N$, which in turn affects the rate at which planetary systems experience encounters. The choices for the initial virial ratio ($Q=0.4$, $0.5$, and $0.6$) correspond to clusters that are contracting, in virial equilibrium, and expanding, respectively. The latter values are motivated by various computational and observational studies of young star-forming regions \citep[e.g.,][]{goodwin2004, philipp2012, parker}. We simulate an ensemble in order to reduce statistical errors.

\subsection{Initial conditions - planetary systems}

Our purpose is to model the evolution of Solar system analogs in star clusters. For this purpose, we identify the set of $\nhosts$ cluster members with masses nearest to $1~\msun$ as planet-hosting stars. The resulting host star masses all have masses in the range $[0.93,1.03]~\msun$. Each host star is assigned to a planetary system similar to our own Solar System. For reasons of accuracy and computational cost, we have excluded Mercury, Venus, and the minor Solar-system bodies. The latter bodies have a small affect on the dynamical evolution of the more massive bodies. We adopt the present-day masses and orbital elements of the six planets in our model (Earth, Mars, Jupiter, Saturn, Uranus, and Neptune) as our initial conditions.

\subsection{Numerical method} \label{sec:style}

Planetary systems and star clusters are very different dynamical systems, and modelling their evolution thus requires a different numerical approach. Planetary systems are AU-scale hierarchical few-body systems, whereas star clusters are parsec-scale non-hierarchical many-body systems. The orbital periods of planets vary from hours to several hundred years, whereas the crossing timescale of star clusters are typically in the order of a million years. A simulation of planetary systems is generally considered satisfactory when the relative energy error is $\Delta E/E \la 10^{-10}$  whereas for star cluster simulations this criterion is usually relaxed to $\Delta E/E \la 10^{-5}$ \citep[see, e.g.,][]{1i}. The latter limit is often adopted when the Hermite scheme is used to integrate star clusters; it is however not fundamental to the scheme itself, which can be used to obtain accuracies of $\Delta E/E < 10^{-10}$ \citep[see, e.g.,][]{kokubo1998, yoshinaga1999}, and thus is sufficient for integrating isolated planetary system using the Hermite scheme as well \citep[see, e.g.,][]{mikkola1998}.

For this reason we combine two separate numerical integrators in order to calculate the dynamical evolution of planetary systems in star clusters. We follow the numerical approach of \cite{1i}.  This methodology can be summarised in four steps: (i) initialisation of the star clusters and planetary systems; (ii) numerical modelling of the evolution of the star cluster; (iii) identifying the close encounters experienced by the planet-hosting stars; and (iv) modelling the evolution of the planetary systems under the influence of these stellar encounters.

We use \nb{} \citep{1p, wang2016} to model the evolution of the stellar population in the star clusters. \nb{} is based on the earlier $N$-body codes \nbody{} \citep{1j} and \nbodyplus{} \citep{1l}, but the main difference is its ability to use graphical processing units (GPUs). The parallelisation is achieved via \mpi \ (Message Passing Interface) \citep{mpi}, where both regular and irregular forces are parallelised. The GPU implementation in \nb{} provides a significant acceleration, especially for the long-range (regular) gravitational forces.  Stellar evolution is modelled following the models of \cite{hurley2000, hurley2002, hurley2005} using their algorithms for single stellar evolution \citep{hurley2013sse} and binary stellar evolution \citep{hurley2013bse}, and fallback and kicks for the formation of stellar remnants is modelled using the prescriptions of \cite{belczynski2002}.

The simulation data are stored in the \hdf{} format \citep[see, e.g.,][]{hdf5}, which is a highly efficient storage scheme that can be used for reconstructing the dynamical properties of the star clusters with high temporal and spatial accuracy for further analysis \citep{bts}.

We use the approach described in \cite{1i} to model the evolution of the planetary systems in \rebound{} \citep{1o}, by including the tidal force of the nearest neighbour to each planetary system at any time.
The \gpu-accelerated pseudo-gravitational dynamics interface \caigpu{} \citep{1i} loads the output data stored in the \hdf{} files. Subsequently, \caigpu{} loads the snapshots at two adjacent times during integration, while the particle states are interpolated in parallel on the GPU using a set of seventh-order septic splines. Finally, \rebound{} transfers the perturber data to \caigpu{} and vice-versa, and the planetary systems are evolved until the simulation ends.

As the star cluster evolves, stars can escape from the cluster through tidal evaporation or dynamical ejection. Following \cite{gnbs}, we identify stars with a cluster-centric radius of $r > 2\rtidal$ as escapers, where $\rtidal$ is the tidal radius of the star cluster. We follow the approach \cite{1i} for the identification of planets escaping from their host star: a planet is considered as an escaper when its eccentricity (relative to the host star) is sufficiently large ($e > 0.995$). Due to current limitations of the code, we do not follow the further evolution of free-floating planets in this study.
Physical collisions may occur when two bodies in a planetary system experience a sufficiently close approach. Whenever the distance between two bodies is smaller than the sum of their radii, the two bodies merge. In such events, the two bodies are replaced by a merger product with a position and velocity of the centre of mass of the two bodies, and a mass equal to that of the sum of the masses of the two bodies.

\section{Results}

In this section we describe the star clusters evolution, the close encounters properties that affect planetary systems, and the planetary systems evolution. Unless mentioned otherwise, we describe results for our reference model C051E4.

\subsection{Star cluster evolution}

\begin{figure}
\includegraphics[width=0.5\textwidth,height=!]{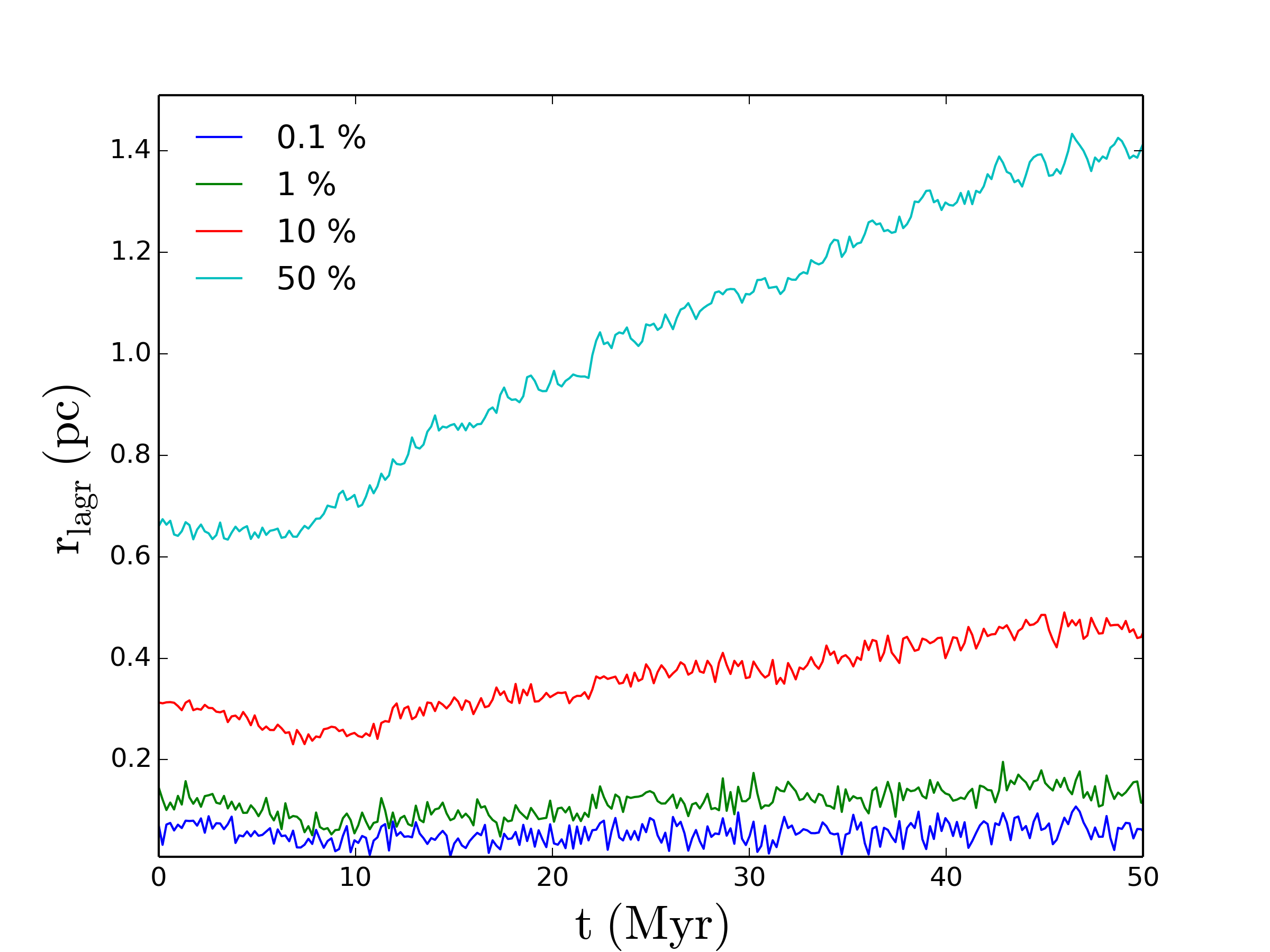}
\caption{The Lagrangian radii evolution containing 0.1\%, 1\%, 10\%, and 50\% of the {\em initial} star cluster mass for model C051E4.   \label{figure:clusterevolution}  }
\end{figure}

The dynamical evolution of star clusters is primarily characterised by the crossing time ($\crossingtime$), the half-mass relaxation time ($\relaxationtime$), the mass segregation timescale ($\segregationtime$), the stellar evolution timescale, and by the Lagrangian radii relative to the tidal radius ($\rtidal$). 
The crossing time corresponds to the time in which a star with a typical velocity travels through the cluster under the assumption of virial equilibrium: 
\begin{equation}
\crossingtime = \sqrt{\frac{2 \halfmass^3}{G \mcluster }}
\end{equation}
\citep[e.g.,][]{spitzer, lamers2005, 3b}. Here, $\halfmass$ is the half-mass radius, $G$ is the gravitational constant, and $\mcluster$ the total mass of the star cluster. 
The half-mass relaxation time corresponds to the time over which the cumulative effect of stellar encounters on the velocity of a star becomes comparable to the star's velocity itself:
\begin{equation}
\relaxationtime \approx \frac{0.138\nstars}{\ln(\Lambda)} \sqrt{\frac{\halfmass^3 }{G \mcluster }}
\end{equation}
\citep[e.g.,][]{spitzer, khalisi2007}. The parameter $\Lambda$ depends on the stellar density distribution in the cluster  \citep[see][]{spitzer}, and a value of $\Lambda = 0.4 \ \nstars$ is often considered appropriate for intermediate-mass star clusters.
Finally, the mass segregation time scale (or energy equipartition timescale) describes how fast energy is exchanged between stellar population of different masses \citep[see, e.g.,][]{spurzem1995, 3a}. 
This energy exchange results in the gradual migration of massive stars to the cluster core, and of low-mass stars to the outskirts of the star cluster. The mass segregation timescale is proportional to the relaxation time,
\begin{equation}
\segregationtime = \frac{m}{\langle m \rangle} \relaxationtime 
\end{equation}
\citep{khalisi2007}, where $m$ is the stellar mass under consideration, and $\langle m \rangle$ is the average stellar mass. 

The relevant dynamical timescales are listed in Table~\ref{table:initialconditions}, and the evolution of the Lagrangian radii of cluster model C051E4 is shown in Figure~\ref{figure:clusterevolution}. Any spatial or kinematic substructure is removed on the order of several crossing times, and the clusters also obtain a state of virial equilibrium on these timescales \citep[e.g.,][]{allison2009}. The clusters expand on the order of an initial half-mass relaxation time, resulting in the cluster members to escape. At the same timescale, more massive stars sink to the centre, while lower-mass stars tend to migrate to the outskirts. Stellar evolution also affects the star cluster at a similar timescale, and reduces the mass spectrum and the gravitational potential of the cluster. The latter leads to an expansion of the cluster and a reduction of the tidal radius, which facilitates additional escape beyond the tidal radius, which is $\rtidal=25.5$~pc for our reference model C051E4.

\begin{figure}
\begin{tabular}{c}
  \includegraphics[width=0.5\textwidth,height=!]{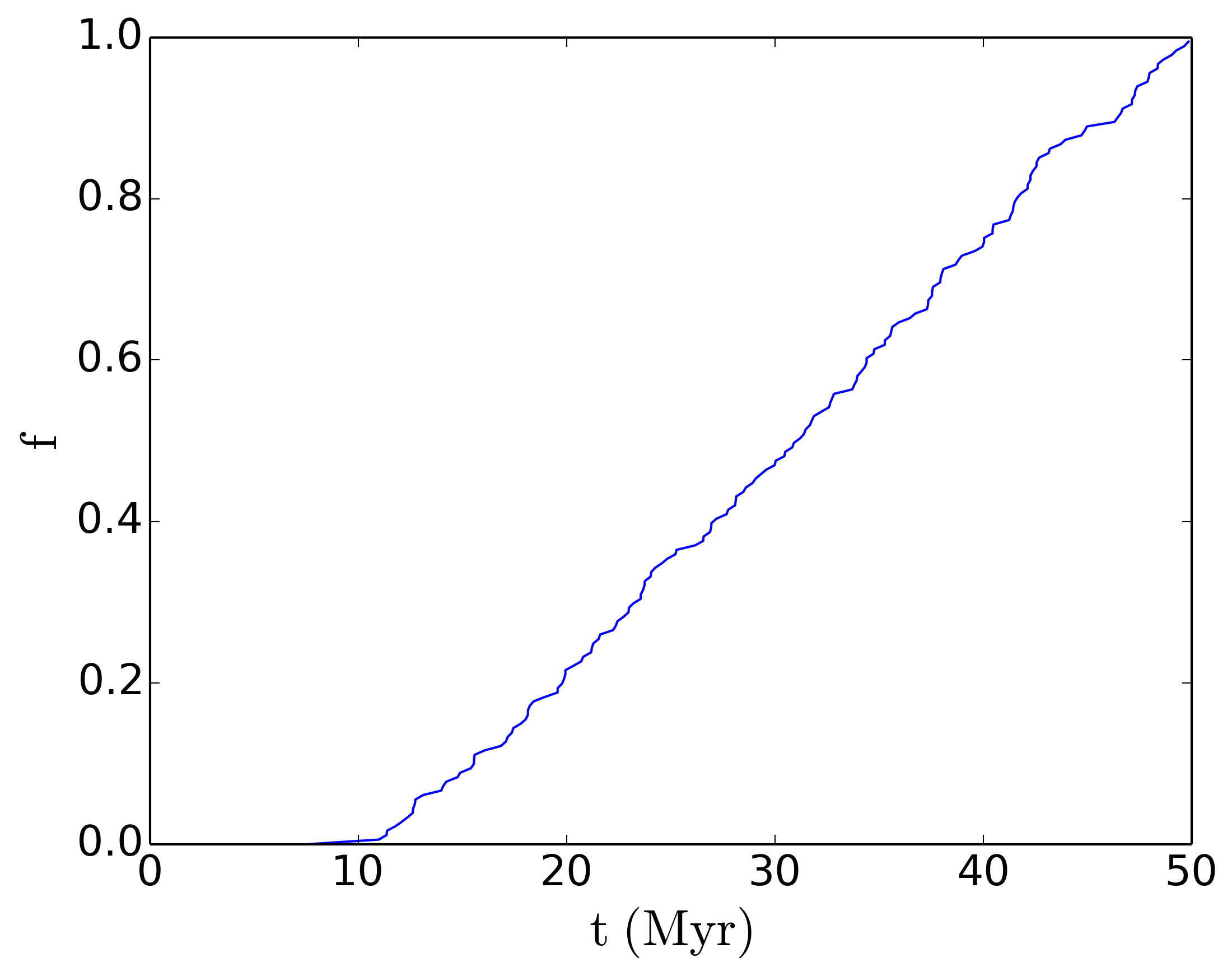} \\
  \includegraphics[width=0.5\textwidth,height=!]{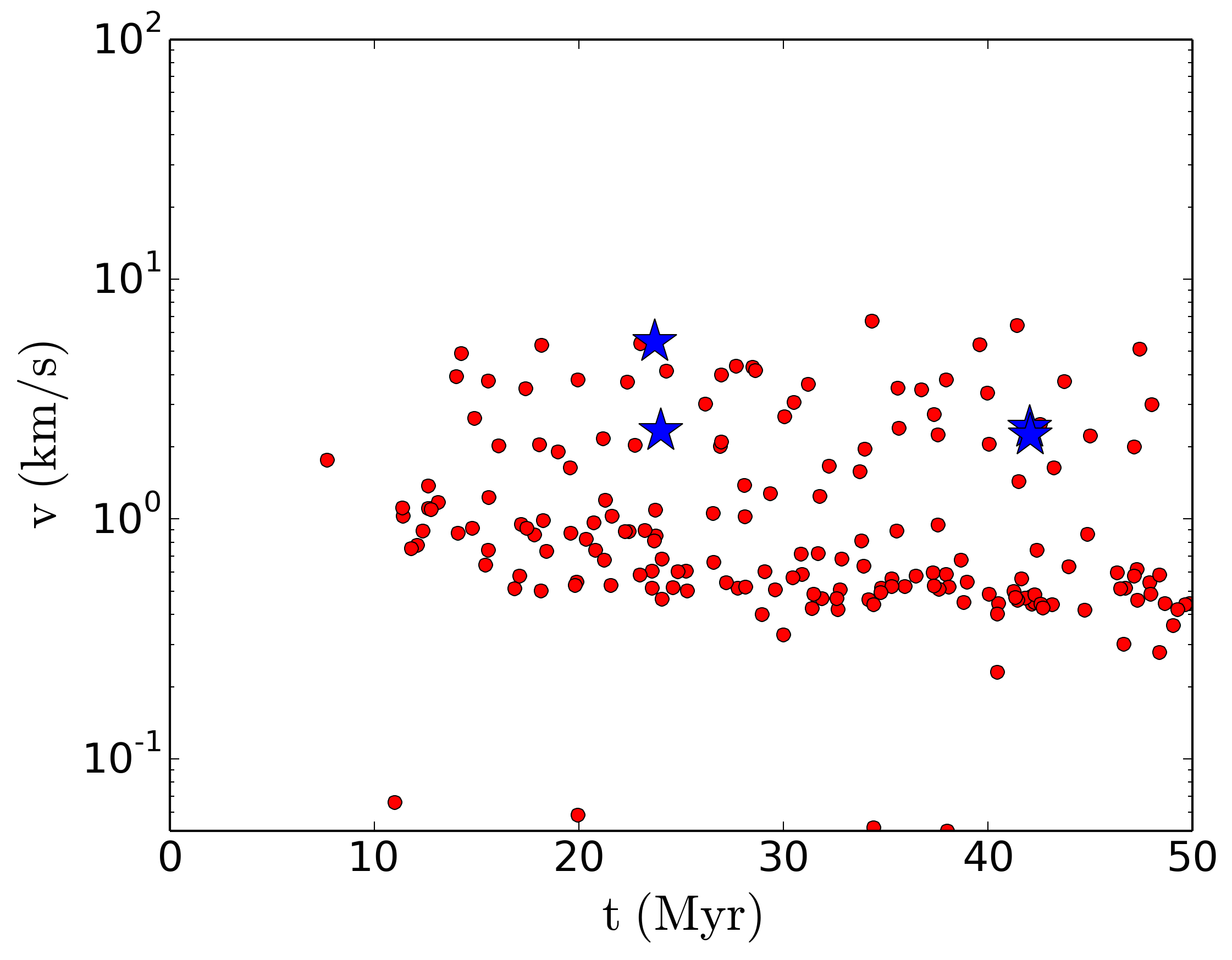}
\end{tabular}
\caption{{\em Top}: cumulative distribution of the stars escaping time, for model C051E4. {\em Bottom}: velocity-at-infinity distribution of escaping stars for model C051E4. Blue stars indicate escaping planet-hosting stars and red circles indicate other escaping stars. The initial three-dimensional velocity dispersion at the half-mass radius is 3.6\,\kms{} for this model.}
\label{figure:escapers}
\end{figure}

Escaping stars can be roughly categorised into two categories: (i) the high-velocity single stars or binaries ejected from the cluster core \citep[e.g.,][]{3c}, and (ii) the stars evaporated from the star cluster outskirts. The properties of the escaping stars in model C051E4 are shown in Figure~\ref{figure:escapers}. As the cluster expands and fills its Roche lobe during its first $\sim 10$~Myr, the tidal field gradually strips off stars from its outskirts. The first stars escape at $t\approx 10$~Myr, followed by a nearly constant escape rate for the remaining simulation time. Escape velocities range from near-zero to roughly 10~\kms. As the star cluster's mass decreases following stellar evolution and escape of cluster members, the gravitational potential is reduced and consequently the typical escape velocity decreases slightly over time. Among the escapers are several planet-hosting stars. During the early phases of evolution, these planetary systems remain mostly intact, while at later stages, high-velocity escaping planetary systems tend to be  perturbed (see below).

\subsection{Stellar encounters} \label{stenco}

We model the evolution of the planetary systems under the influence of the tidal force of neighbouring stars. As the tidal force on a planetary system decreases strongly with distance, $r$, between the host star and the neighbour star ($\propto r^{-3}$), we only model the effect of the nearest neighbour, following the approach of \cite{1i}. As we do not include primordial stellar binaries in our simulations, and as stellar binary formation through capture is rare, the trajectory of (almost) all neighbouring stars can be approximated with hyperbolic orbits. In reality, small deviations may be present due to (i) the presence of the other stars in the cluster, and to (ii) the gravitational pull of the other planets in the planetary system. In most cases, however, these two contributions can be neglected, and a hyperbolic trajectory is a good approximation during each close encounter.

\subsubsection{Quantifying the effect of stellar encounters}

Consider a close encounter between a neighbouring star with mass $m_n$, and a host star with mass $m_h$ that hosts a planet of mass $m_p\ll m_h$. The total energy of such a close encounter is
\begin{equation}
E = \frac{1}{2} \mu v^{2}_{rel} - \frac{G m_{hp} m_n}{r_{rel}}
\end{equation}
where $m_{hp}=m_h+m_p$, $\mu=m_{hp} m_n(m_{hp}+m_n)^{-1}$  is the reduced mass, and $r_{rel}$ and $v_{rel}$ are the relative position and velocity, and $G$ is the gravitational constant. The two principal parameters that describe the trajectory are the hyperbolic semi-major axis,
\begin{equation}
a = - \frac{G m_{hp} m_n}{2 E}  \ ,
\end{equation}
and the hyperbolic eccentricity,
\begin{equation}
e = \left[ \left(1 - \frac{r_{rel}}{a}\right)^{2} + \frac{(\vec{r}_{rel} \cdot \vec{v}_{rel})^2}{a G (m_{hp}+m_n)}\right]^{\frac{1}{2}}  \ .
\end{equation}
Long before the encounter occurs, the relative velocity-at-infinity, $v_{\infty}$, is
\begin{equation}
v_{\infty} = \sqrt{\frac{G (m_{hp}+m_n)}{p} (e-1)} \ ,
\label{eq:vinfty}
\end{equation}
and the impact parameter $b$ of the encounter is
\begin{equation}
b = p \sqrt{1+ \frac{2 G (m_{hp}+m_n)}{p \, v_{\infty}^{2}}} \ .
\end{equation}
The closest approach distance between the two stars (the periastron distance) is evaluated as $p =|a|(e-1)$, and the velocity during periastron passage is 
\begin{equation}
v_p = \sqrt{\frac{G (m_{hp}+m_n)}{|a|} \frac{e+1}{e-1}} = \sqrt{v^2_{\infty} + \frac{2 G (m_{hp}+m_n)}{p}}  \ .
\label{eq:periastronvelocity}
\end{equation}
The strength of a stellar encounter on a planetary orbit is determined by the time-dependent distance of the perturber, the perturber mass, and the semi-major axis of the planet. The encounter strength parameter, $k_p$, is often used for characterising the effect of a close encounter on a binary system \citep[see, e.g.,][]{k3,k2,k1,1g}:
\begin{equation}
k_p = \sqrt{\frac{2 m_{hp}}{m_{hp}+m_n} \left(\frac{p}{a_p}\right)^3  } 
\approx  \left(\frac{p}{a_p}\right)^{3/2} 
\  .
\label{eq:kparameter}
\end{equation}
Here we have added a subscript to indicate the dependence on semi-major axis. The approximation in the right-hand side of Eq.~({\ref{eq:kparameter}) is valid for equal-mass stars ($m_{hp}\approx m_n$).
For a given stellar encounter, the $k_p$ parameter decreases with the planetary semi-major axis as $k_p\propto a_p^{-3/2}$. Smaller values of $k_p$ imply a stronger effect of the encountering star on the planetary system. 
Different planets experience difference encounter strengths. For example, for a given stellar encounter, the $k_p$ parameter for Earth is roughly a factor $165$ larger than that of Neptune.

The effect of different stellar encounters on a planetary can be compared using (i) the tidal encounter strength, (ii) the hyperbolic eccentricity of the orbital trajectory of the encountering star, and (iii) the duration of the encounter. These parameters can be expressed using the encounter strength parameter $k_p$ and the dimensionless relative velocity-at-infinity, $\tilde{v}_{\infty}$,
\begin{equation}
 \tilde{v}_{\infty} = v_{\infty} \left( \frac{G(m_{hp}+m_n)}{a_p} \right)^{-1/2} \ .
 \label{eq:vinftilde}
 \end{equation}
\citep[cf.][]{k1,1g}.
Tidal encounters cause weak, long-lasting perturbations on a planetary system, while for impulsive encounters, the interaction timescale is strong, and usually shorter than the planetary orbital period. When $p \gg a_p$ the regime is {\em tidal} and when $p \le a_p$ is {\em impulsive}. Setting $p = a_p$ in Eq.~(\ref{eq:kparameter}) gives the condition 
\begin{equation}
k_p = \sqrt{\frac{2 m_{hp}}{m_{hp}+m_n}  } \ \approx  1\  ,
\end{equation}
where the approximation is valid for encounters between equal-mass stars ($m_{hp}\approx m_n$). 
Encounters with hyperbolic eccentricity $e>1$ are {\em hyperbolic}, while those with $e=1$ are {\em parabolic}. Following \cite{k1} we distinguish near-parabolic ($1<e<2$) from hyperbolic orbits ($e\ge 2)$. Inserting $e=2$ in Eq.~(\ref{eq:vinfty}) and substituting the resulting expression for $v_\infty$ in Eq.~(\ref{eq:vinftilde}) gives 
\begin{equation}
\tilde{v}_{\infty}=\sqrt{ \frac{a_p}{p}} \ \approx \ k_p^{-\frac{1}{3}}  \ , 
\end{equation}
where the approximation on the right-hand side is obtained using Eq.~(\ref{eq:kparameter}) for the case when the host star and the encountering neighbour are of equal mass ($m_{hp}\approx m_n$). 
Finally, a comparison between the encounter timescale and the orbital timescale defines the {\em adiabatic} and {\em non-adiabatic} regimes. When the duration of the stellar encounter is much longer than the planetary orbital period, the encounter is {\em adiabatic}, and otherwise it is {\em non-adiabatic}. When equating the timescale of close approach, $p/v_p$, to the orbital timescale, $a_p (Gm_{hp}/a)^{-1/2}$, Eq.~(\ref{eq:vinftilde}) gives
\begin{equation}
\tilde{v}_{\infty} = 
\sqrt{2}
\left( \frac{p}{a_p} \right)
 \sqrt{ 
\frac{m_{hp}}{2(m_{hp}+m_n)} 
- \left(\frac{p}{a_p}\right)^{-3} 
}
\end{equation}
for the boundary that separates adiabatic and non-adiabatic encounters. In the case of equal-mass stars ($m_{hp}\approx m_n$), the above expression reduces to the curve shown in figure~1 of  \cite{1g}. When expressed in terms of $k_p$, the criterion separating these regimes is described by the curve
\begin{equation}
\tilde{v}_{\infty} = k_p^{2/3} \left(\frac{m_{hp}+m_n}{2m_{hp}}\right)^{\frac{1}{3}}  \left( \frac{m_{hp}}{m_{hp}+m_n} - \frac{4m_{hp}}{k_p^2(m_{hp}+m_n)} \right)^{\frac{1}{2}}
\end{equation}
which, in the case of nearly equal-mass stars, reduces to 
\begin{equation}
\tilde{v}_{\infty} \approx k_p^{2/3}   \left( \frac{1}{2} - \frac{2}{k_p^2} \right)^{\frac{1}{2}} \ .
\end{equation}

We remind the reader that the analysis above was developed under the assumption that only three bodies are involved in the interaction. In our study, this is in the  majority of the cases a good assumption, i.e., in those cases where (i) gravitational planet-planet interactions can be ignored during the encounter, and (ii) the tidal influence of the other cluster member stars is small compared to that of the encountering stars.

\subsubsection{Analysis of stellar encounters} 

\begin{figure}
\includegraphics[width=0.5\textwidth,height=!]{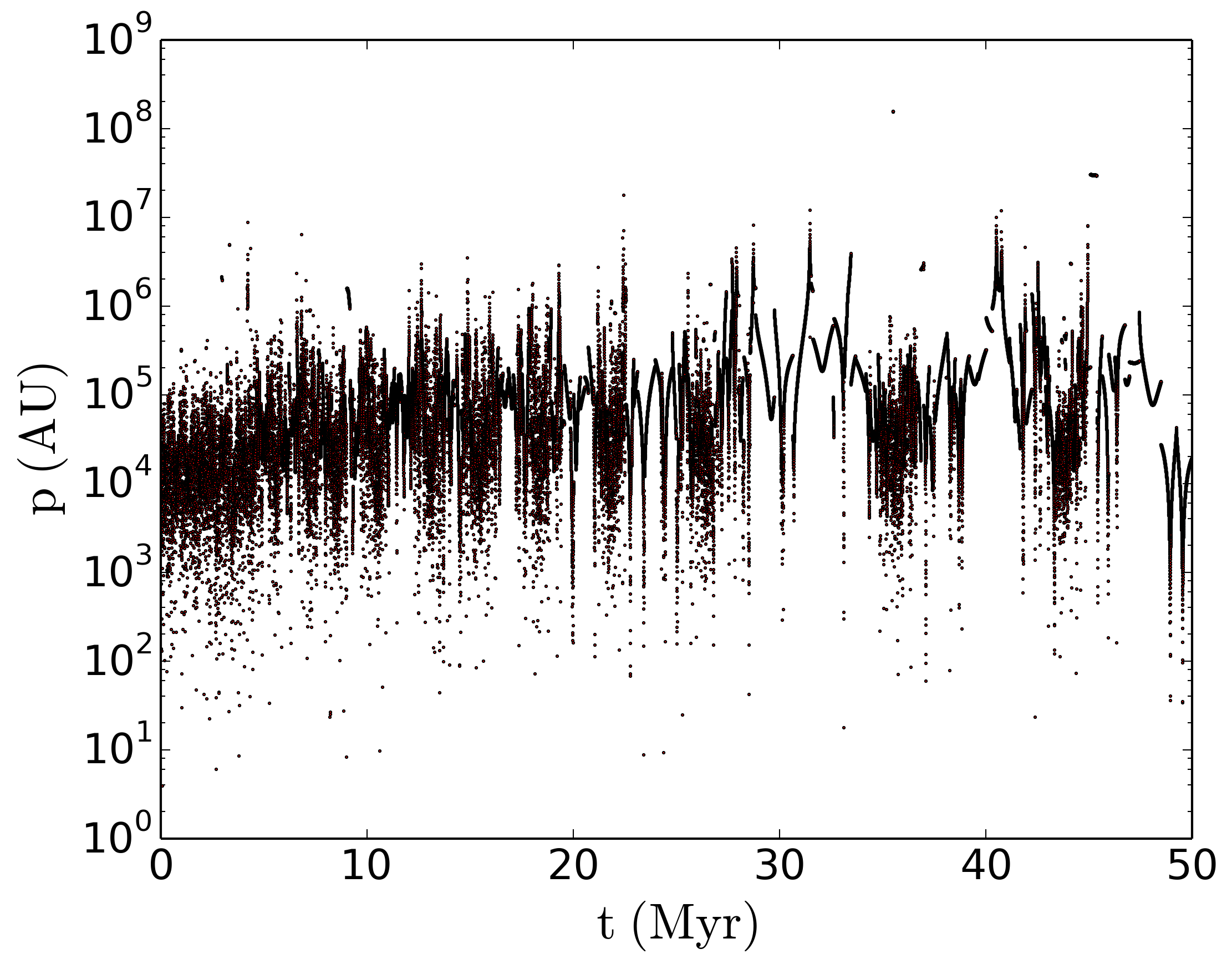}
\caption{Temporal distribution of the instantaneous periastron distance $p$ for stellar encounters with nearest neighbours experienced by planetary system P010 in  model~C051E4.}
\label{figure:tvp}
\end{figure}

\begin{figure}
\includegraphics[width=0.5\textwidth,height=!]{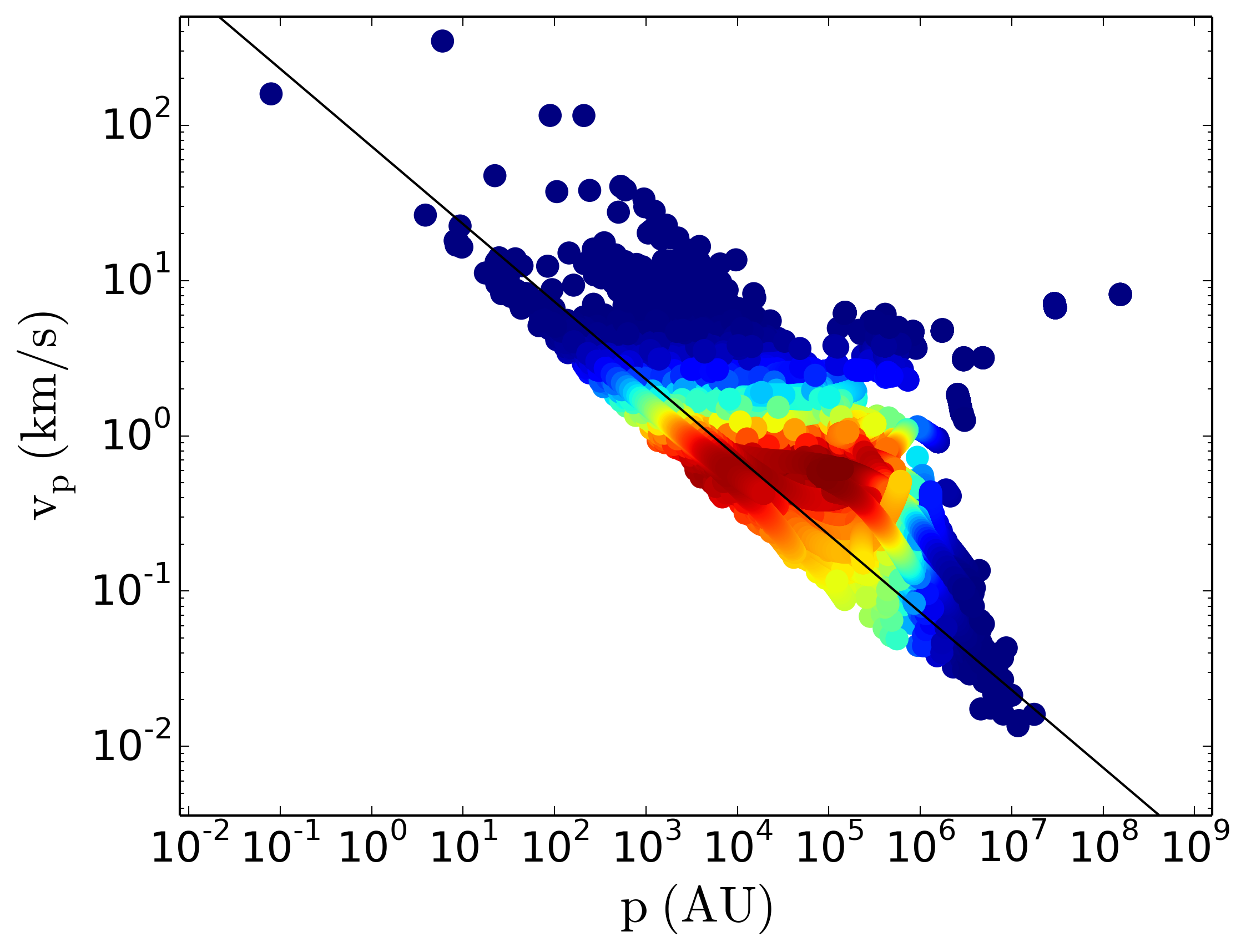}
\caption{Distribution of the instantaneous periastron distance $p$ and periastron velocity $v_p$ computed for the nearest neighbour of planetary system P010 in  model~C051E4. The blue curve separates near-parabolic encounters (below the curve) and hyperbolic encounters (above the curve). The density distribution is indicated with colours, where the most frequent types of encounters are indicated in red. } 
\label{figure:pvp}
\end{figure}

\begin{figure}
\begin{tabular}{c}
   \includegraphics[width=0.45\textwidth,height=!]{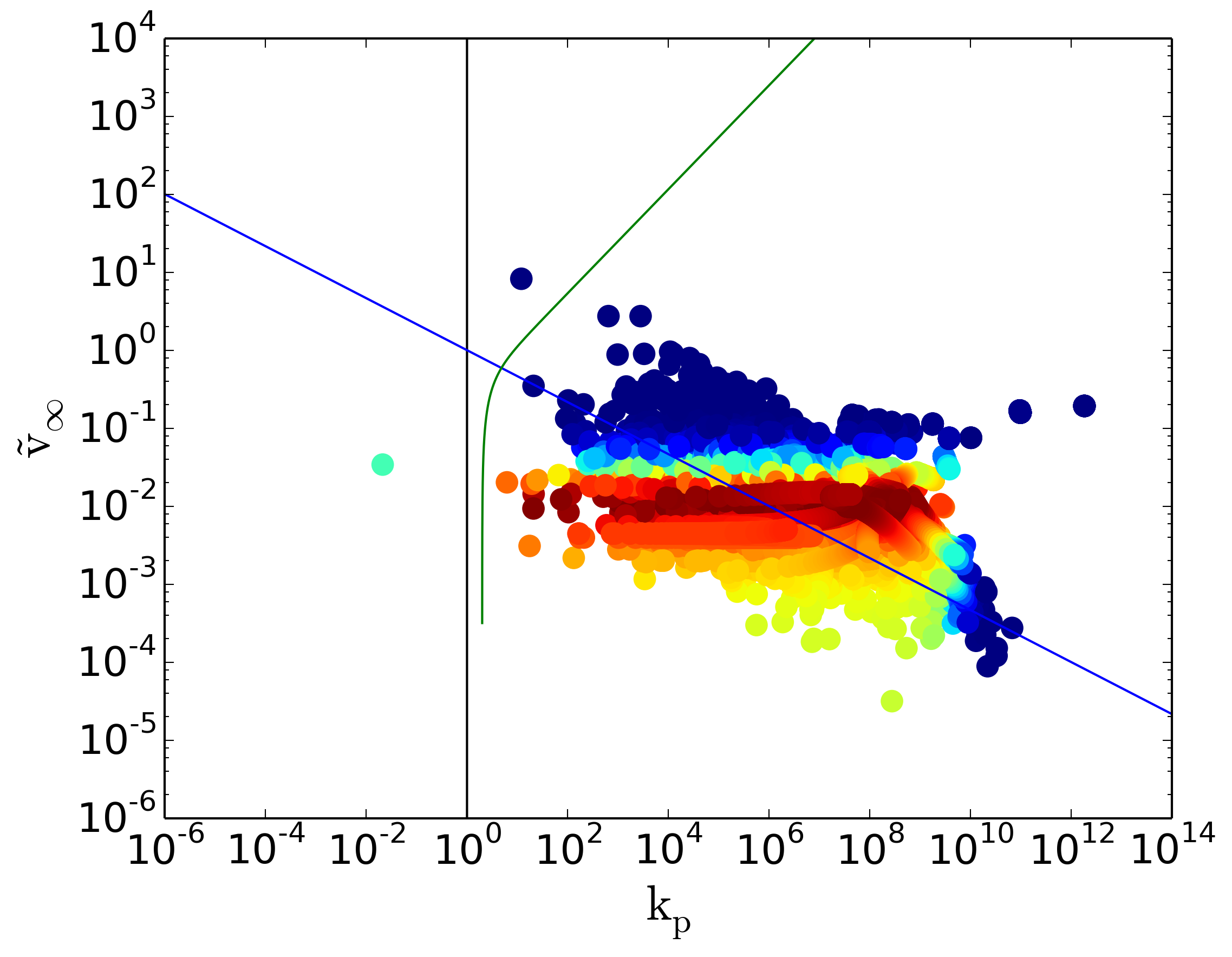} \\
   \includegraphics[width=0.45\textwidth,height=!]{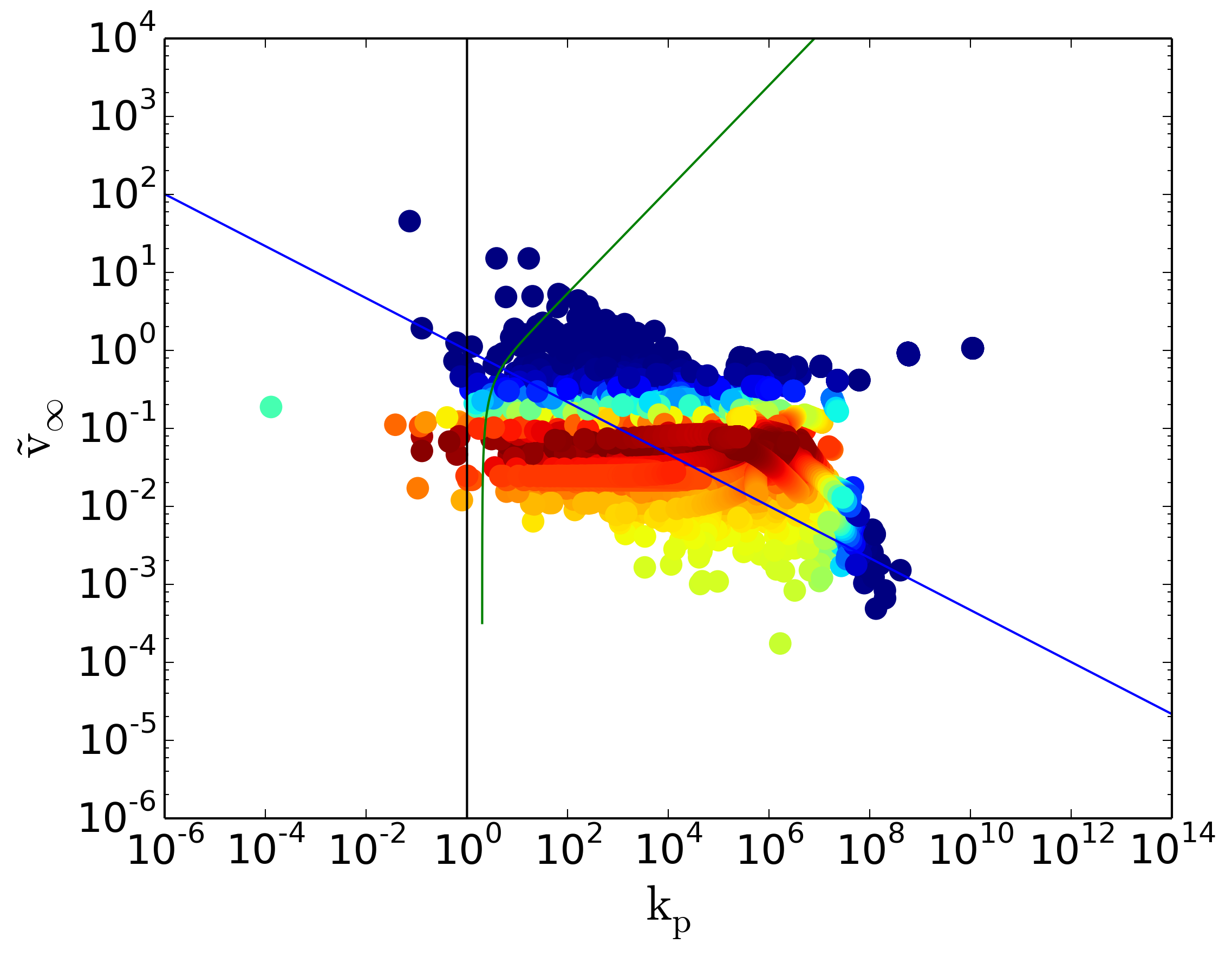}
\end{tabular}
\caption{The distributions of the encounter strength parameter $k$ of the nearest stellar encounters versus time, experienced by planetary system P010 in  model~C051E4, for the planets Earth ({\em top}) and Neptune ({\em bottom}). The black curve separates the {\em tidal} (right) and {\em impulsive} (left) encounters, the blue curve separates the  {\em hyperbolic} (above the curve) and {\em near-parabolic} encounters (below the curve), and the green curve separates  {\em adiabatic} (right) and {\em non-adiabatic} (left) encounters. The curves are obtained from the properties of individual stellar encounters. The different colours indicate the distribution density, with the most frequent encounters indicated in red. }
\label{figure:kpar}
\end{figure}

The encounter properties experienced by a typical planetary system in star cluster model~C051E4 are illustrated in Figures~\ref{figure:tvp}, \ref{figure:pvp}, and~\ref{figure:kpar}. Encountering stars with small values of $p$, $v_p$, $\tilde{v}_{\infty}$, and $k_p$ are typically the most effective perturbers, but uncommon.

During the first relaxation time, the average periastron distance slightly increases due to expansion of the star cluster. The planetary system illustrated in Figure~\ref{figure:tvp} experiences several epochs of durations $1-5$~Myr in relative isolation, while orbiting in the star cluster outskirts. The instantaneous hyperbolic periastron distance varies between 10~AU (during a close encounter) and $10^6$~AU (the size of the star cluster halo). Note that the values illustrated in this figure represent the values of $p$ as calculated for the hyperbolic encounter at each time; this does not necessarily imply that all neighbour stars reach the periastron distance in their approach, as they can be perturbed by other neighbouring stars. 

The distribution over hyperbolic periastron distance and periastron velocity for the same system is shown in Figure~\ref{figure:pvp}. For an encounter of the host star system ($m_{hp}\approx 1~\msun$) with typical neighbour star of mass $m_n\approx 1~\msun$, Eq.~(\ref{eq:periastronvelocity}) reduces to $v_p \approx (3GM/p)^{-1/2}$, so that $v_p/$\kms$\,\approx 73\,(p/{\rm AU})^{-1/2}$. As most encounters have a periastron distance in the range of $10^4-10^5$~AU (Figure~\ref{figure:tvp}), stars typically experience periastron velocities of $0.2-0.8$~\kms{} during their encounter, as shown in Figure~\ref{figure:pvp}. The curve for $e=2$ is shown in Figure~\ref{figure:pvp}, for encounters between equal-mass stars. The encounters below this curve are near-parabolic ($1<e<2$). As mentioned above, not all neighbours complete their hyperbolic orbit while in near the star, occasionally resulting in measurements of $p$ and $v_p$ that are not realised. The data points in the upper-right region of the envelope result from unrelated flyby's while the host star resides in the outskirts of the star cluster.

The distribution over $k_p$ and $\tilde{v}_{\infty}$ is illustrated in Figure~\ref{figure:kpar} for all close encounters with planetary system P010 in star cluster model C051E4, with the three curves separating different types of stellar encounters.
The hyperbolic periastron distances are typically much larger than the planetary semi-major axis, so that almost all encounters in our simulations are tidal. 
Most encounters are hyperbolic, although the fraction of near-parabolic encounters is substantial. 
A large majority of the stellar encounters is adiabatic, but non-adiabatic encounters occasionally occur. Note that, when compared to the terrestrial planets, the outer planets experience more frequent non-adiabatic and impulsive encounters due to their larger semi-major axes. These outer planets experience stronger perturbations from neighbouring stars.

\subsection{Planetary system evolution}

\subsubsection{Isolated planetary systems}

\begin{figure}
\begin{tabular}{c}
  \includegraphics[width=0.45\textwidth,height=!]{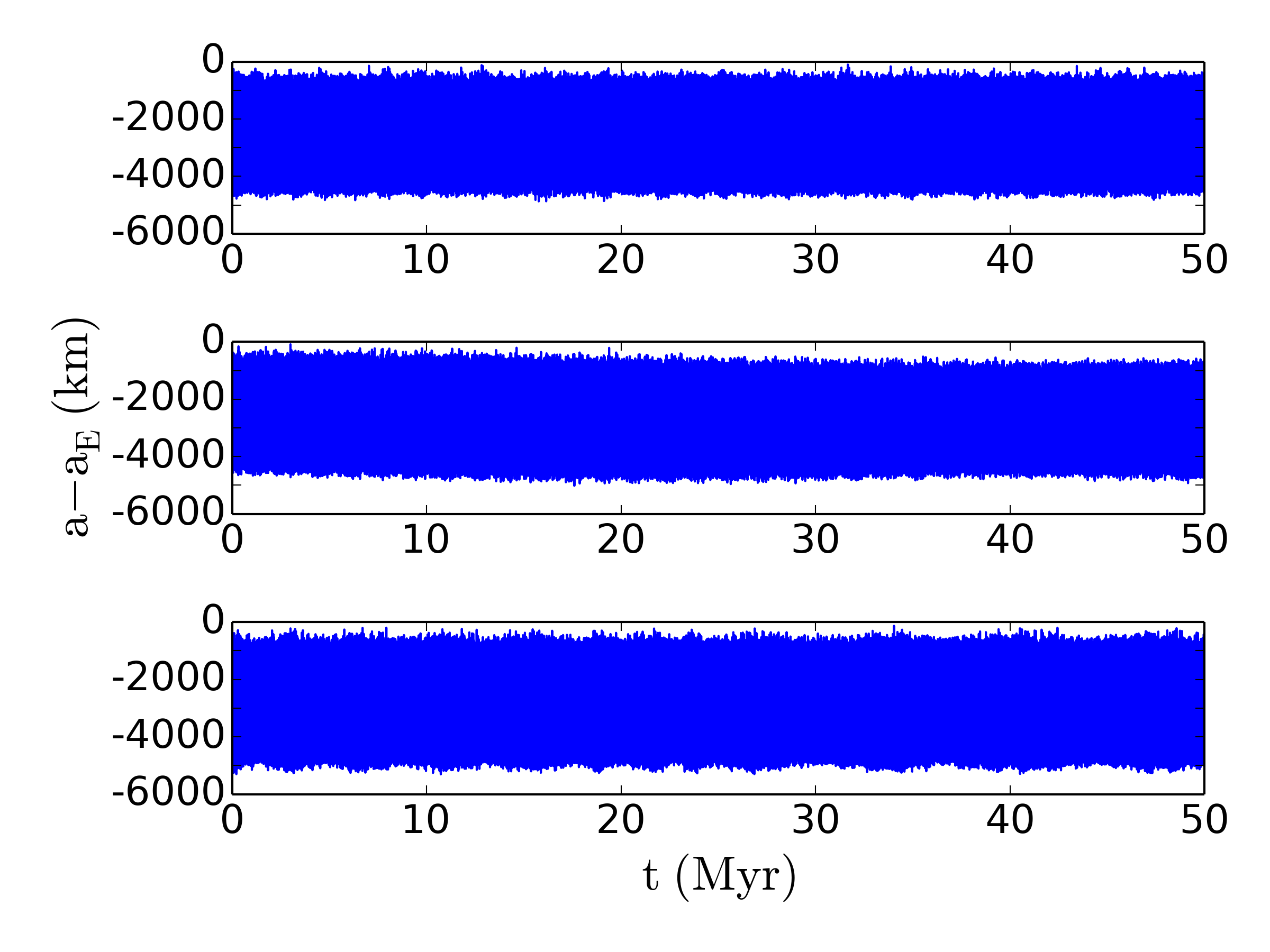} \\
  \includegraphics[width=0.45\textwidth,height=!]{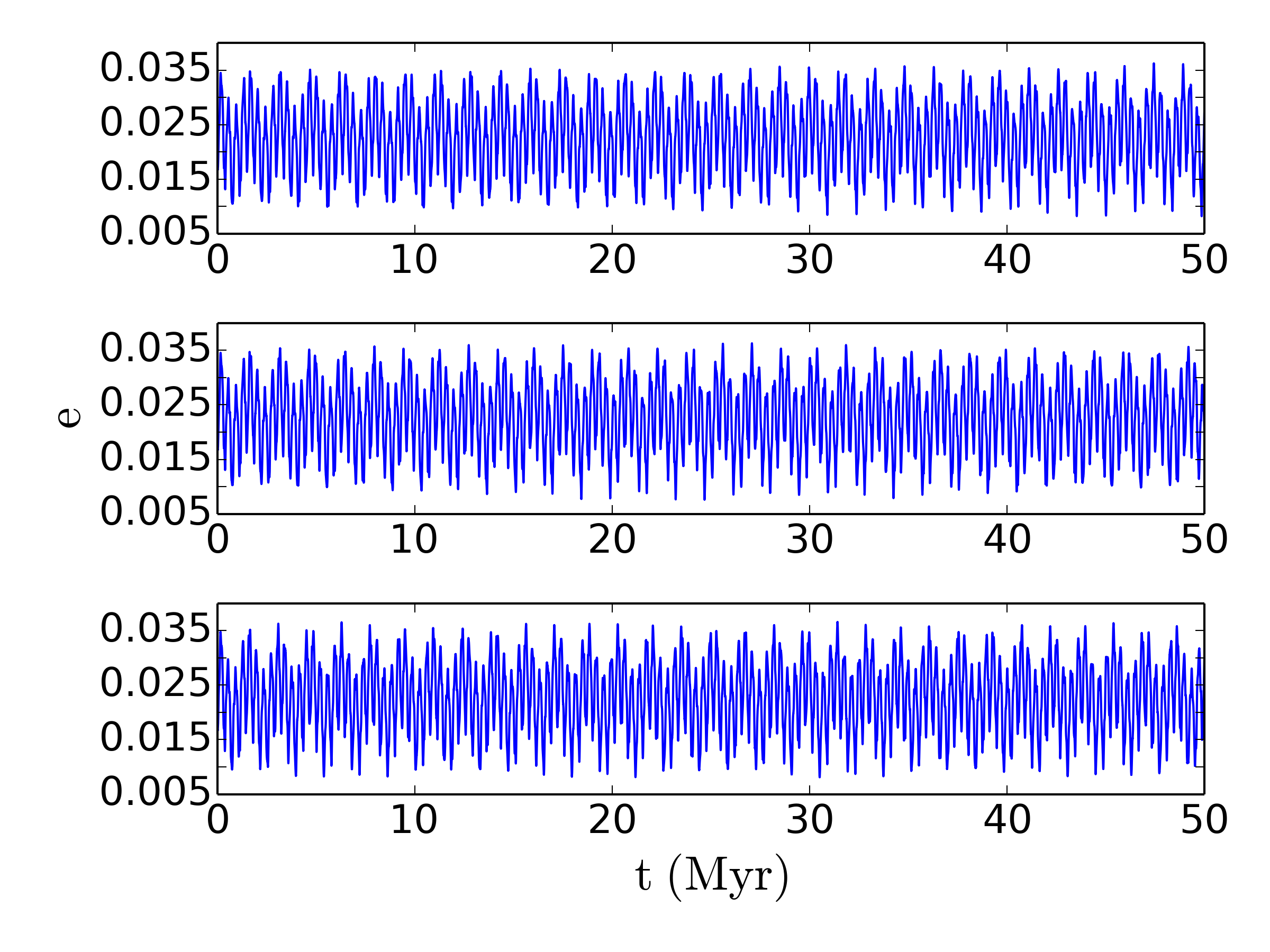}
\end{tabular}
\caption{Evolution of the semi-major axis and eccentricity of planet Earth under the influence of the other planets, when ignoring the effects of external perturbations, for multi-planet systems with a host star mass of $0.93~\msun$ ({\em top}),  $1~\msun$ ({\em middle}), and $1.03~\msun$ ({\em bottom}).}
\label{figure:noclu}
\end{figure}

Planetary systems evolve over time due to instabilities caused by internal and external effects. To distinguish between these two effects, we carry out simulations of isolated planetary systems for an identical simulation time (50~Myr) using \rebound.  The initial conditions for the planetary system is identical to those mentioned above, but with host star masses of $0.93~\msun$, $1.0~\msun$, and $1.03~\msun$, encompassing the range of host stars masses in our star cluster simulations. The results are shown in Figure~\ref{figure:noclu}, and indicate that there are no significant changes; the planetary system architectures remain unchanged for 50~Myr, apart from periodic secular evolution. The evolution seen for the planetary systems in the cluster is thus a direct or indirect result of close encounters with neighbouring stars and not related to internal perturbations as of the planetary system alone.

\subsubsection{Survival rates of planetary systems in clusters}

\begin{figure}
\includegraphics[width=0.5\textwidth,height=!]{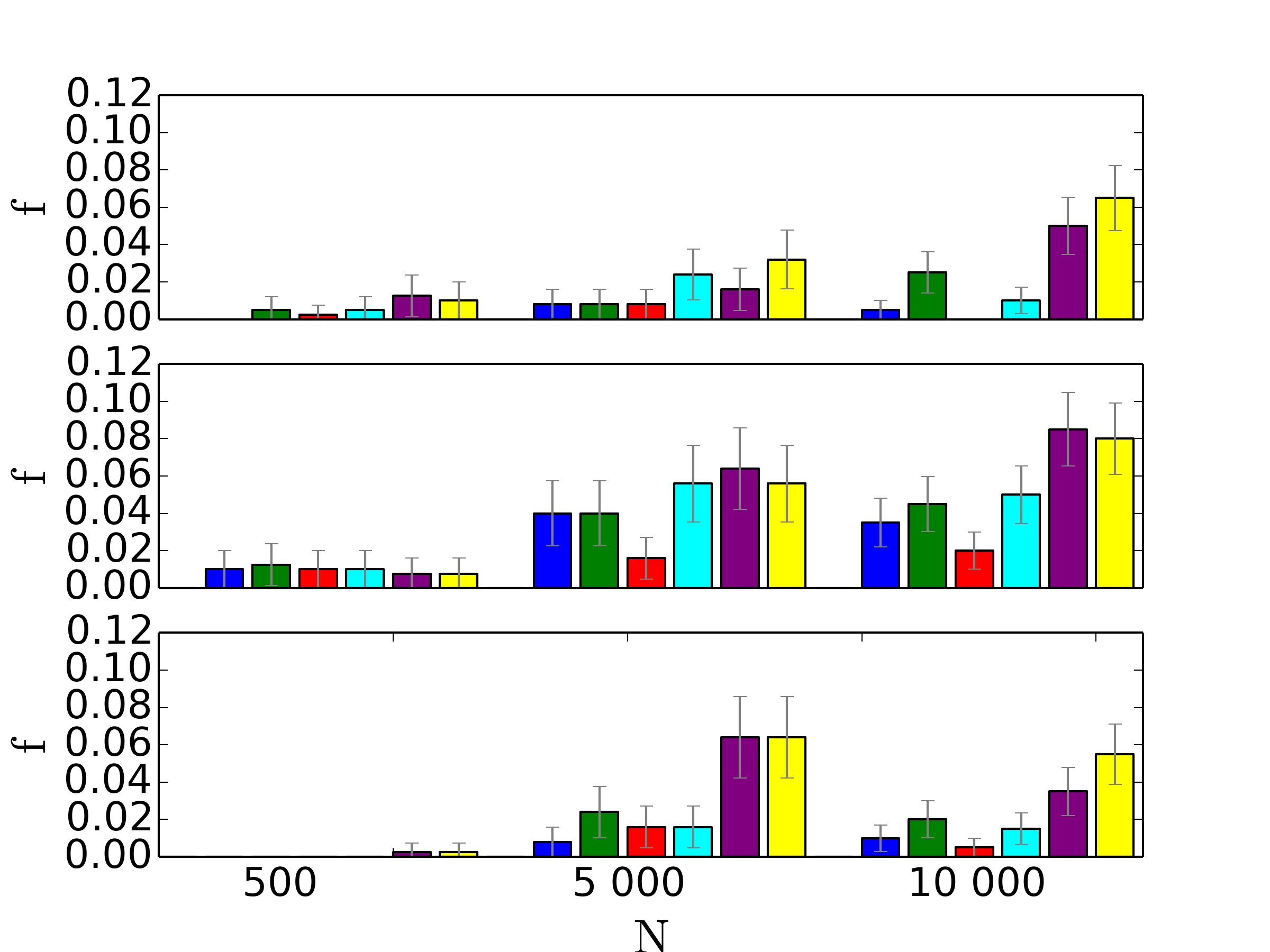}
\caption{The fraction of planets that escapes from their planetary system through an during the first 50~Myr, for star cluster models with different initial masses and different virial ratios. The top, middle, and bottom panels show the results for models with $Q=0.4$, 0.5, and 0.6, respectively. The different planets are indicated, from left to right, with dark blue (Earth), green (Mars), red (Jupiter), light blue (Saturn), purple (Uranus) and yellow (Neptune). }
\label{figure:survival1}
\end{figure}

The local stellar density at a given time is the main factor that determines the encounter frequency and encounter properties, and hence the determining factor for the planetary escapes.
Planets in the outer regions of the planetary system experience the strongest perturbations from encountering stars, as the have the largest semi-major axes and therefore the smallest values for $k_p$. Jupiter and Saturn have the highest masses in the system, and are therefore the most stable against planet-planet scattering, although in some cases they also escape. 
In accordance to similar results for equal-mass systems \citep[e.g.,][]{1i}, our results show a tendency for wider-orbit planets to have lower survival rates, although a noticeable mass dependence on planetary mass is also seen.

Figure~\ref{figure:survival1} shows the fraction $f_p=N_{e,p}/N_p$ among the ensemble of planets that experiences an escape event within $t=50$~Myr, for star clusters with different initial conditions. Here, the subscript $p$ refers to the planet for which the fraction is calculated, $N_{e,p}$ is the number of planets that have escaped and $N_p$ is the total number of planetary systems in the ensemble of simulations. Figure~\ref{figure:survival1} clearly shows the combined effect of mass and semi-major axis, and a strong correlation between neighbouring planets of similar mass, in particular for the star cluster of higher mass. Planetary siblings (Earth-Mars, Jupiter-Saturn, and Neptune-Uranus) exhibit similar trends, and this is a direct consequence of the effect of their masses.

The average probability for a planet to escape from its host star ranges from 0.3\% for low-mass, super-virial star clusters to 5.3\% for high-mass star clusters that start out in virial equilibrium. The most common types of escaper involves Uranus and Neptune. These escape events are either a direct consequence of a stellar encounter, or the indirect result of an encounter that induced a perturbation of Neptune's orbit and subsequent secular interactions with Uranus.  The vast majority of escape events of the planets Earth and Mars are a consequence of Jupiter being directly perturbed by a close encounter with a neighbouring star, while only a small fraction of the planets Earth and Mars escapes as the result of an even closer encounter with a neighbouring star or an interaction with a perturbed outer planet. The survival rate of terrestrial habitable-zone planets can, for Solar system-like architectures, thus be linked to properties of Jupiter-like planets at a larger semi-major axis.

Under otherwise identical conditions, star clusters that initialised in virial equilibrium are most hostile to the survival of planetary system, as compared to those with initial sub-virial and super-virial states. Star clusters that are virially cool or hot tend to evolve towards virial equilibrium within one or two crossing times. In both cases, this process results in an expansion of the cluster, which increases the half-mass relaxation time, reduces the close-encounter rates, and hence to some degree increases the survival rates of planetary systems.

Similar to the results of \cite{1h}, we find no evidence for planet-planet collisions in our simulations. We observe several cases where a terrestrial planet merges with the host star; these all occur in the smaller ($\nstars=500$) star clusters. In each of these three cases, the merger results from orbital perturbations of the inner planets by Jupiter, typically several million years after a close stellar encounter has substantially altered the outer regions of the planetary system.

\subsubsection{Escaping planetary systems} 

\begin{table}
\caption{The fraction of planetary systems that have escaped from the star cluster at $t=50$~Myr, for all star cluster models.   }
\begin{tabular}{lccc}
\hline
Model & $N= 500$ & $N= 5000$ & $N= 10\,000$ \\
 \hline
$Q = 0.4$ & 5.0 $\pm$ 3.9  \% & $0.0^{+0.7}_{-0.0} \%$  & 1.0 $\pm$ 0.7 \%  \\
$Q = 0.5$ & 6.0  $\pm$ 4.2 \% & 0.8 $\pm$ 0.8 \% & 2.0 $\pm$ 0.9 \% \\
$Q = 0.6$ & 9.0 $\pm$ 5.7 \% & 3.2 $\pm$ 1.6 \% & 3.0 $\pm$ 1.2 \%\\
\hline
\end{tabular}
\label{table:eschs}
\end{table}

On several occasions, intact planetary systems escape from the star cluster, while most of the planet-hosting stars that escape from the cluster have strongly-perturbed planetary systems. The fraction of planet-hosting stars that escape from the star cluster during the first 50~Myr are listed in Table~\ref{table:eschs} for the different star cluster models. The fraction of escaping planetary systems decreases for larger $N$ due to the longer relaxation time \citep[e.g.,][]{3b,lamers2005}; see Table \ref{table:initialconditions}. As mentioned in the previous section, star clusters that are initially sub- or super-virial tend to rapidly evolve towards virial equilibrium, and experience expansion in the process, which also increases the relaxation time and decreases the escape rate over longer periods of time. Note however, that for virially hot clusters, this initial evolution towards virial equilibrium also involves a rapid loss of stars (and planetary systems) in the outskirts of the star cluster at early times. 

The results listed in Table~\ref{table:eschs} and corresponding trends should be interpreted with caution} because of the small number of planetary system escape events. Higher escape rates are seen for super-virial star clusters ($Q=0.6$), particularly during the first few million years. Whether or not a planetary system survives intact depends on the properties of close stellar encounter prior to, and during the escape event. Planetary systems that escape at earlier times, and those that escape with smaller speeds, tend to have a higher probability of escaping the star cluster intact (see also Section~\ref{section:trends}).

\subsection{Planetary system evolution}

\subsubsection{Time dependence}

\begin{figure}
\includegraphics[width=0.52\textwidth,height=!]{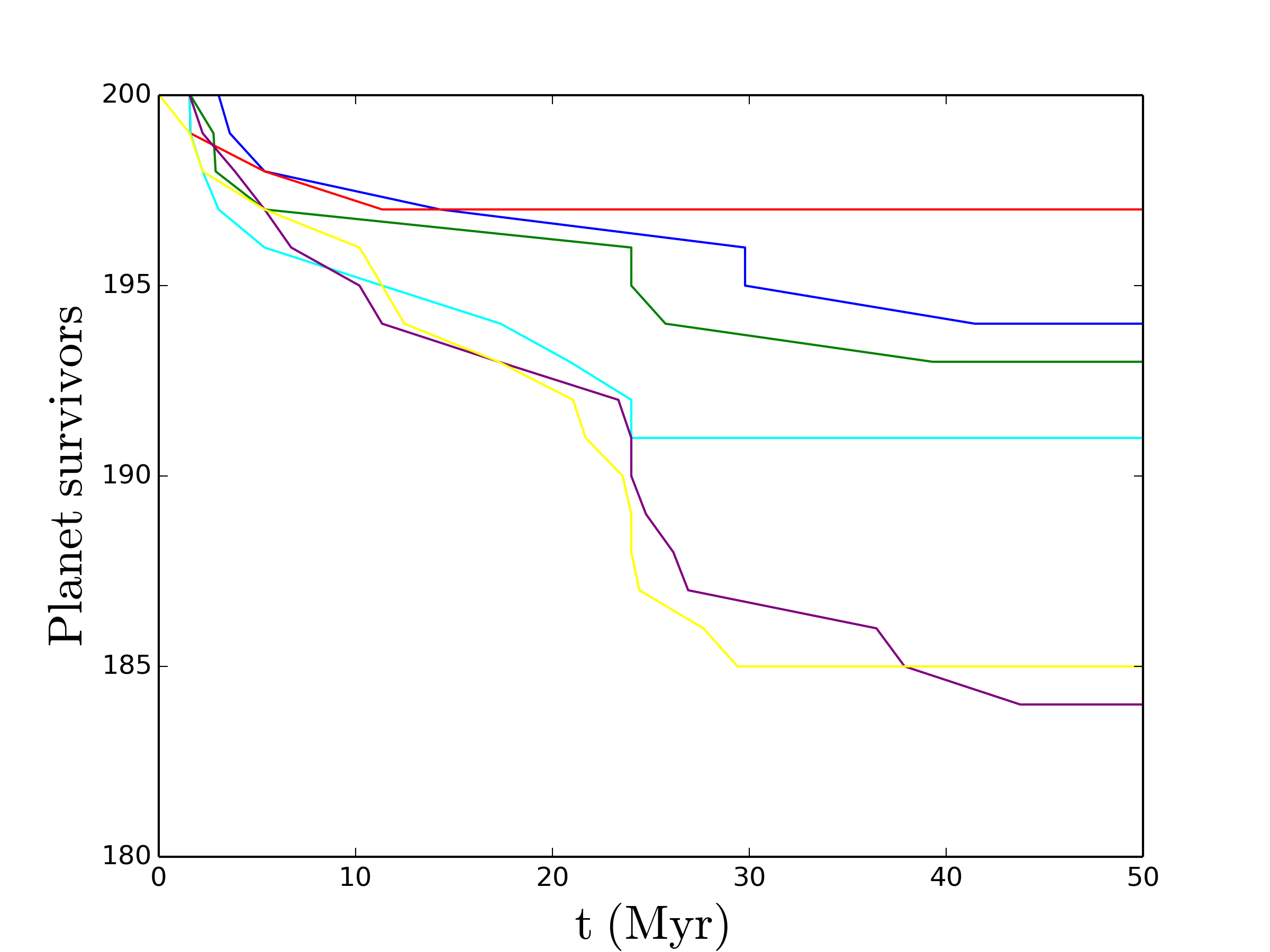}
\caption{Number of planets remaining bound to their host star as a function of time, for star cluster model C051E4. The curves show the remaining number of planets Earth (dark blue), Mars (green), Jupiter (red), Saturn (light blue), Uranus (purple) and Neptune (yellow). }
\label{figure:survival3}
\end{figure}

\begin{table}
\caption{Fraction of escapers among all the planets that were initially present in the planetary systems anymore at time 50~Myr, for model C051E4.  }
\begin{tabular}{lc}
\hline
Planet & Escape fraction \\
\hline
Earth    &  $3.0 \pm 1.2$\,\% \\
Mars     &  $4.5 \pm 1.5$\,\% \\
Jupiter  &  $1.5 \pm 0.9 $\,\% \\
Saturn   &  $5.0 \pm 1.6$\,\% \\
Uranus   &  $8.5 \pm 2.0$\,\% \\
Neptune  &  $8.0 \pm 1.9$\,\% \\
\hline
All planets  &  $5.3 \pm 0.6  \,\%$ \\
\hline
\end{tabular}
\label{table:escapefractions}
\end{table}

\begin{table}
\caption{The fraction of planetary systems with a certain number planets remaining in orbit around the star at time $t=50$~Myr, for star cluster model C051E4. The majority of the planetary systems remain intact (6 planets), although they may be somewhat perturbed by encounters, while none of the planetary systems in this model lose all their planets.} 
\begin{tabular}{ll}
\hline
Planets remaining & Fraction \\
\hline
6 planets    &  $88.5 \pm 2.3$\,\% \\
5 planets     &  $5.0 \pm 1.5$\,\% \\
4 planets  &  $1.0 \pm 0.7 $\,\% \\
3 planets   &  $0.0^{+0.7}_{-0.0}$\,\% \\
2 planets   &  $2.5 \pm 1.1$\,\% \\
1 planets  &  $3.0 \pm 1.2$\,\% \\
No remaining planets   &  $0.0^{+0.7}_{-0.0}$\,\% \\
\hline
\end{tabular}
\label{table:remains}
\end{table}

Figure~\ref{figure:survival3} shows the evolution of the bound number of planets, for cluster model C051E4, over time. This figure demonstrates the complexity of the evolution of unequal-mass multi-planet systems in star clusters.  The rate at which the planets are removed from the planetary systems depends on the semi-major axis, but, notably also on the planetary masses. The average escape rates of all planets are listed in Table \ref{table:escapefractions}.
As the star clusters expands over time (see Figure~\ref{figure:clusterevolution}), the escape rates drop. This transition is typically seen at a timescale comparable to the relaxation time ($t \approx 27$~Myr for the models with $\nstars=10^4$). 

Among all modelled planets, Jupiter has the lowest probability of escape, as it is considerably more massive than the other planets. Jupiter can therefore only be strongly perturbed by external interactions (i.e., stars). Neptune and Uranus have the highest escape fractions, as they are more prone to external perturbations due to their large semi-major axes. Although Uranus has a smaller semi-major axis than Neptune, its lower mass makes it more prone to planet-planet scattering, and therefore the escape rate of Uranus tends to be slightly higher than that of Neptune. Saturn, although comparatively massive, experiences a somewhat higher escape fraction than Jupiter because of its wider orbit, and due to occasional scattering with Jupiter. The planets Earth and Mars rarely experience a strong perturbation by neighbouring stars. In addition, they are to some degree  protected from perturbed planets in the outer region of the planetary system by Jupiter. In the vast majority of the cases where Earth or Mars are ejected, we see that these events are accompanied by a strong perturbation of Jupiter by neighbouring stars are earlier times.

As time passes, planetary systems may lose one or more planets through ejection. The number of planets remaining in orbit at $t=50$~Myr ranges from zero (when all planets are ejected) to six (with all planets remaining in orbit). The distribution over the number of surviving planets at $t=50$~Myr for model C051E4 is listed in  Table~\ref{table:remains}. As is illustrated in Figure~\ref{figure:survival3}, a large majority of the  planetary systems remain intact (although a fraction of the systems has their orbital configuration is altered). Planetary systems that have four or five planets remaining at the end of the simulations are almost exclusively those that have lost Neptune and/or Uranus after a close encounter with neighbouring star. When planetary systems only have one planet remaining at the end of the simulations, this planet is always Jupiter, and when two planets are remaining, these are always Jupiter and Saturn. Situations where all six planets are ejected are rare, as this requires Jupiter to be directly ejected through interaction with a stellar encounter (as planet-planet scattering is unable to provide Jupiter with enough momentum to escape). Such situations did not occur in model C051E4, but occasionally occur for other models (see, e.g., Section~\ref{section:trends} and Appendix~\ref{section:appendix}). Due to the orbital architecture of the systems, it is unlikely to find planetary systems with three remaining planets, and we do not find such cases in any of our models. Such a situation would require one or more strong encounters (which would likely result in the ejection of Uranus and Neptune), and which would result in the ejection of only one of the other four remaining planets, while perturbing the remaining three planets only mildly. Such events regularly occur on equal-mass systems \citep[see, e.g.,][]{caisignatures2018}, but are highly unlikely in our case, where Jupiter and Saturn strongly interact, and where Jupiter's orbital evolution is strongly correlated with the survival chances of the inner planetary system.

Our simulations escape rate depends strongly on the semi-major axis, similar to the findings for the single-planet systems investigated previously by \cite{11z} and \cite{fujii2019}. Both studies find an escape rate that increases with increasing semi-major axis. \cite{1i, caisignatures2018} and \cite{vanelteren2019} find similar results for equal-mass multi-planet systems, with and without proto-planetary disks.
In our simulations we find a roughly linear dependence on semi-major axis, but this dependence is not monotonic due to the presence of a mass spectrum amongst the planets. The outermost planets, Uranus and Neptune, differ in semi-major axis by almost a factor 50\%, but they experience very similar escape rates due to their comparatively large semi-major axes. This behaviour is supported by \cite{fujii2019}, who also studied the behaviour of star-planet binaries with $a > 10$ AU.

\subsubsection{Orbital elements evolution}

\begin{figure}
\begin{tabular}{c}
 \includegraphics[width=0.45\textwidth,height=!]{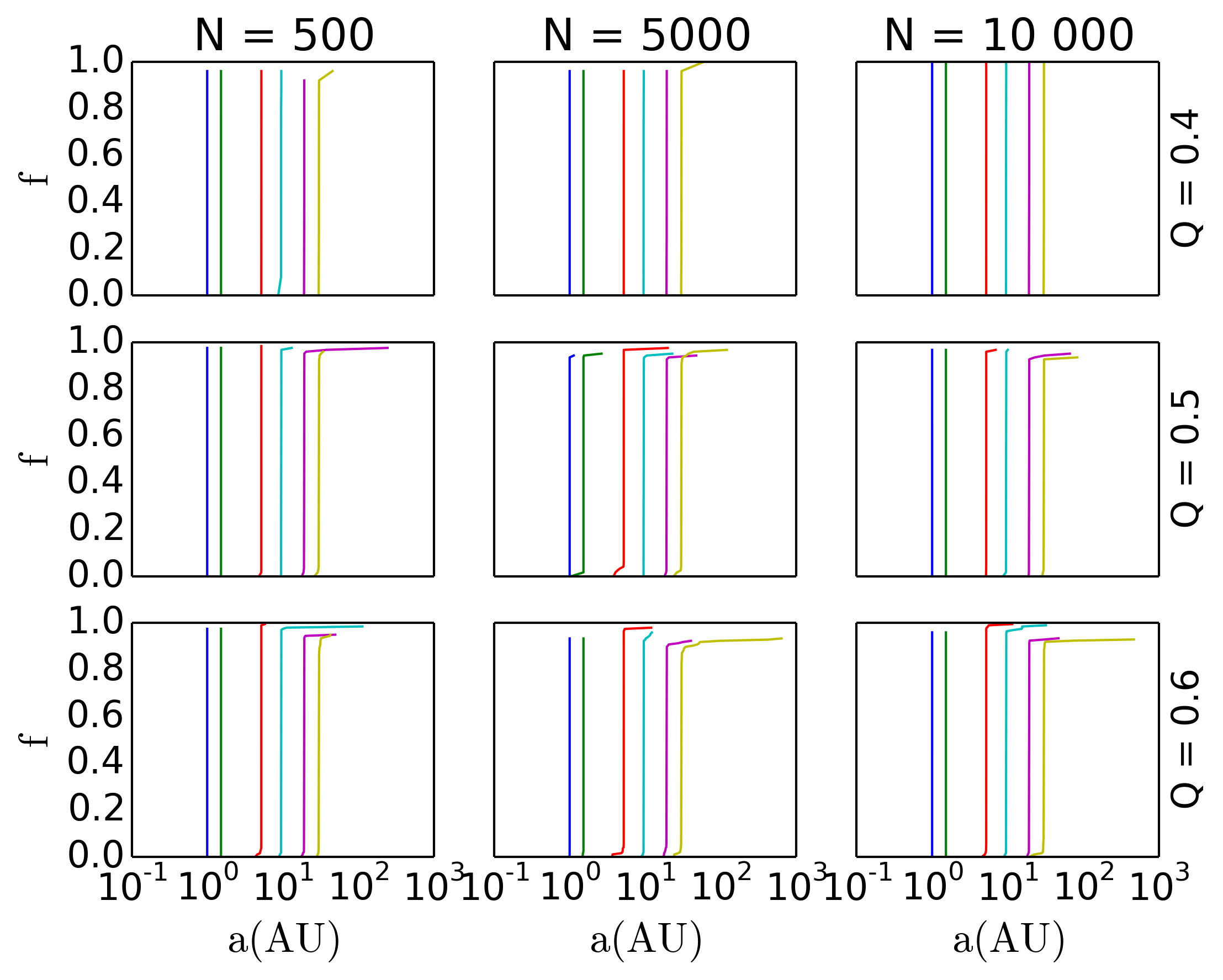} \\
 \includegraphics[width=0.45\textwidth,height=!]{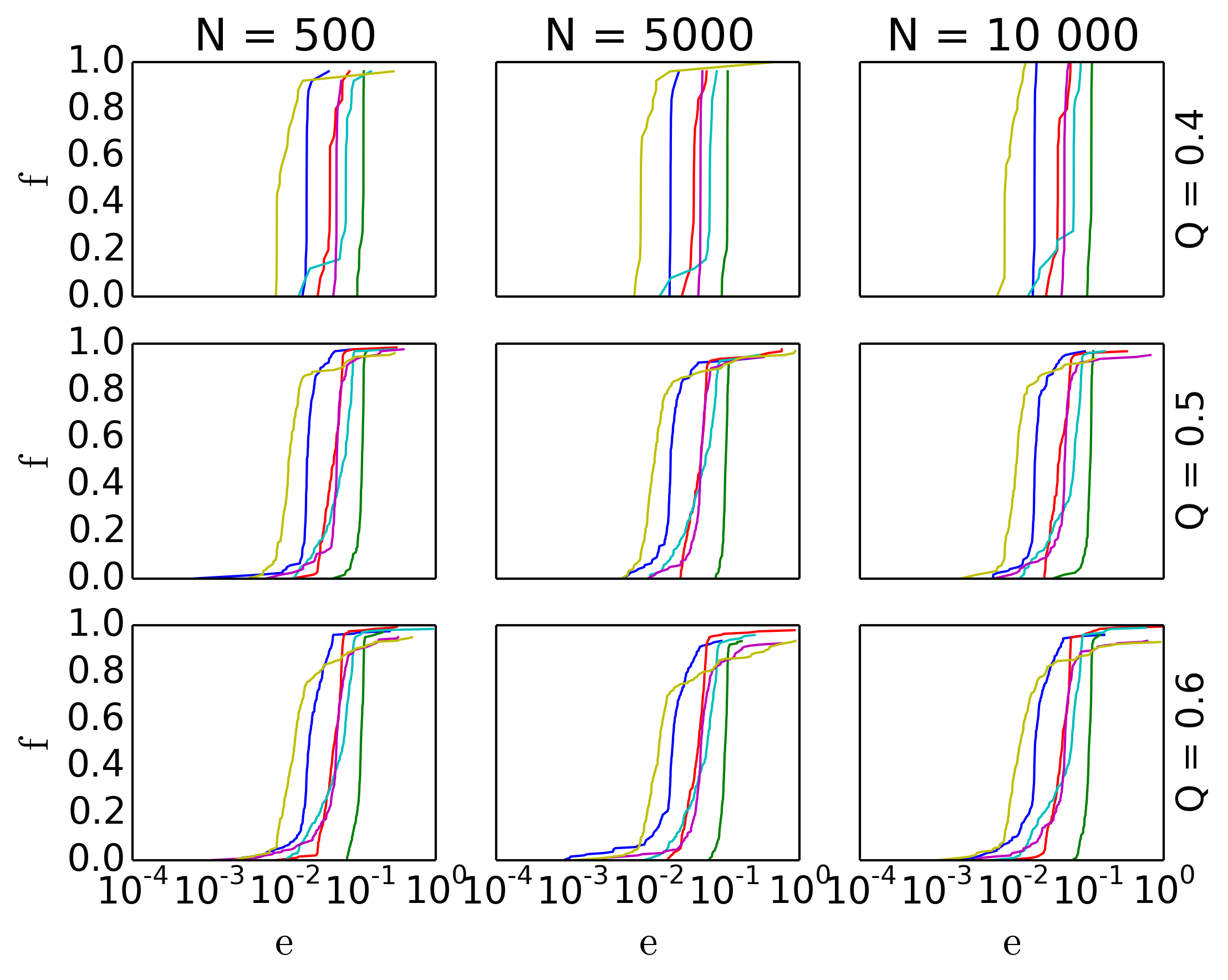} \\
 \includegraphics[width=0.45\textwidth,height=!]{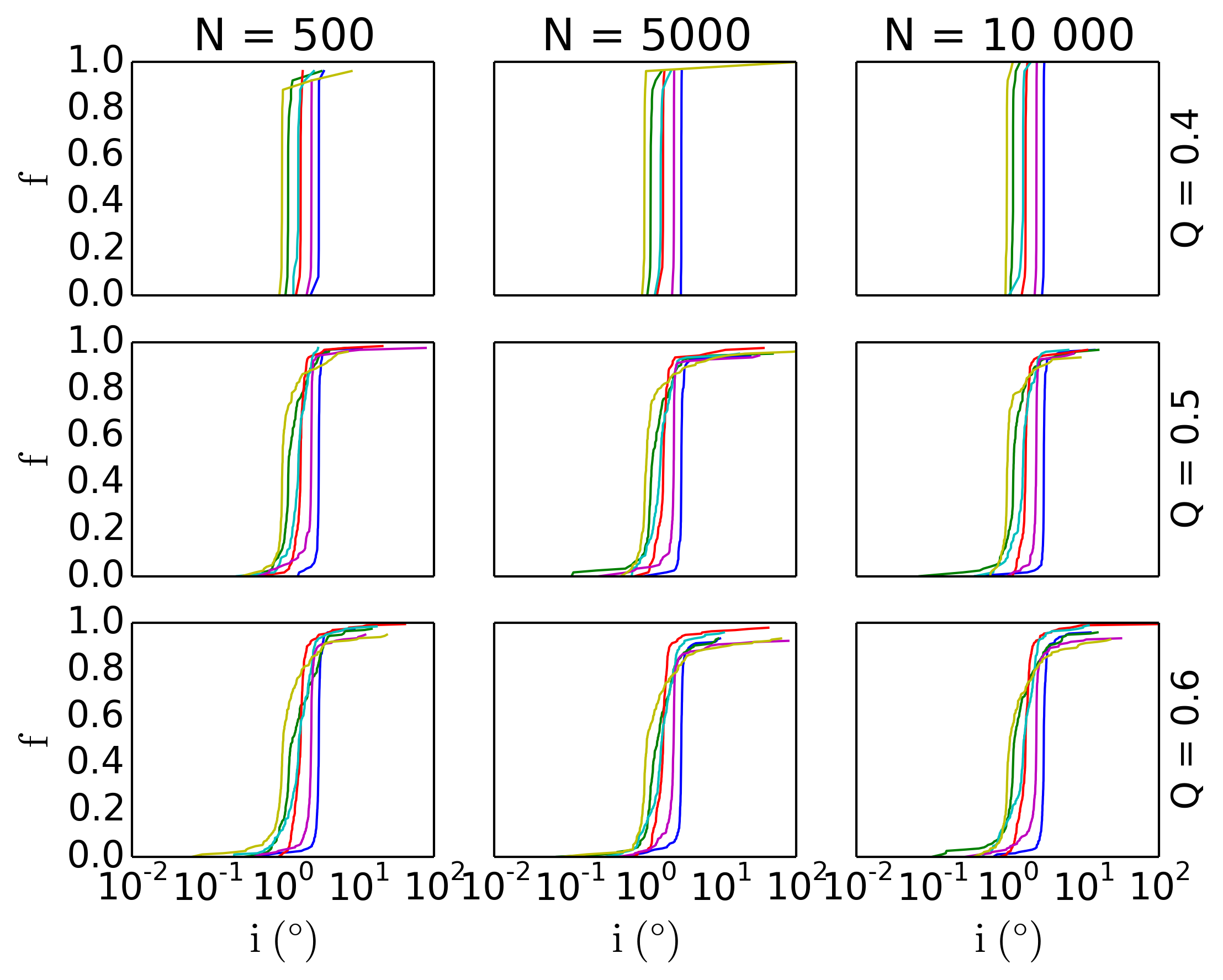} \\
\end{tabular}
\caption{Cumulative distribution of the semi-major axis ({\em top}), eccentricity ({\em middle}), and inclination ({\em bottom}) of all planets in all the planetary systems at 50~Myr, for all cluster models. The different panels represent models with $N=500$ ({\em left}), $N=5000$ ({\em middle}), and $N=10\,000$ ({\em right}), and with $Q=0.4$ ({\em top}), $Q=0.5$ ({\em centre}), and $Q=0.6$ ({\em bottom}). Planetary orbital elements distribution are indicated with dark blue (Earth), green (Mars), red (Jupiter), light blue (Saturn), purple (Uranus) and yellow (Neptune). }
\label{figure:cumdist}
\end{figure}

\begin{figure}
\begin{tabular}{c}
 \includegraphics[width=0.5\textwidth,height=!]{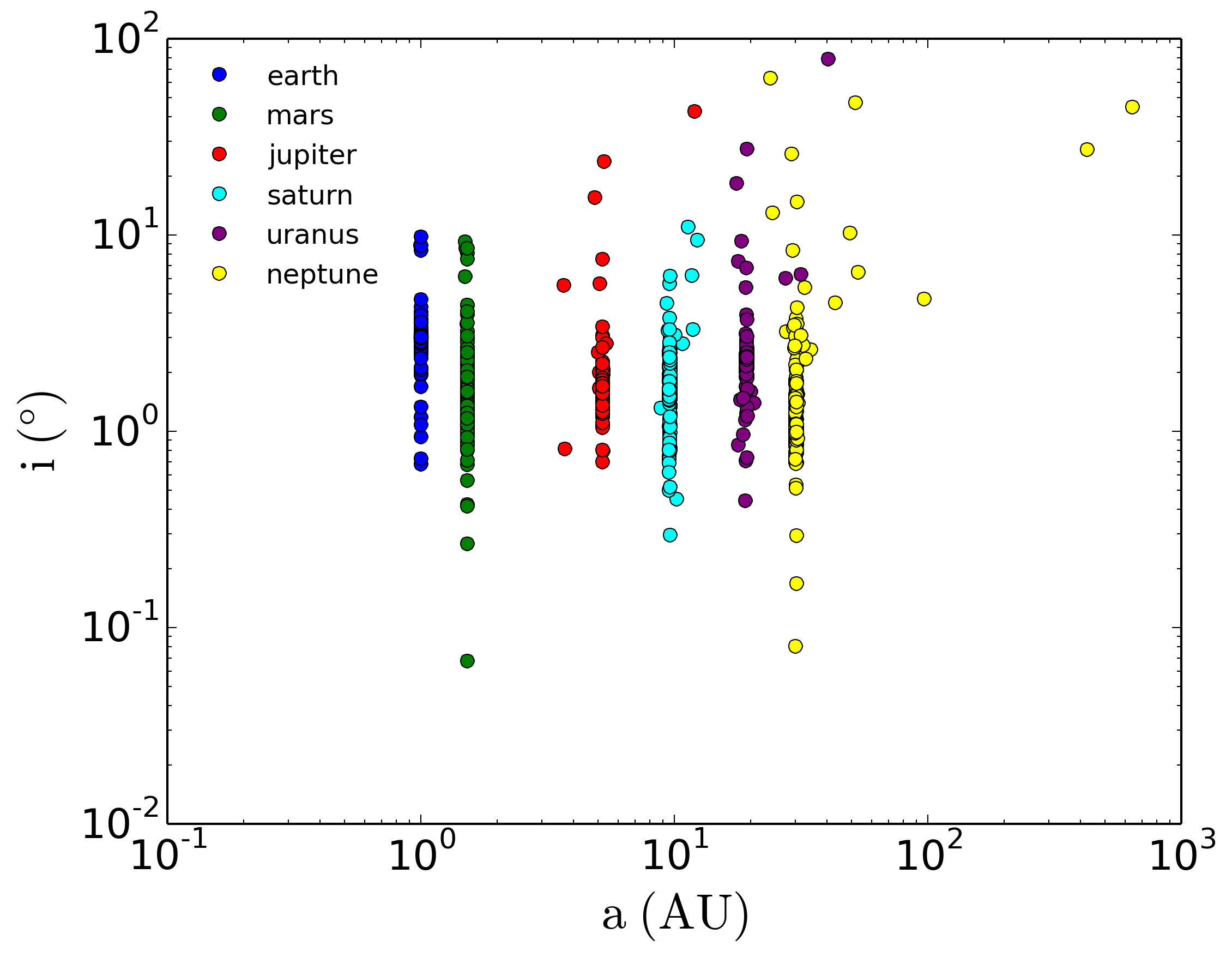} \\
 \includegraphics[width=0.5\textwidth,height=!]{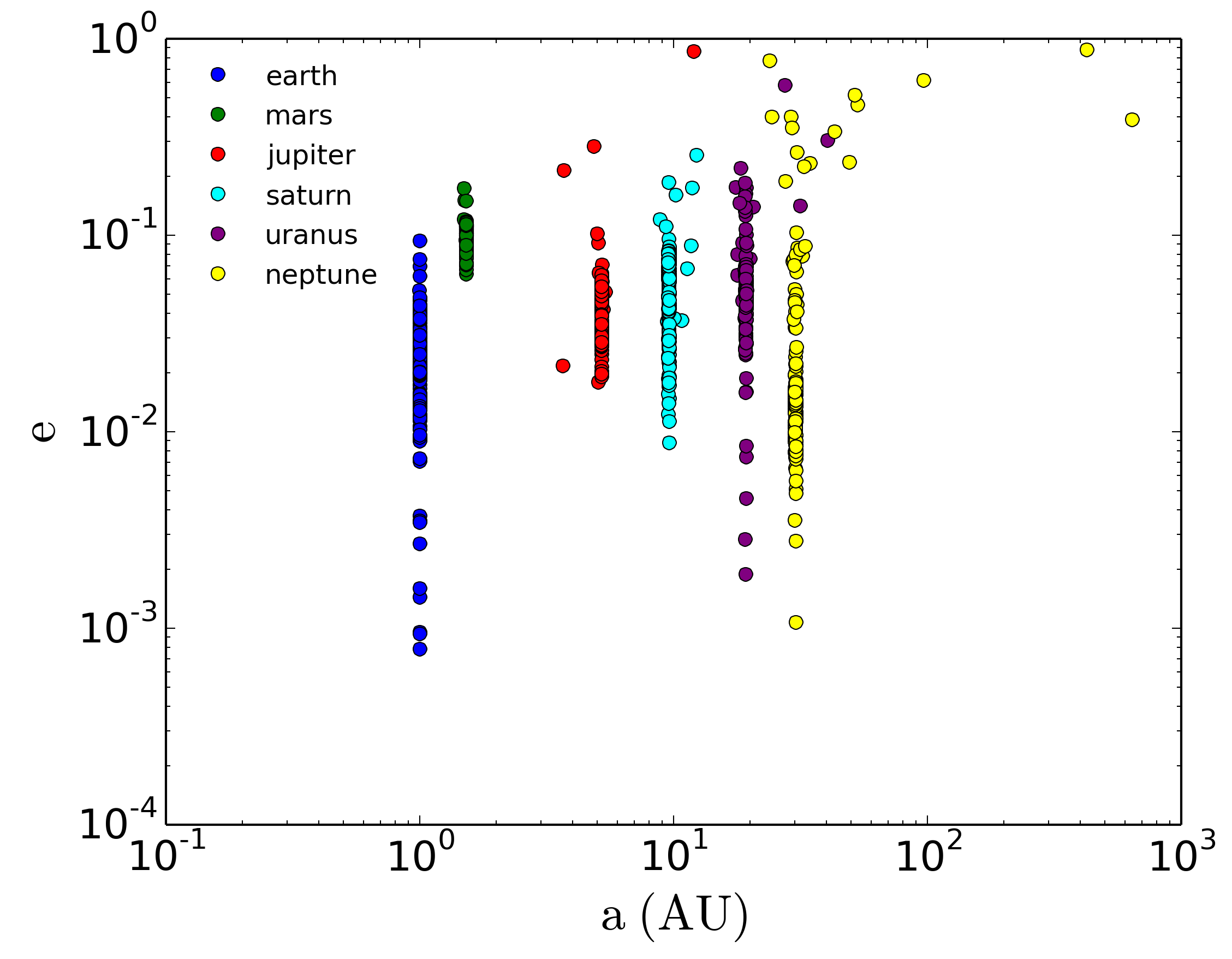} \\
 \includegraphics[width=0.5\textwidth,height=!]{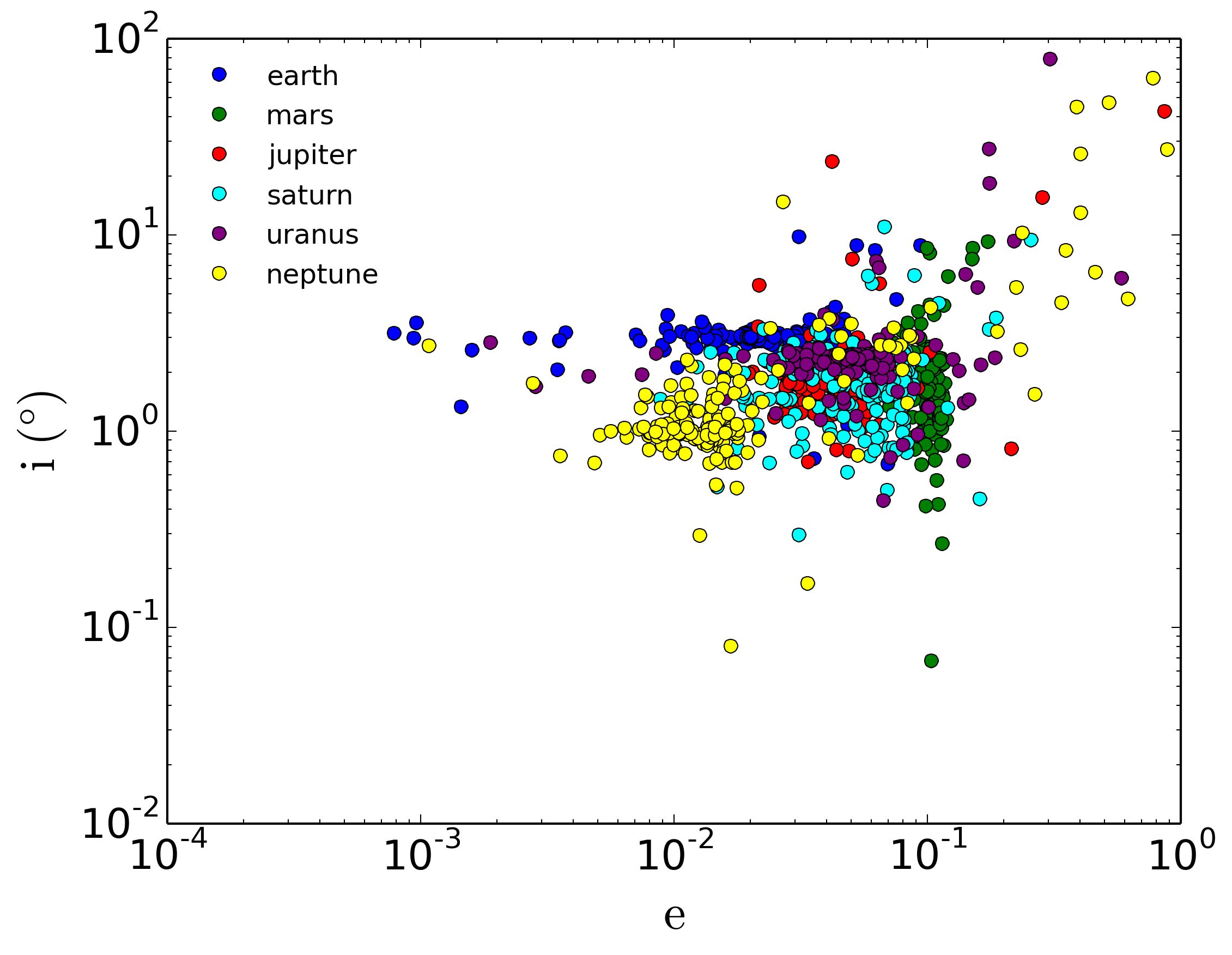} \\
\end{tabular}
\caption{Distribution of the semi-major axis, eccentricity, and inclination of all planets in all the planetary systems at $t=50$~Myr, for the reference star cluster (model C051E4). }
\label{figure:parpar}
\end{figure}

The planetary escape rates are strongly correlated with the changes in the orbital elements of the planets in each system.
Figure~\ref{figure:cumdist} shows the cumulative distribution the planetary semi-major axes, eccentricities, and inclinations after 50~Myr, for all star cluster models. The cumulative distributions are normalised to the initial number of planets (including the planets that may have escaped or merged with the host star at a later stage). Most planetary systems remain unperturbed, and the planets in these system retain their original orbital elements, apart from small changes due to secular evolution of the systems. In general, the largest changes in the orbital elements are seen for Uranus and Neptune, which indicates the effect of close stellar encounters. The changes in the remaining orbital elements are most notable in the planetary systems with $Q=0.5$. As shown in many earlier works, the changes in eccentricity and inclination are substantially larger than those in semi-major axis. 

The relations between the orbital elements of the remaining planets at 50~Myr is shown in Figure~\ref{figure:parpar} for model C051E4. All orbital elements change due to the combined effect of stellar encounters and planet-planet interactions. Changes in the semi-major axes are generally much smaller than changes in eccentricity and inclination, indicating that exchange of energy is less prominent than exchange of angular momentum. Occasionally, some planets (Uranus and Neptune in particular) obtain inclinations above the critical Kozai angle, or even obtain retrograde orbits.

\subsubsection{Trends and notable planetary systems} \label{section:trends}

\begin{figure*}
\begin{tabular}{cc}
 \includegraphics[width=0.5\textwidth,height=!]{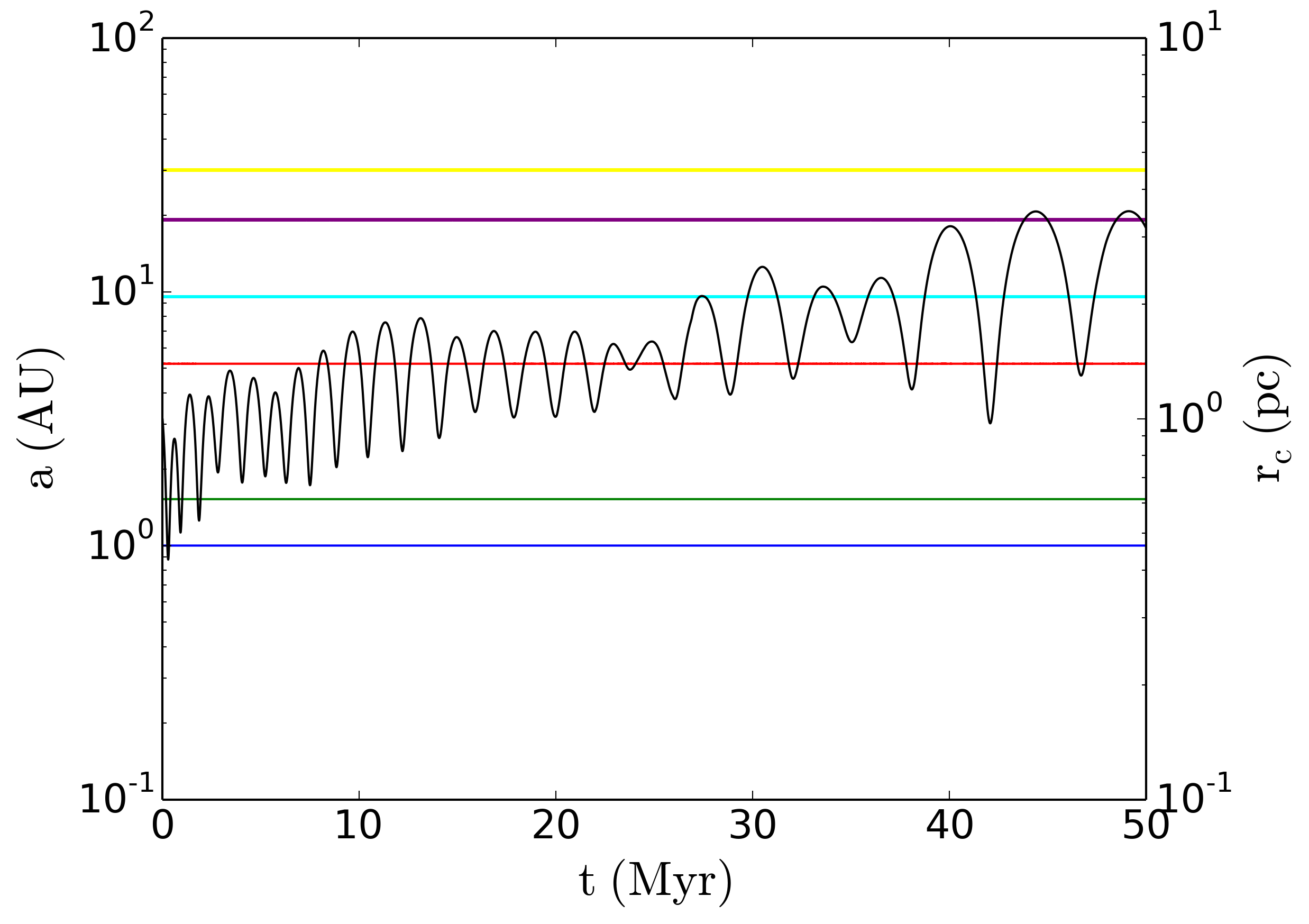} &
 \includegraphics[width=0.5\textwidth,height=!]{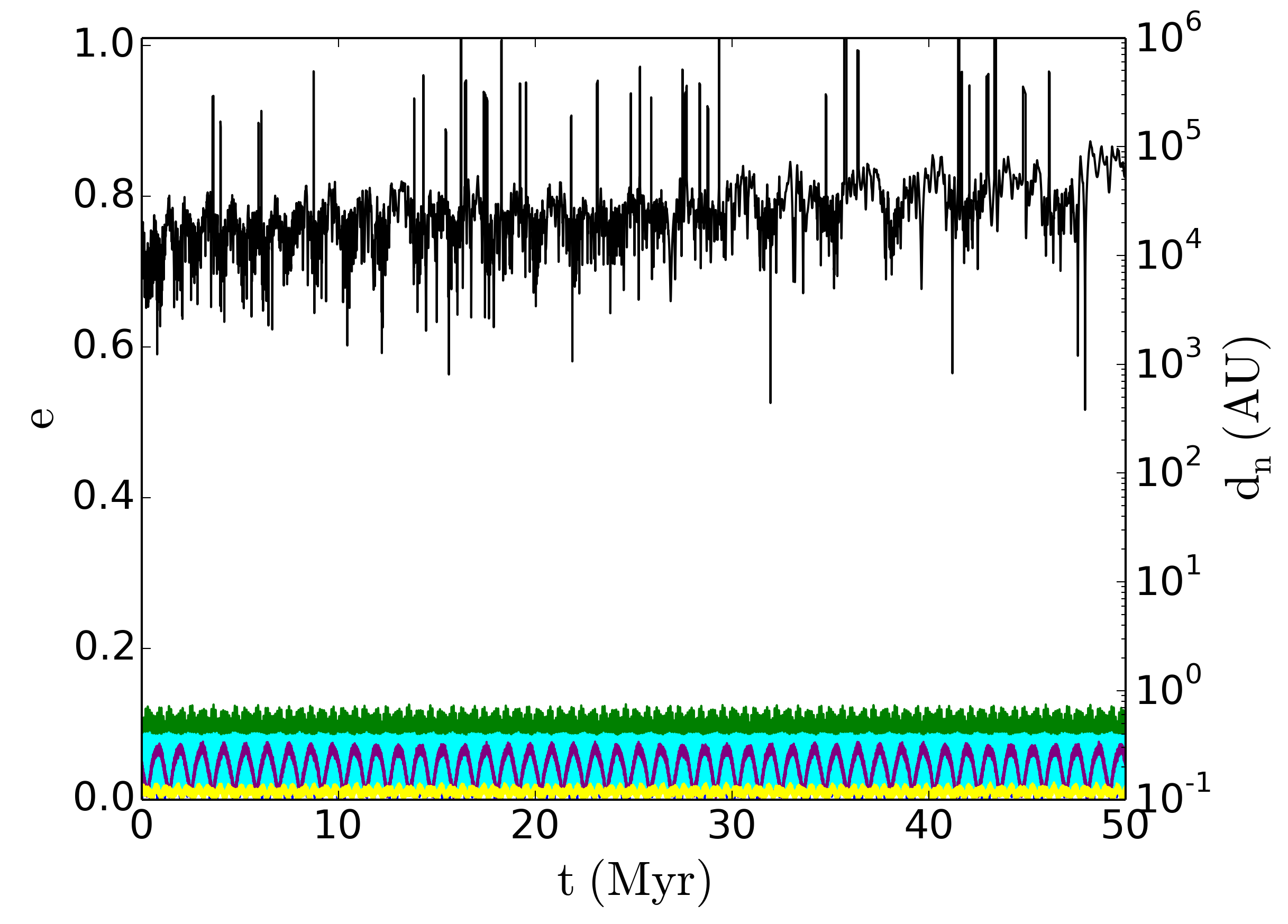} \\
 \includegraphics[width=0.5\textwidth,height=!]{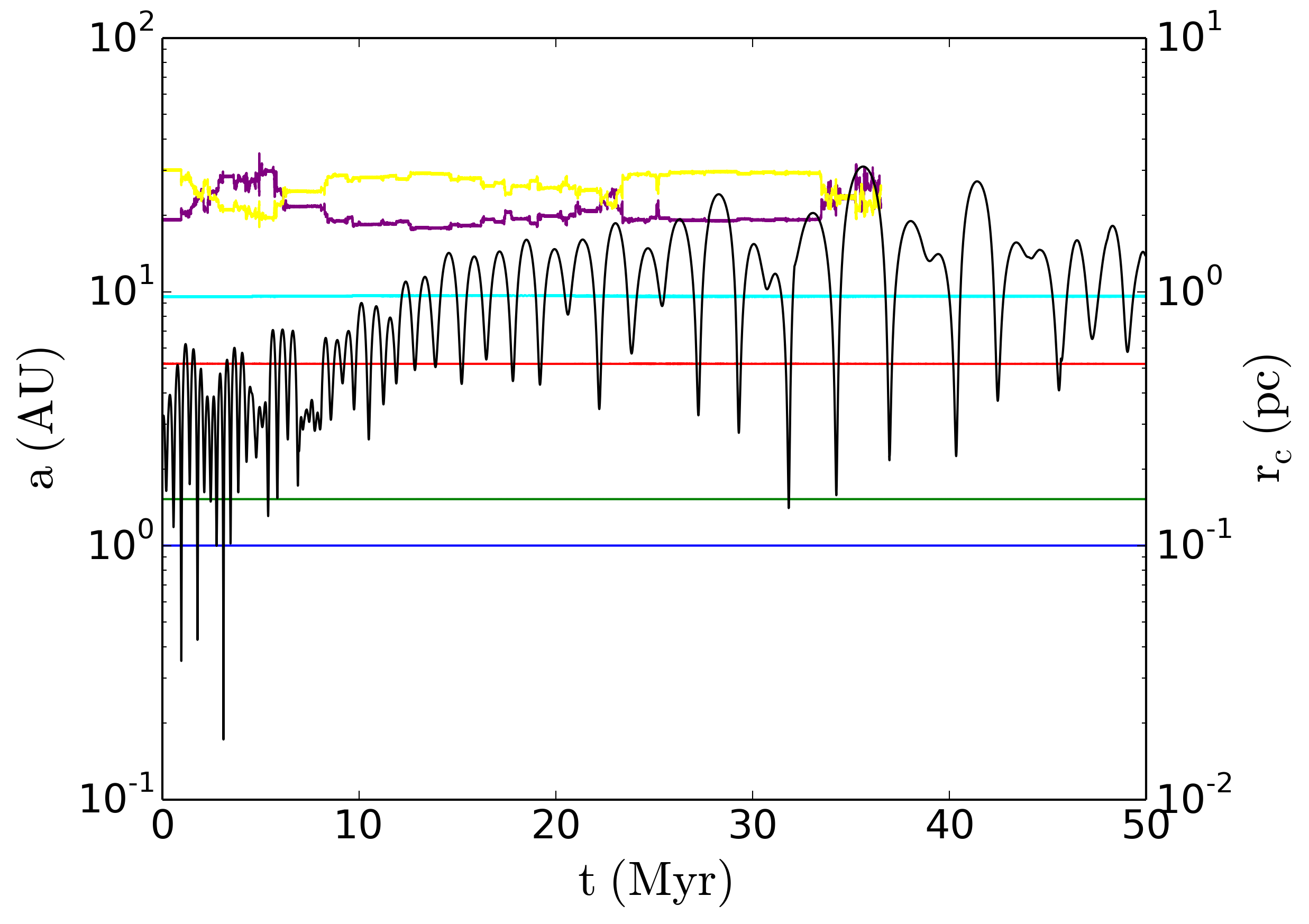} &
 \includegraphics[width=0.5\textwidth,height=!]{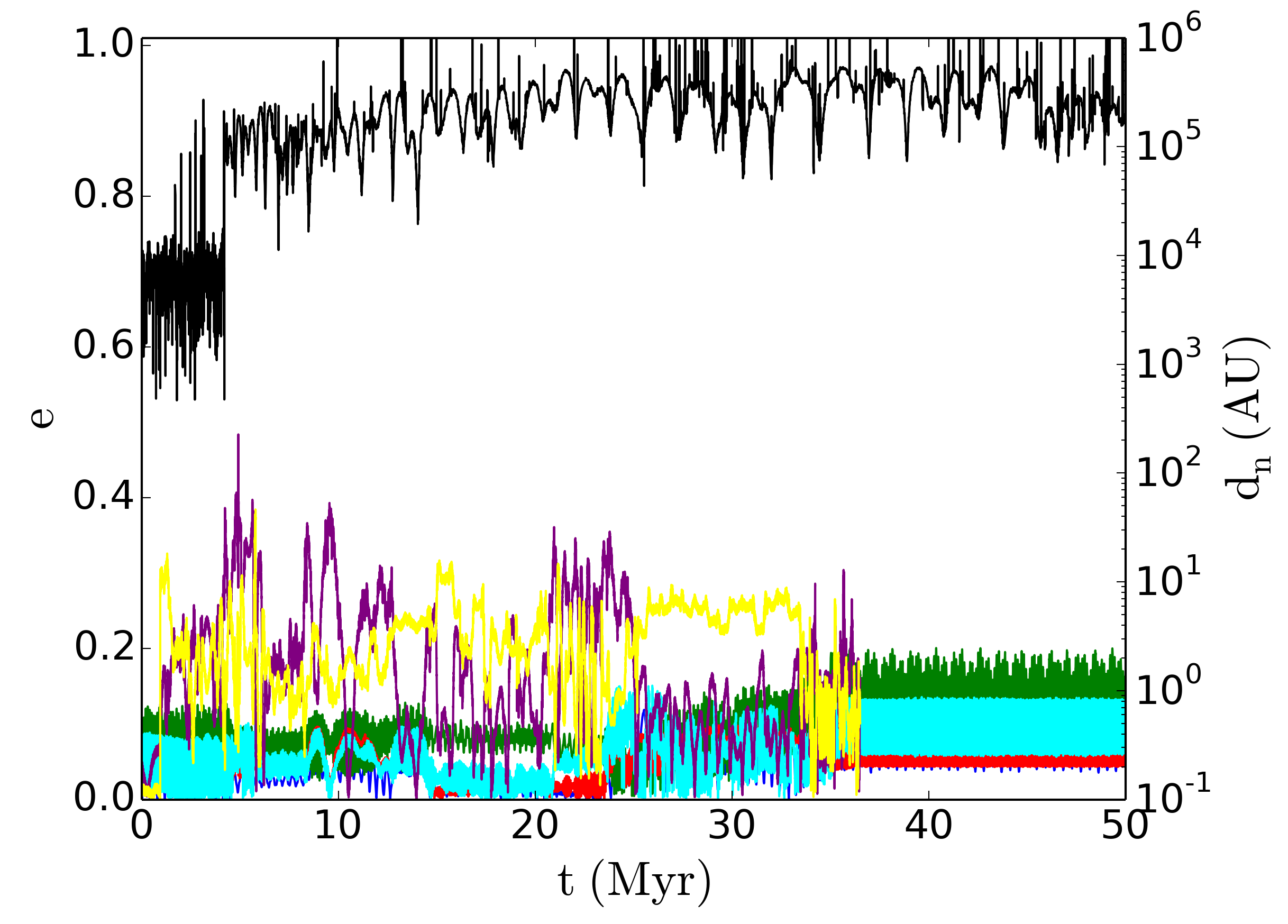} \\
\end{tabular}
\caption{{\em Left}: evolution of the planetary semi-major axes and distance from the cluster centre (black curve). {\em Right}: evolution of the planetary eccentricities and distance to the nearest neighbour star (black curve). Results are shown for planetary systems P194 ({\em top}) in star cluster model C041E4 and P161 ({\em bottom}) in star cluster model C061E4. These two planetary systems both remain star cluster members. System P194 does not experience strong encounters and remains intact, while system P161 loses both Uranus and Neptune, approximately 35~Myr after a close encounter. }
\label{figure:comandchan}
\end{figure*}

\begin{figure*}
\begin{tabular}{cc}
 \includegraphics[width=0.5\textwidth,height=!]{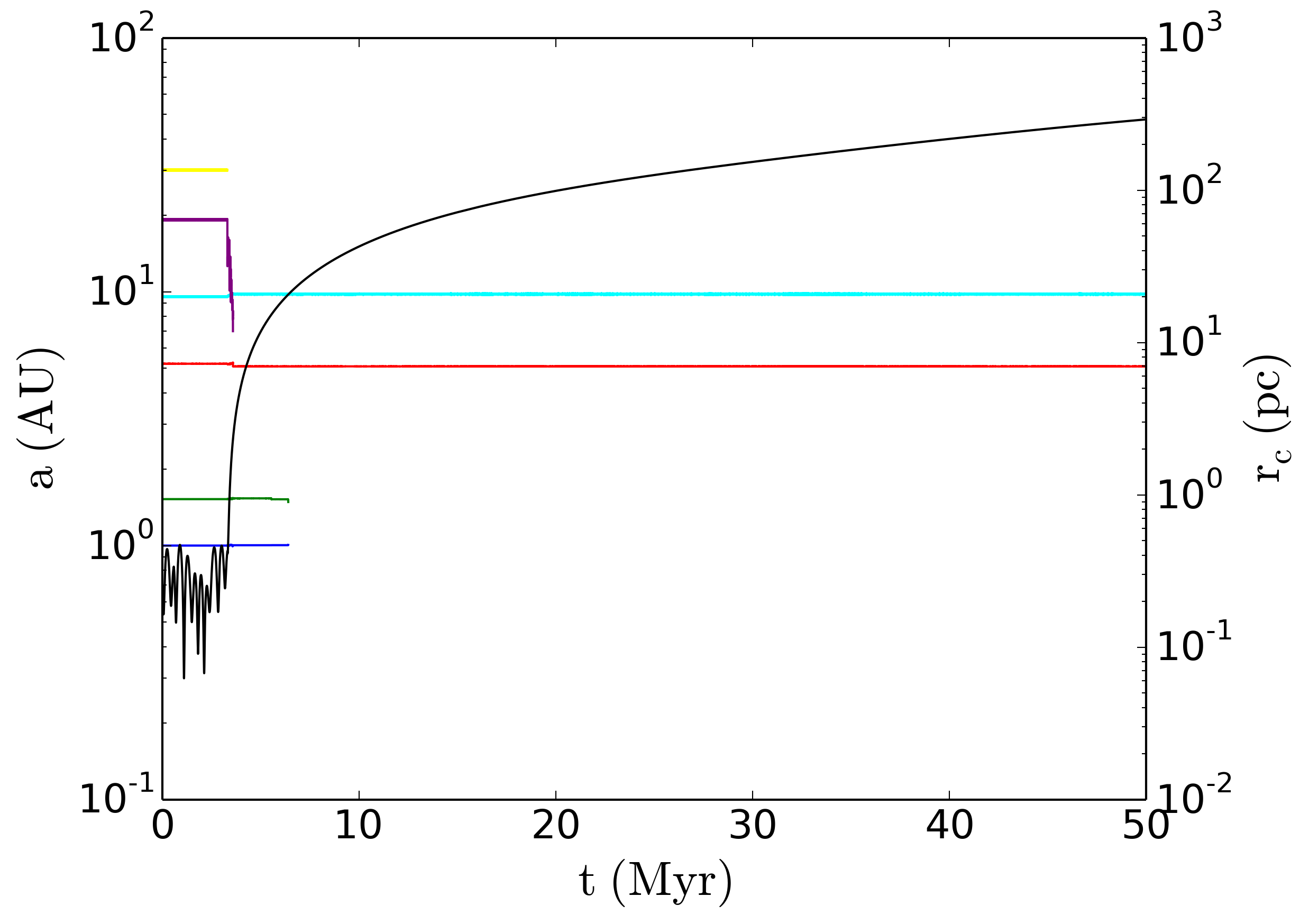} &
 \includegraphics[width=0.5\textwidth,height=!]{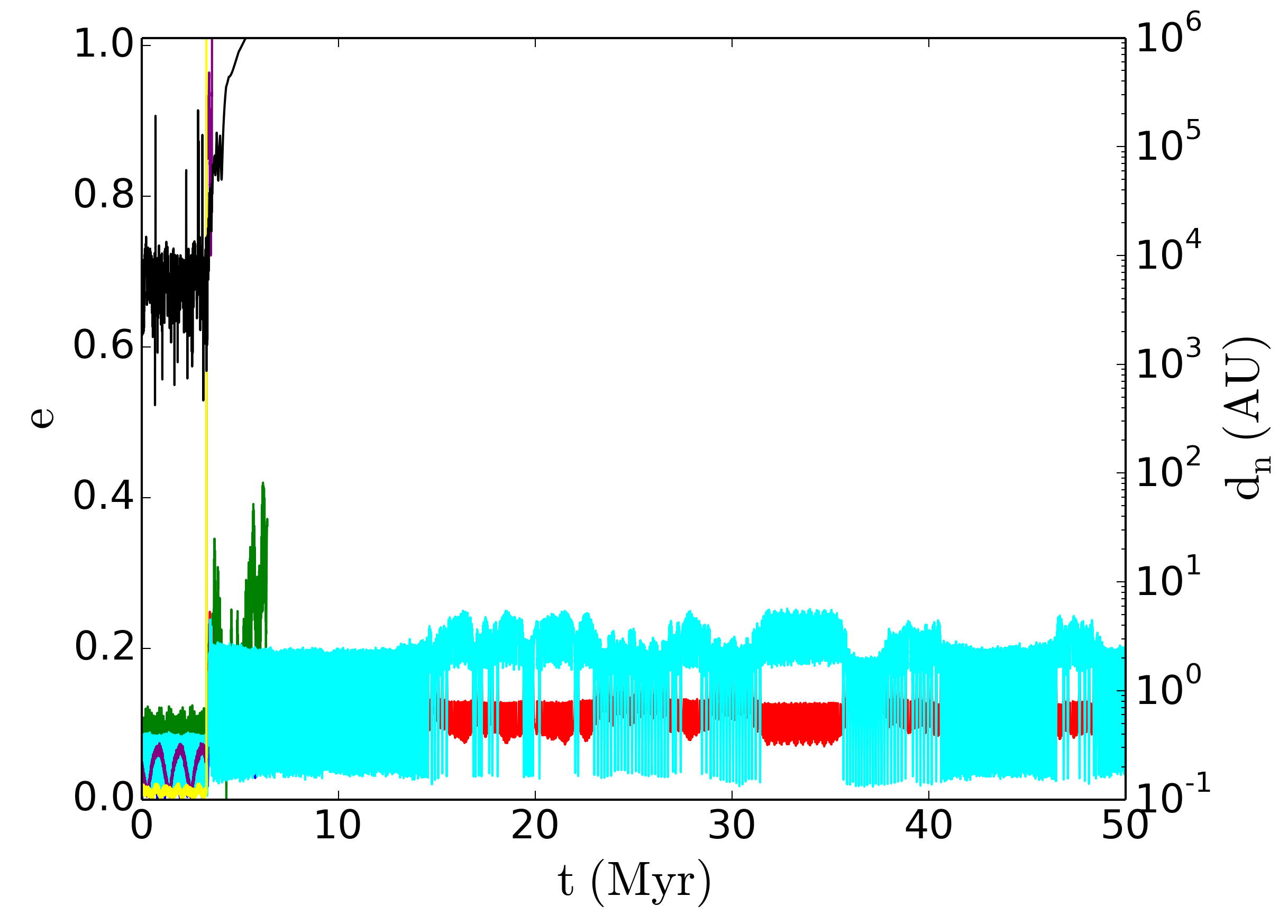} \\
 \includegraphics[width=0.5\textwidth,height=!]{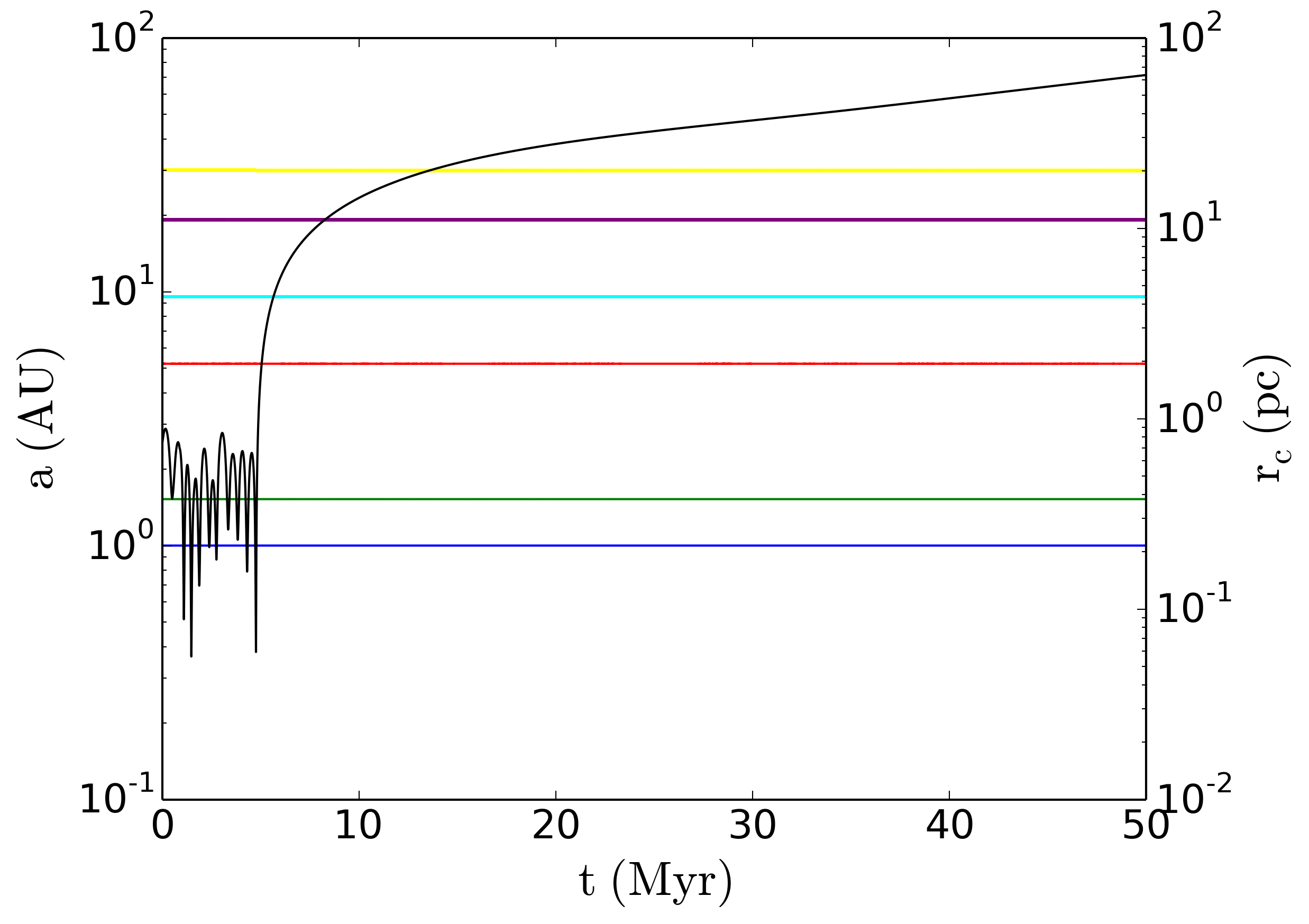} &
 \includegraphics[width=0.5\textwidth,height=!]{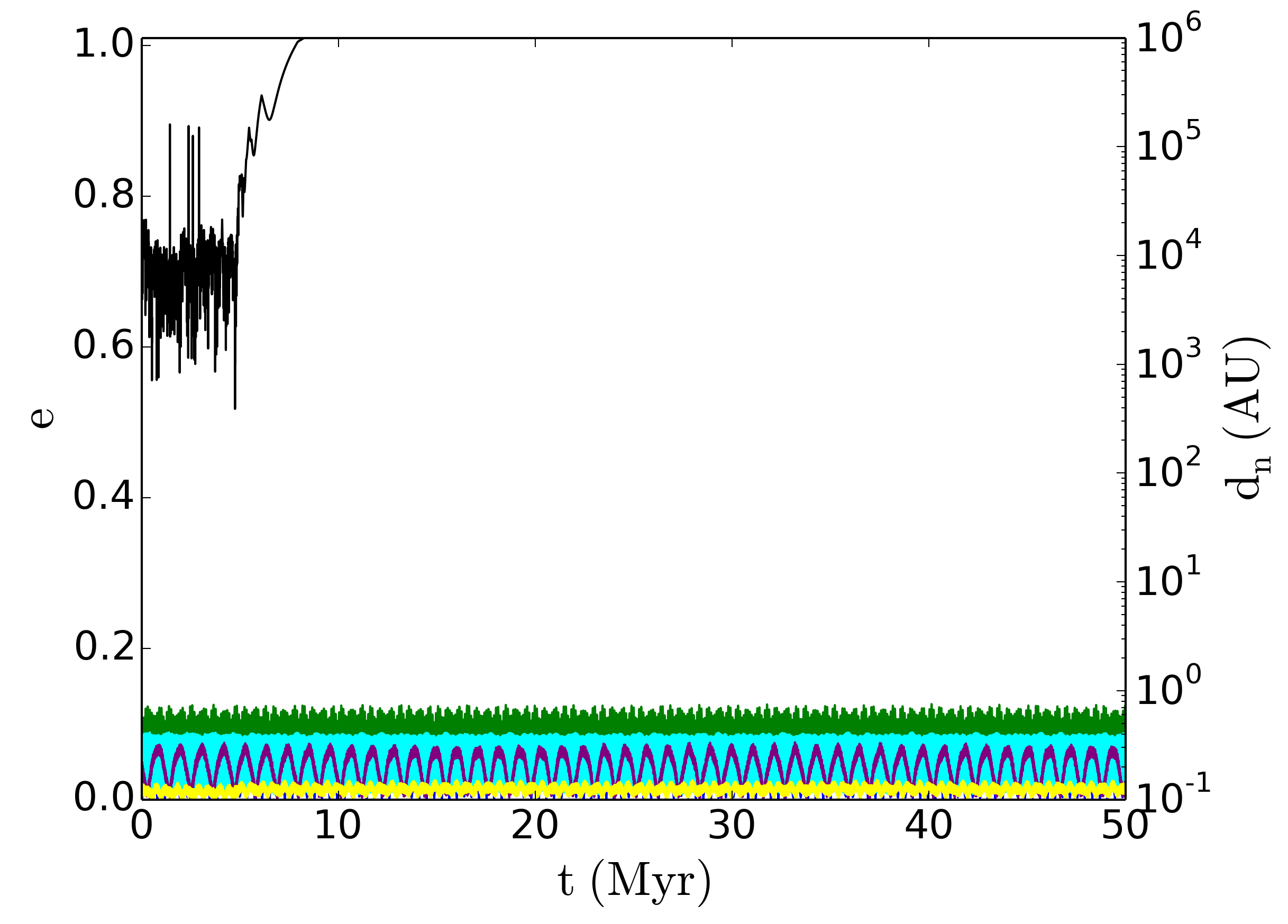} \\
\end{tabular}
\caption{Same as in Figure~\ref{figure:comandchan}, P165 ({\em top}) in star cluster model C061E4 and P191 ({\em bottom}) in star cluster model C051E4.
Systems P165 and P191 both escape from the star cluster. System P165 is strongly perturbed during the ejection process, and loses several of its planets during and after it is ejected from the cluster with a speed of $\sim 5$~\kms. System P191, on the other hand, suffers only from weak encounters, and is ejected intact with a speed of $\sim 2$~\kms. }
\label{figure:highandlowvel}
\end{figure*}

\begin{figure*}
\begin{tabular}{cc}
 \includegraphics[width=0.5\textwidth,height=!]{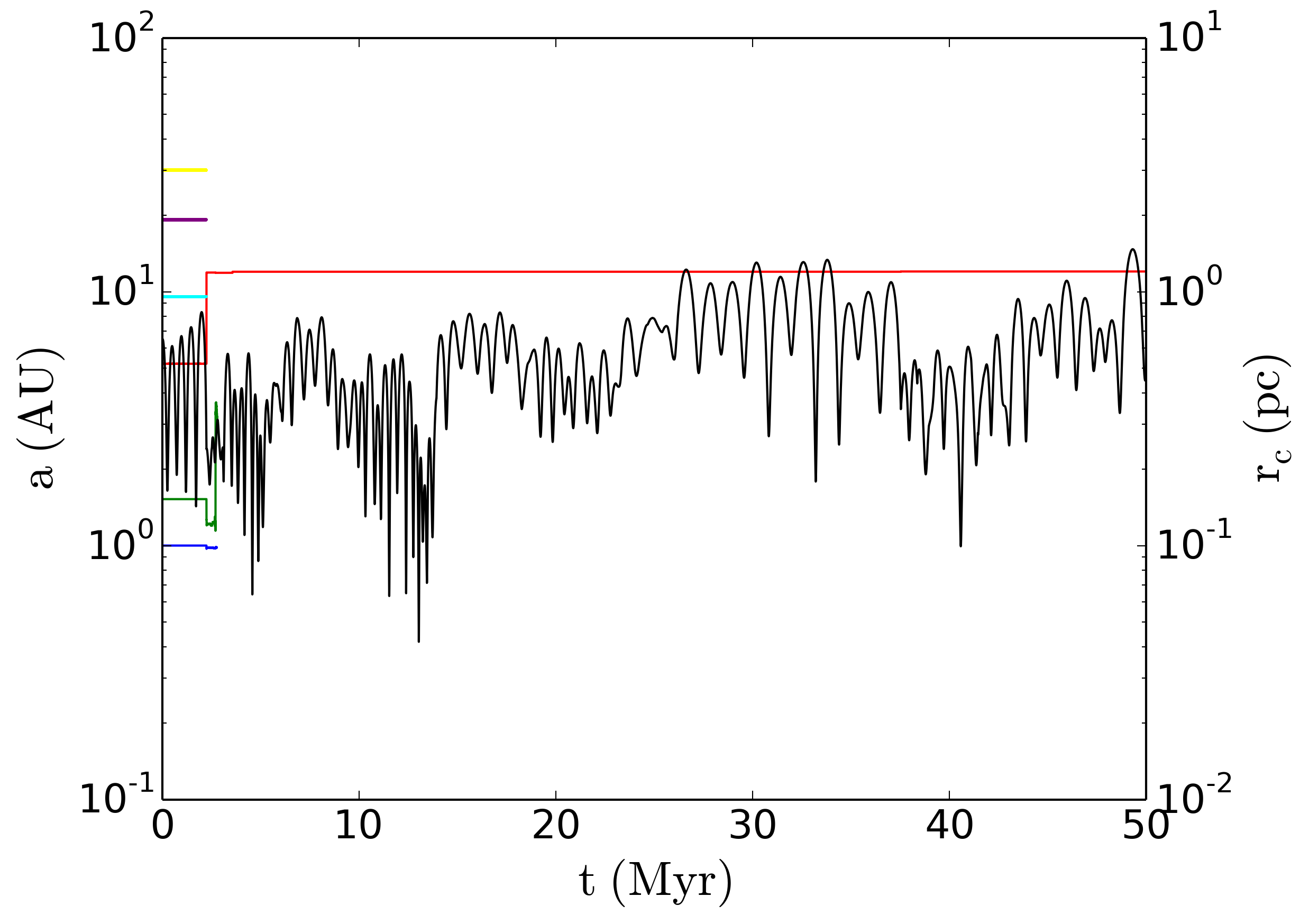} &
 \includegraphics[width=0.5\textwidth,height=!]{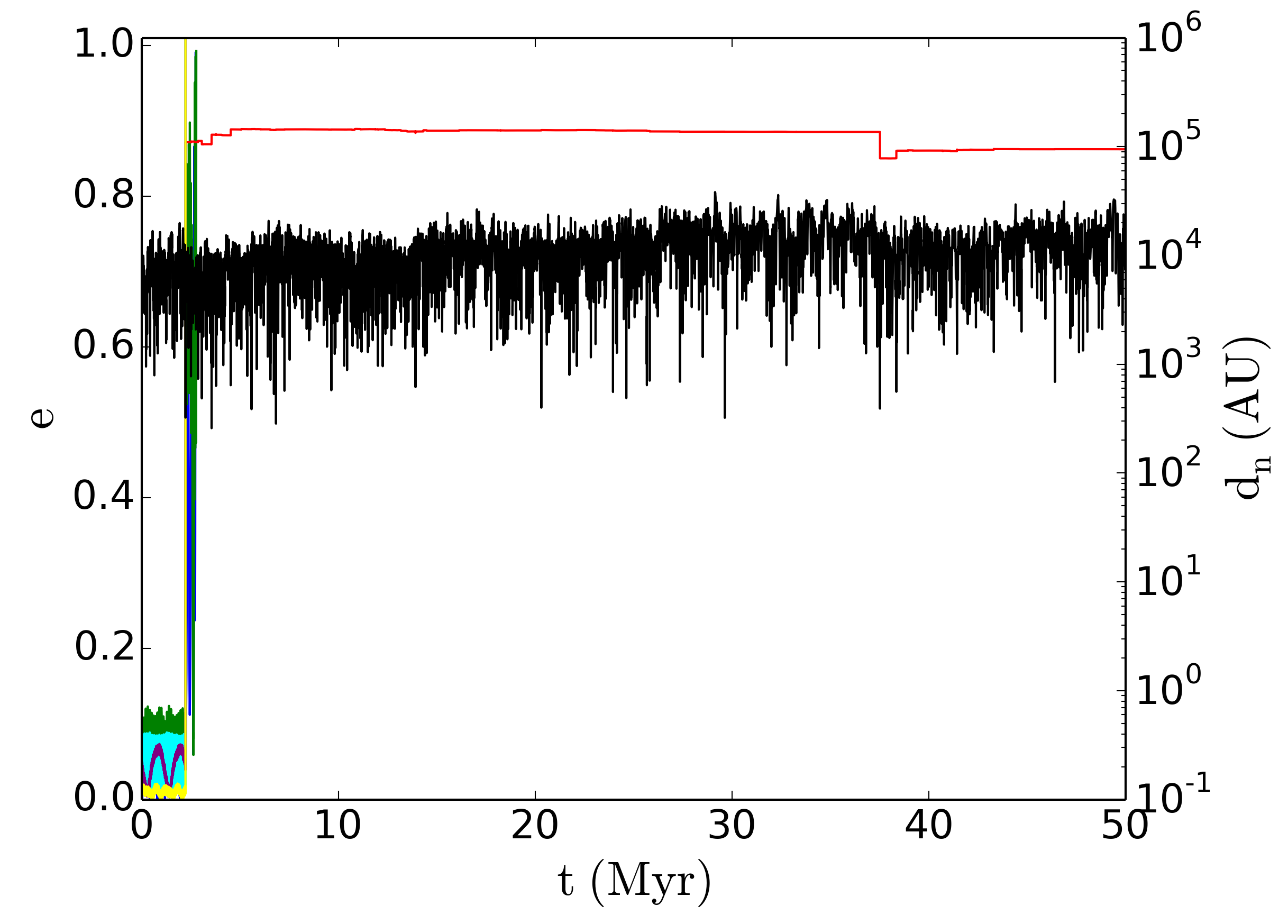} \\
 \includegraphics[width=0.5\textwidth,height=!]{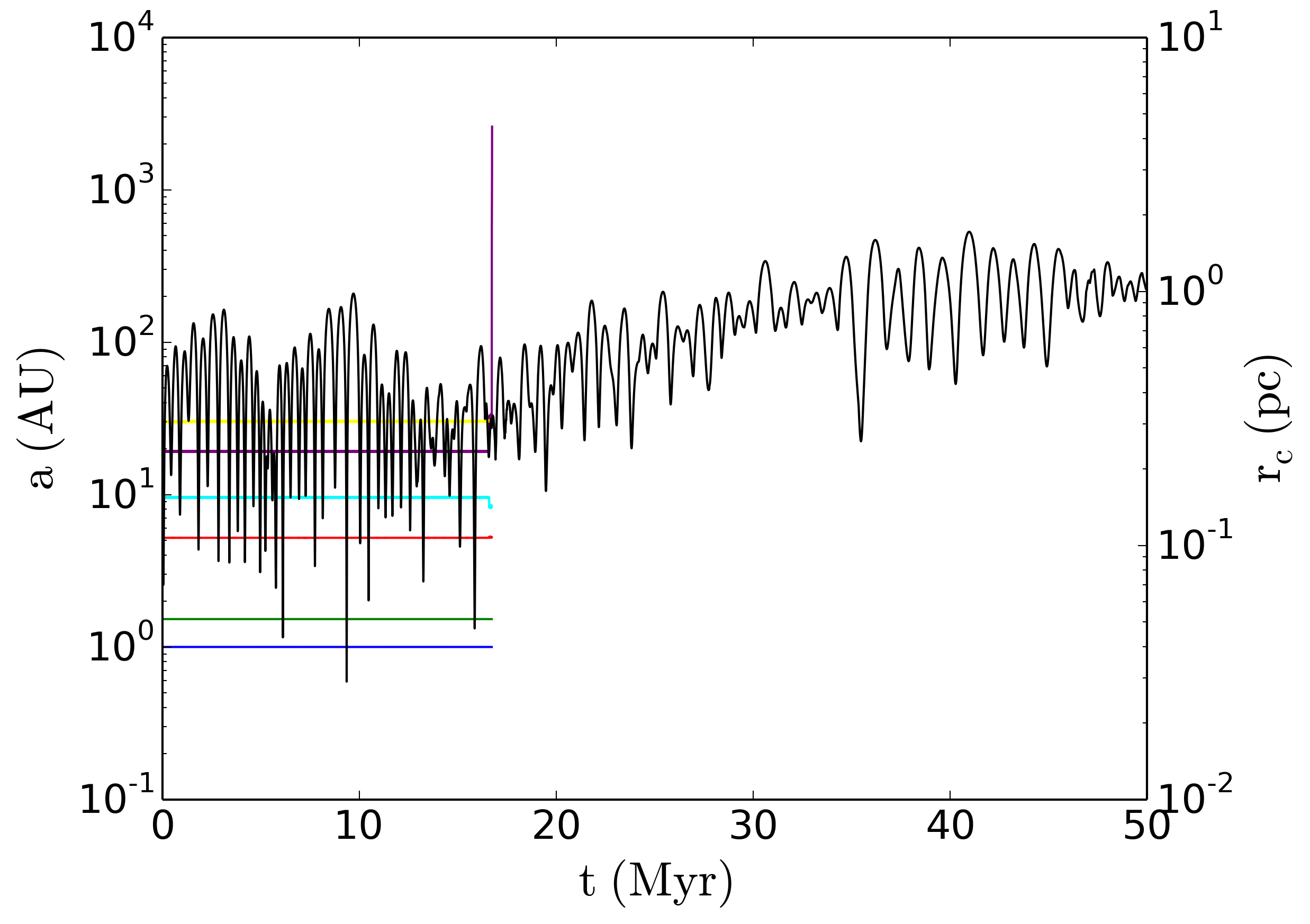}&
 \includegraphics[width=0.5\textwidth,height=!]{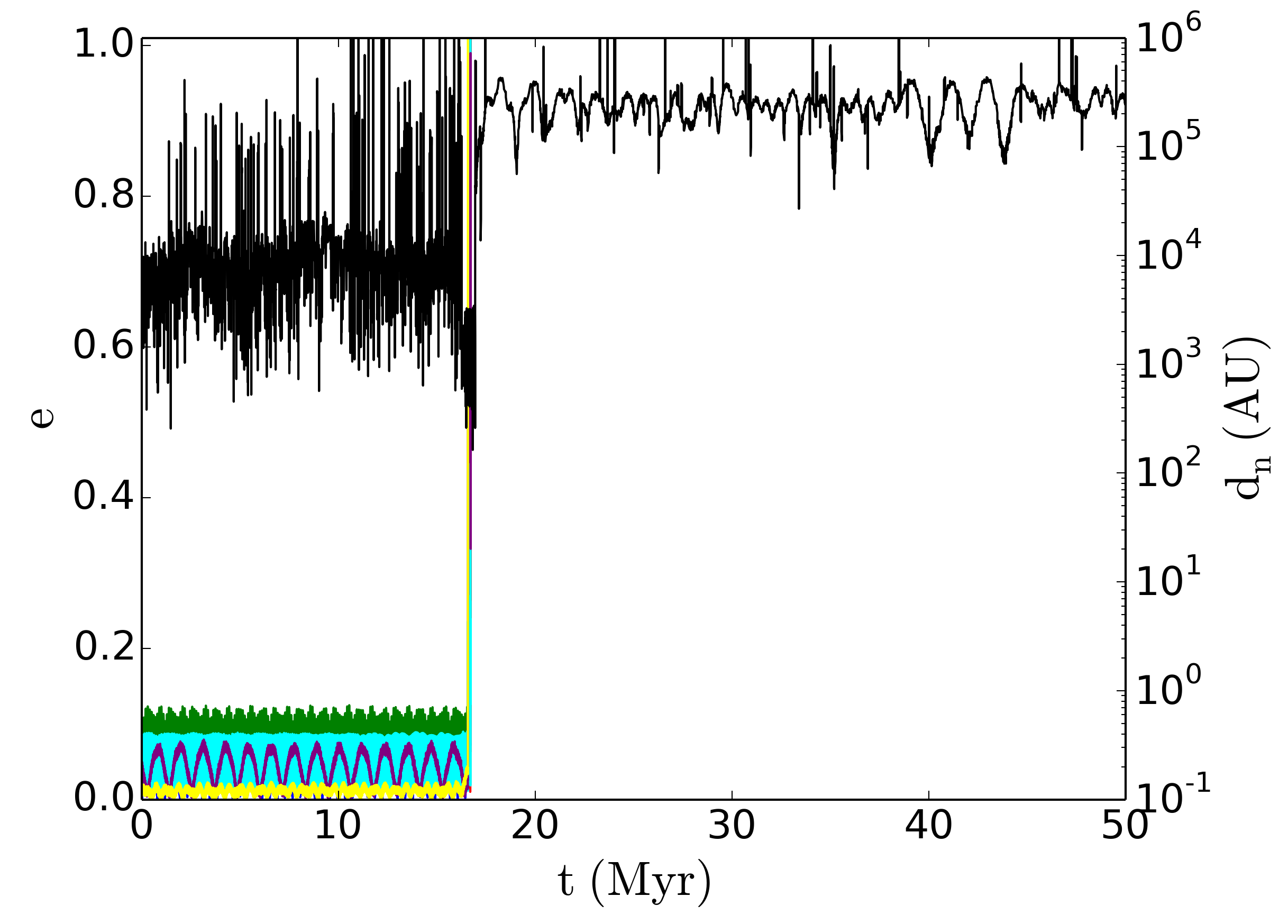} \\
\end{tabular}
\caption{Same as in Figure~\ref{figure:comandchan}, planetary system P024 ({\em top}) in star cluster model C051E4 and planetary system P187 ({\em bottom}) in star cluster model C061E4. Both planetary systems are strongly perturbed by nearby stars, resulting in (near-)complete destruction of the planetary systems. At $t=50$~Myr, system P024 only has one planet (Jupiter) remaining in a wide, highly-eccentric orbit, while all planets have escaped from system P187. }
\label{figure:extraesc}
\end{figure*}

We illustrate some of the most common types of evolution amongst planetary systems in our simulations in Figures~\ref{figure:comandchan}--\ref{figure:extraesc}. Additional, higher-resolution figures are presented in Appendix~\ref{section:appendix}. These figures show the evolution of the orbital parameters, the distance between the planetary system and the star cluster centre, and the distance between the neighbourhood star and the planetary system.

The top panels in Figure~\ref{figure:comandchan} show the evolution of the orbital semi-major axes and eccentricities of planetary system P194 in model C061E4. This planetary system experiences no disruptive stellar encounters, and all planets  retain their original orbits. This system resides primarily in the outskirts of the star clusters, initially at a distance of $0.5-1.0$~pc from the cluster centre, and after $\approx 20$~Myr at a distance of $1.0-3.0$~pc from the cluster centre. Due to the low stellar density in the halo, this system rarely experience a strong encounter. 
All encounters experienced by system P194 are tidal ($k\gg 1$). At time $t=47.9$~Myr, the closest encountering star has a mass of $0.14~\msun$ and approaches within a distance of 380~AU with a velocity of 3.8~\kms, with $k$ parameters of 1875 and 128 for Jupiter and Neptune, respectively. Despite the relatively close approach, the low-mass neighbour star is unable to perturb system P194 significantly. As system P194 spends most time in the outskirts of the star cluster, encounters with neighbouring stars are thus relatively infrequent, and typically with lower-mass stars.

The bottom panels of  Figure~\ref{figure:comandchan} illustrate a relatively common situation  where close encounters substantially perturb the outer planetary system. Planetary system P161 in model C061E4 passes through the cluster three times during the first 5~Myr, during which the orbits of Uranus and Neptune are perturbed by frequent close encounters with stars approaching the system within 500~AU. Planet-planet scattering results in Uranus and Neptune to exchange orbits. Although planetary system P161 is ejected from the core into the the lower-density cluster halo, the interactions between Uranus and Neptune continues, and ultimately results in an ejection at 35~Myr. This system shows how planet-planet scattering can cause delayed ejections, tens of millions of years after a close encounter has occurred, which is consistent with the earliest findings of \cite{malmberg2011}. During the time between the encounter and the ejection, the outer planets obtain eccentricities of up to $e=0.4$ and therefore interact with both Jupiter and Saturn. Although the four innermost planets retain their original semi-major axes, their eccentricities are pumped up due to these encounters. 

Figure~\ref{figure:highandlowvel} shows two cases of planetary systems escaping from the star cluster: system P165 in model C061E4 ({\em top panels}) is disrupted before escaping, whereas system P191 in model C051E4 ({\em bottom panels}) retains intact while escaping from the star cluster. Both systems are initially located in the star cluster (P191 somewhat farther from the cluster centre than system P165), and both are ejected from the star cluster at around 5~Myr. System P165 is ejected from the star cluster with a speed of $\sim 5$~\kms{} following a scattering event. During this scattering event, Uranus and Neptune are highly perturbed and are ejected from the system after strong interactions with Jupiter and Saturn. Although Jupiter and Saturn remain part of the planetary system, their eccentricities and inclinations have substantially increased, resulting in a delayed scattering event with Earth and Mars about three million years later, and an ejection of the both terrestrial planets. For the remaining simulation time, Jupiter and Saturn experience strong and chaotic exchanges of angular momentum, which may result in a merger or ejection events at later times. System P191, on the other hand, suffered a much weaker encounter and escapes the cluster intact with a speed of $\sim 2$~\kms{}. 

We illustrate extreme cases of planetary system perturbations in Figure~\ref{figure:extraesc}. Both system shown in this figure experience a strong (periastron distance $< 1000$ AU) encounter with a neighbouring star. In the top panels, planetary system P024 experiences a  with a neighbouring star of mass $0.87~\msun$ that disrupts most of the system. The $k$-parameter corresponding to the encounter is 85 for the planet Neptune, indicating a relatively high encounter strength.Although highly perturbed, the planet Jupiter survives the close encounter and obtains a wider orbit ($a\approx 11$ AU), while Saturn, Uranus and Neptune are expelled immediately at $t=2.231$~Myr. As the perturbed planet Jupiter attains a highly eccentric orbit ($e\approx 0.87$), it interacts with both Mars and Earth, which are eventually expelled at 2.78~Myr and 3.61~Myr, respectively. Jupiter's wider orbit makes it more prone to perturbations by encountering stars, but it remains bound to its host star. Unlike system P194 (shown in the top panels of Figure~\ref{figure:comandchan}), which remains intact, planetary system P024 is highly perturbed by the neighbouring star. Although the closest encounter distance for P194 (for P194) is somewhat larger than the 320~AU (for P024), the differences in dynamical outcomes are also determined by the large difference in the mass of the encountering stars, the presence of consecutive close encounters at similar times in the case of P024, the speeds at which the encountering stars approach the planetary systems, and the orbital phases of the planets during the encounter. These somewhat stronger perturbations in system P024 result in strong planet-planet interactions that ultimately result in the ejection of all but one of the planets from the system.

In the bottom panels, a passing star almost instantaneously destroys the entire planetary system P187.  The close encounter at $t=16.581$~Myr has a periastron distance of $p\approx 422$~AU, and perturbs all the planetary system, resulting in the ejection of Neptune. A subsequent close  encounter at t=16.71~Myr with a periastron distance of $\approx 490$~AU eject all remaining planets within 2000~years. System P187 is a good example of a situation in which the expulsion of terrestrial planets is caused directly by the stellar encounter, instead of the more common indirect case in which a perturbed Jupiter is responsible for the ejections. 

The complex interaction between planetary systems and neighbouring stars and the gravitational interaction between the planets in each system results in a large diversity in evolutionary outcomes. The systems shown above illustrates the most common types of evolution, but represent merely a small subset of the observed outcomes. This demonstrates that even in the case where all planetary systems are formed with identical orbital architectures, the diversity among planetary systems in the Galactic field is large.

\section{Discussion and conclusions}

We have carried out a set of comprehensive $N$-body simulations in order to characterise the dynamical evolution of planetary systems similar to our own Solar system in different star cluster environments. We follow the evolution of hundreds of planetary systems containing a Solar-mass star and six planets (Earth, Mars, Jupiter, Saturn, Uranus, and Neptune) and analyse their dynamical fates. Our main results can be summarised as follows:

\begin{enumerate}
\item The star cluster environments affects planetary systems through the direct effect of stellar neighbours on planetary systems, and through the subsequent gravitational interaction between planets in perturbed planetary systems. The latter can result in planetary escape events or star-planet mergers immediately after the stellar encounter, or up to tens of millions of years later. 
\item The large majority of encounters between planetary systems and neighbours stars is tidal, hyperbolic, and adiabatic in nature, although near-parabolic encounters also occur frequently. The effect of a stellar encounter depends on the planetary semi-major axis, the degree in which an encounter is adiabatic and the degree in which an encounter is tidal. The outermost planets in the system occasionally experience non-adiabatic and impulsive encounters that strongly perturb their orbits. Terrestrial planets, on the other hand are rarely directly affected by stellar encounters. Instead, terrestrial planets are primarily affected by Jupiter perturbations
\item Most planetary systems remain stable as the star clusters evolve for 50~Myr, despite perturbations in the orbital elements of the outer planets, while others are partially or completely disrupted.   
\item Planetary systems that escape the star cluster with low speeds tend to remain intact, while those ejected with high speeds tend to be severely disrupted prior to, or during the interaction that led to the ejection from the star cluster. This suggests that our Solar system, if indeed it formed in a star cluster, left the star cluster with a comparatively low speed.
\item Overall planetary escape rate (from their host stars) range from 0.3\% to 5.3\%. Escape rates tend to increase for denser star clusters, and for star clusters that are initialised closer to virial equilibrium. We do not find evidence for physical collisions between planets, while star-planet mergers rarely occur.
\item The probability that a planet remains part of its host planetary system depends strongly on the orbital architecture of the planetary system, and in particular on its semi-major axis and mass. In our simulations, in which we model Solar system analogues, we find that the retention rate increases to some extent with semi-major axis, while the dependence on planetary mass is also significant. The latter dependence is a result of planet-planet interactions, and hence a direct consequence the orbital architecture.
\item Due to its high mass, Jupiter often acts as a dynamical barrier in Solar system-like systems, protecting the inner planetary system from external perturbations. However, a strong perturbation of Jupiter itself may result in chaos in the inner parts of the planetary system. In low-density environments, Jupiter thus provides protection against perturbations in the outer planetary system, while in high-density environments, perturbations of Jupiter can easily result in the disruption of short-period planets in the habitable zone. Note that this result is model-dependent: planetary systems with different architectures have different survival probabilities for terrestrial habitable-zone planets.
\item Planetary systems in star clusters evolve differently depending on the frequency and properties of stellar encounters. After dynamical processing in the star cluster, the diversity of planetary system architectures is large, even for planetary systems with identical initial conditions. 
\end{enumerate}

Our study provides insights into how multi-mass, multi-planet systems evolve in star clusters. For simplicity we have not included primordial binary stars in our star clusters. Binary systems have a substantially larger collisional cross section, and may therefore further contribute to the disruption of planetary systems. As the vast majority of stars, particularly the higher-mass stars, forms as part of a binary or higher-order multiple system \citep[e.g.,][]{shatsky2002, kouwenhoven2005, kouwenhoven2007, kobulnicky2007}, it is necessary to further investigate the evolution of planetary systems in star cluster that contain realistic binary fractions.

\section*{Acknowledgements}

We are grateful to the anonymous referee for providing comments and suggestions that helped to improve this paper.
M.B.N.K. acknowledges support from the National Natural Science Foundation of China (grant 11573004). This research was supported by the Research Development Fund (grant RDF-16-01-16) of Xi'an Jiaotong-Liverpool University (XJTLU). 
We acknowledge the support of the DFG priority program SPP 1992 "Exploring the Diversity of Extrasolar Planets (Sp 345/20-1)".  R.S. acknowledges support from National Natural Science Foundation of China (grant 11673032).
M.X.C. acknowledges support from SURFsara (the Dutch National Supercomputing Center) and the EU Horizon 2020 grant No. 671564 (COMPAT project). 
We are grateful to Martin Gorbahn and Qi Shu for discussions that helped to improve the paper. \\

\bibliography{Paper.bib}

\begin{thebibliography}{}
\makeatletter
\relax
\def\mn@urlcharsother{\let\do\@makeother \do\$\do\&\do\#\do\^\do\_\do\%\do\~}
\def\mn@doi{\begingroup\mn@urlcharsother \@ifnextchar [ {\mn@doi@}
  {\mn@doi@[]}}
\def\mn@doi@[#1]#2{\def\@tempa{#1}\ifx\@tempa\@empty \href
  {http://dx.doi.org/#2} {doi:#2}\else \href {http://dx.doi.org/#2} {#1}\fi
  \endgroup}
\def\mn@eprint#1#2{\mn@eprint@#1:#2::\@nil}
\def\mn@eprint@arXiv#1{\href {http://arxiv.org/abs/#1} {{\tt arXiv:#1}}}
\def\mn@eprint@dblp#1{\href {http://dblp.uni-trier.de/rec/bibtex/#1.xml}
  {dblp:#1}}
\def\mn@eprint@#1:#2:#3:#4\@nil{\def\@tempa {#1}\def\@tempb {#2}\def\@tempc
  {#3}\ifx \@tempc \@empty \let \@tempc \@tempb \let \@tempb \@tempa \fi \ifx
  \@tempb \@empty \def\@tempb {arXiv}\fi \@ifundefined
  {mn@eprint@\@tempb}{\@tempb:\@tempc}{\expandafter \expandafter \csname
  mn@eprint@\@tempb\endcsname \expandafter{\@tempc}}}

\bibitem[\protect\citeauthoryear{{Aarseth}}{{Aarseth}}{1999}]{1j}
{Aarseth} S.~J.,  1999, \mn@doi [Celest. Mech. Dyn. Astron.]
  {10.1023/A:1008390828807}, \href
  {http://adsabs.harvard.edu/abs/1999CeMDA..73..127A} {73, 127}

\bibitem[\protect\citeauthoryear{{Aarseth}}{{Aarseth}}{2010}]{gnbs}
{Aarseth} S.~J.,  2010, {Gravitational N-Body Simulations}.
Cambridge Monographs on Mathematical Physics

\bibitem[\protect\citeauthoryear{{Adams}}{{Adams}}{2010}]{adams2010}
{Adams} F.~C.,  2010, \mn@doi [\araa] {10.1146/annurev-astro-081309-130830},
  \href {http://adsabs.harvard.edu/abs/2010ARA%26A..48...47A} {48, 47}

\bibitem[\protect\citeauthoryear{{Allison}, {Goodwin}, {Parker}, {de Grijs},
  {Portegies Zwart}  \& {Kouwenhoven}}{{Allison} et~al.}{2009}]{allison2009}
{Allison} R.~J.,  {Goodwin} S.~P.,  {Parker} R.~J.,  {de Grijs} R.,  {Portegies
  Zwart} S.~F.,   {Kouwenhoven} M.~B.~N.,  2009, \mn@doi [\apjl]
  {10.1088/0004-637X/700/2/L99}, \href
  {http://adsabs.harvard.edu/abs/2009ApJ...700L..99A} {700, L99}

\bibitem[\protect\citeauthoryear{{Belczynski}, {Kalogera}  \&
  {Bulik}}{{Belczynski} et~al.}{2002}]{belczynski2002}
{Belczynski} K.,  {Kalogera} V.,   {Bulik} T.,  2002, \mn@doi [\apj]
  {10.1086/340304}, \href {http://adsabs.harvard.edu/abs/2002ApJ...572..407B}
  {572, 407}

\bibitem[\protect\citeauthoryear{{Binney} \& {Tremaine}}{{Binney} \&
  {Tremaine}}{2008}]{3b}
{Binney} J.,  {Tremaine} S.,  2008, {Galactic Dynamics: Second Edition}.
Princeton University Press

\bibitem[\protect\citeauthoryear{{Cai}, Meiron, Kouwenhoven, Assmann  \&
  Spurzem}{{Cai} et~al.}{2015}]{bts}
{Cai} M.~X.,  Meiron Y.,  Kouwenhoven M.,  Assmann P.,   Spurzem R.,  2015,
  ApJ, 219, 12

\bibitem[\protect\citeauthoryear{{Cai}, {Kouwenhoven}, {Portegies Zwart}  \&
  {Spurzem}}{{Cai} et~al.}{2017}]{1i}
{Cai} M.~X.,  {Kouwenhoven} M.~B.~N.,  {Portegies Zwart} S.~F.,   {Spurzem} R.,
   2017, \mn@doi [\mnras] {10.1093/mnras/stx1464}, \href
  {http://adsabs.harvard.edu/abs/2017MNRAS.470.4337C} {470, 4337}

\bibitem[\protect\citeauthoryear{{Cai}, {Portegies Zwart}  \& {van
  Elteren}}{{Cai} et~al.}{2018}]{caisignatures2018}
{Cai} M.~X.,  {Portegies Zwart} S.,   {van Elteren} A.,  2018, \mn@doi [\mnras]
  {10.1093/mnras/stx3064}, \href
  {http://adsabs.harvard.edu/abs/2018MNRAS.474.5114C} {474, 5114}

\bibitem[\protect\citeauthoryear{{Cai}, {Portegies Zwart}, {Kouwenhoven}  \&
  {Spurzem}}{{Cai} et~al.}{2019}]{cai2019}
{Cai} M.~X.,  {Portegies Zwart} S.,  {Kouwenhoven} M.~B.~N.,   {Spurzem} R.,
  2019, arXiv e-prints, \href
  {http://adsabs.harvard.edu/abs/2019arXiv190302316C} {}

\bibitem[\protect\citeauthoryear{{Chambers}}{{Chambers}}{1999}]{1n}
{Chambers} J.~E.,  1999, \mn@doi [\mnras] {10.1046/j.1365-8711.1999.02379.x},
  \href {http://adsabs.harvard.edu/abs/1999MNRAS.304..793C} {304, 793}

\bibitem[\protect\citeauthoryear{{Flammini Dotti}, {Cai}, {Spurzem}  \&
  {Kouwenhoven}}{{Flammini Dotti} et~al.}{2018}]{flamminidotti2018}
{Flammini Dotti} F.,  {Cai} M.~X.,  {Spurzem} R.,   {Kouwenhoven} M.~B.~N.,
  2018, arXiv e-prints, \href
  {http://adsabs.harvard.edu/abs/2018arXiv181112660F} {}

\bibitem[\protect\citeauthoryear{{Fujii} \& {Hori}}{{Fujii} \&
  {Hori}}{2019}]{fujii2019}
{Fujii} M.~S.,  {Hori} Y.,  2019, \mn@doi [\aap] {10.1051/0004-6361/201834677},
  \href {https://ui.adsabs.harvard.edu/abs/2019A&A...624A.110F} {624, A110}

\bibitem[\protect\citeauthoryear{{Fujii} \& {Portegies Zwart}}{{Fujii} \&
  {Portegies Zwart}}{2015}]{fujii2015}
{Fujii} M.~S.,  {Portegies Zwart} S.,  2015, \mn@doi [\mnras]
  {10.1093/mnras/stv293}, \href
  {http://adsabs.harvard.edu/abs/2015MNRAS.449..726F} {449, 726}

\bibitem[\protect\citeauthoryear{Fujii, Saitoh  \& Portegies~Zwart}{Fujii
  et~al.}{2012}]{fujii2012}
Fujii M.~S.,  Saitoh T.~R.,   Portegies~Zwart S.~F.,  2012, Apjl, 753, 12

\bibitem[\protect\citeauthoryear{{Giersz} \& {Spurzem}}{{Giersz} \&
  {Spurzem}}{2000}]{gierszspurzem2000}
{Giersz} M.,  {Spurzem} R.,  2000, \mn@doi [\mnras]
  {10.1046/j.1365-8711.2000.03727.x}, \href
  {http://adsabs.harvard.edu/abs/2000MNRAS.317..581G} {317, 581}

\bibitem[\protect\citeauthoryear{{Giersz} \& {Spurzem}}{{Giersz} \&
  {Spurzem}}{2003}]{mac1}
{Giersz} M.,  {Spurzem} R.,  2003, \mn@doi [\mnras]
  {10.1046/j.1365-8711.2003.06717.x}, \href
  {http://adsabs.harvard.edu/abs/2003MNRAS.343..781G} {343, 781}

\bibitem[\protect\citeauthoryear{{Girichidis}, {Federrath}, {Allison},
  {Banerjee}  \& {Klessen}}{{Girichidis} et~al.}{2012}]{philipp2012}
{Girichidis} P.,  {Federrath} C.,  {Allison} R.,  {Banerjee} R.,   {Klessen}
  R.~S.,  2012, \mn@doi [\mnras] {10.1111/j.1365-2966.2011.20250.x}, \href
  {http://adsabs.harvard.edu/abs/2012MNRAS.420.3264G} {420, 3264}

\bibitem[\protect\citeauthoryear{{Goodwin} \& {Whitworth}}{{Goodwin} \&
  {Whitworth}}{2004}]{goodwin2004}
{Goodwin} S.~P.,  {Whitworth} A.~P.,  2004, \mn@doi [\aap]
  {10.1051/0004-6361:20031529}, \href
  {http://adsabs.harvard.edu/abs/2004A%26A...413..929G} {413, 929}

\bibitem[\protect\citeauthoryear{{Gould} et~al.,}{{Gould} et~al.}{2014}]{1b}
{Gould} A.,  et~al., 2014, \mn@doi [Science] {10.1126/science.1251527}, \href
  {http://adsabs.harvard.edu/abs/2014Sci...345...46G} {345, 46}

\bibitem[\protect\citeauthoryear{{Gvaramadze}, {Gualandris}  \& {Portegies
  Zwart}}{{Gvaramadze} et~al.}{2009}]{3c}
{Gvaramadze} V.~V.,  {Gualandris} A.,   {Portegies Zwart} S.,  2009, \mn@doi
  [\mnras] {10.1111/j.1365-2966.2009.14809.x}, \href
  {http://adsabs.harvard.edu/abs/2009MNRAS.396..570G} {396, 570}

\bibitem[\protect\citeauthoryear{Hao, Kouwenhoven  \& Spurzem}{Hao
  et~al.}{2013}]{1h}
Hao W.,  Kouwenhoven M. B.~N.,   Spurzem R.,  2013, \mn@doi [MNRAS]
  {10.1093/mnras/stt771}, 433, 867

\bibitem[\protect\citeauthoryear{{Heggie}}{{Heggie}}{2006}]{k1}
{Heggie} D.~C.,  2006, {Gravitational Scattering}.
University of Turku, p.~20

\bibitem[\protect\citeauthoryear{{Heggie} \& Rasio}{{Heggie} \&
  Rasio}{1996}]{k3}
{Heggie} D.~C.,  Rasio F.~A.,  1996, MNRAS, 282, 1064

\bibitem[\protect\citeauthoryear{Hester, Healy  \& Desch}{Hester
  et~al.}{2004}]{11c}
Hester J.,  Healy K.,   Desch S.,  2004, AAS, 36, 1516

\bibitem[\protect\citeauthoryear{{Hillenbrand}}{{Hillenbrand}}{2005}]{1d}
{Hillenbrand} L.~A.,  2005, STScI Symposium Series, \href
  {http://adsabs.harvard.edu/abs/2005astro.ph.11083H} {19, 20}

\bibitem[\protect\citeauthoryear{{Hurley}, {Pols}  \& {Tout}}{{Hurley}
  et~al.}{2000}]{hurley2000}
{Hurley} J.~R.,  {Pols} O.~R.,   {Tout} C.~A.,  2000, \mn@doi [\mnras]
  {10.1046/j.1365-8711.2000.03426.x}, \href
  {http://adsabs.harvard.edu/abs/2000MNRAS.315..543H} {315, 543}

\bibitem[\protect\citeauthoryear{{Hurley}, {Tout}  \& {Pols}}{{Hurley}
  et~al.}{2002}]{hurley2002}
{Hurley} J.~R.,  {Tout} C.~A.,   {Pols} O.~R.,  2002, \mn@doi [\mnras]
  {10.1046/j.1365-8711.2002.05038.x}, \href
  {http://adsabs.harvard.edu/abs/2002MNRAS.329..897H} {329, 897}

\bibitem[\protect\citeauthoryear{{Hurley}, {Pols}, {Aarseth}  \&
  {Tout}}{{Hurley} et~al.}{2005}]{hurley2005}
{Hurley} J.~R.,  {Pols} O.~R.,  {Aarseth} S.~J.,   {Tout} C.~A.,  2005, \mn@doi
  [\mnras] {10.1111/j.1365-2966.2005.09448.x}, \href
  {http://adsabs.harvard.edu/abs/2005MNRAS.363..293H} {363, 293}

\bibitem[\protect\citeauthoryear{{Hurley}, {Tout}  \& {Pols}}{{Hurley}
  et~al.}{2013b}]{hurley2013bse}
{Hurley} J.~R.,  {Tout} C.~A.,   {Pols} O.~R.,  2013b, {BSE: Binary Star
  Evolution}, Astrophysics Source Code Library (\mn@eprint {ascl} {1303.014})

\bibitem[\protect\citeauthoryear{{Hurley}, {Pols}  \& {Tout}}{{Hurley}
  et~al.}{2013a}]{hurley2013sse}
{Hurley} J.~R.,  {Pols} O.~R.,   {Tout} C.~A.,  2013a, {SSE: Single Star
  Evolution}, Astrophysics Source Code Library (\mn@eprint {ascl} {1303.015})

\bibitem[\protect\citeauthoryear{{Khalisi}, {Amaro-Seoane}  \&
  {Spurzem}}{{Khalisi} et~al.}{2007}]{khalisi2007}
{Khalisi} E.,  {Amaro-Seoane} P.,   {Spurzem} R.,  2007, \mn@doi [\mnras]
  {10.1111/j.1365-2966.2006.11184.x}, \href
  {http://adsabs.harvard.edu/abs/2007MNRAS.374..703K} {374, 703}

\bibitem[\protect\citeauthoryear{Kobulnicky \& Fryer}{Kobulnicky \&
  Fryer}{2007}]{kobulnicky2007}
Kobulnicky H.,  Fryer C.~L.,  2007, ApJ, 670, 747

\bibitem[\protect\citeauthoryear{{Kokubo}, {Yoshinaga}  \& {Makino}}{{Kokubo}
  et~al.}{1998}]{kokubo1998}
{Kokubo} E.,  {Yoshinaga} K.,   {Makino} J.,  1998, \mn@doi [\mnras]
  {10.1046/j.1365-8711.1998.01581.x}, \href
  {http://adsabs.harvard.edu/abs/1998MNRAS.297.1067K} {297, 1067}

\bibitem[\protect\citeauthoryear{{Kouwenhoven}, {Brown}, {Zinnecker}, {Kaper}
  \& {Portegies Zwart}}{{Kouwenhoven} et~al.}{2005}]{kouwenhoven2005}
{Kouwenhoven} M.~B.~N.,  {Brown} A.~G.~A.,  {Zinnecker} H.,  {Kaper} L.,
  {Portegies Zwart} S.~F.,  2005, \mn@doi [\aap] {10.1051/0004-6361:20048124},
  \href {http://adsabs.harvard.edu/abs/2005A%26A...430..137K} {430, 137}

\bibitem[\protect\citeauthoryear{{Kouwenhoven}, {Brown}, {Portegies Zwart}  \&
  {Kaper}}{{Kouwenhoven} et~al.}{2007}]{kouwenhoven2007}
{Kouwenhoven} M.~B.~N.,  {Brown} A.~G.~A.,  {Portegies Zwart} S.~F.,   {Kaper}
  L.,  2007, \mn@doi [\aap] {10.1051/0004-6361:20077719}, \href
  {http://adsabs.harvard.edu/abs/2007A%26A...474...77K} {474, 77}

\bibitem[\protect\citeauthoryear{{Kroupa}}{{Kroupa}}{2001}]{2b}
{Kroupa} P.,  2001, \mn@doi [\mnras] {10.1046/j.1365-8711.2001.04022.x}, \href
  {http://adsabs.harvard.edu/abs/2001MNRAS.322..231K} {322, 231}

\bibitem[\protect\citeauthoryear{{Lada} \& {Lada}}{{Lada} \& {Lada}}{2003}]{1c}
{Lada} C.~J.,  {Lada} E.~A.,  2003, \mn@doi [\araa]
  {10.1146/annurev.astro.41.011802.094844}, \href
  {http://adsabs.harvard.edu/abs/2003ARA%26A..41...57L} {41, 57}

\bibitem[\protect\citeauthoryear{{Lamers}, {Gieles}  \& {Portegies
  Zwart}}{{Lamers} et~al.}{2005}]{lamers2005}
{Lamers} H.~J.~G.~L.~M.,  {Gieles} M.,   {Portegies Zwart} S.~F.,  2005,
  \mn@doi [\aap] {10.1051/0004-6361:20041476}, \href
  {http://adsabs.harvard.edu/abs/2005A%26A...429..173L} {429, 173}

\bibitem[\protect\citeauthoryear{{Looney}}{{Looney}}{2006}]{11d}
{Looney} L.~W.,  2006, in {Backer} D.~C.,  {Moran} J.~M.,   {Turner} J.~L.,
  eds,  Astronomical Society of the Pacific Conference Series Vol. 356,
  Revealing the Molecular Universe: One Antenna is Never Enough. p.~177

\bibitem[\protect\citeauthoryear{{Malmberg}, {de Angeli}, {Davies}, {Church},
  {Mackey}  \& {Wilkinson}}{{Malmberg} et~al.}{2007}]{1f}
{Malmberg} D.,  {de Angeli} F.,  {Davies} M.~B.,  {Church} R.~P.,  {Mackey} D.,
    {Wilkinson} M.~I.,  2007, \mn@doi [\mnras]
  {10.1111/j.1365-2966.2007.11885.x}, \href
  {http://adsabs.harvard.edu/abs/2007MNRAS.378.1207M} {378, 1207}

\bibitem[\protect\citeauthoryear{{Malmberg}, {Davies}  \& {Heggie}}{{Malmberg}
  et~al.}{2011}]{malmberg2011}
{Malmberg} D.,  {Davies} M.~B.,   {Heggie} D.~C.,  2011, \mn@doi [\mnras]
  {10.1111/j.1365-2966.2010.17730.x}, \href
  {http://adsabs.harvard.edu/abs/2011MNRAS.411..859M} {411, 859}

\bibitem[\protect\citeauthoryear{{Mardling}}{{Mardling}}{2008}]{mardling2008}
{Mardling} R.~A.,  2008, in {Vesperini} E.,  {Giersz} M.,   {Sills} A.,  eds,
  IAU Symposium Vol. 246, Dynamical Evolution of Dense Stellar Systems. pp
  199--208, \mn@doi{10.1017/S1743921308015615}

\bibitem[\protect\citeauthoryear{Mayo, Vanderburg  \& et al}{Mayo
  et~al.}{2018}]{1a}
Mayo A.~W.,  Vanderburg A.,   et al 2018, ApJ, 136, 25

\bibitem[\protect\citeauthoryear{{McMillan}, {Portegies Zwart}, {van Elteren}
  \& {Whitehead}}{{McMillan} et~al.}{2012}]{1q}
{McMillan} S.,  {Portegies Zwart} S.,  {van Elteren} A.,   {Whitehead} A.,
  2012, in {Capuzzo-Dolcetta} R.,  {Limongi} M.,   {Tornamb{\`e}} A.,  eds,
  Astronomical Society of the Pacific Conference Series Vol. 453, Advances in
  Computational Astrophysics: Methods, Tools, and Outcome. p.~129

\bibitem[\protect\citeauthoryear{{Mikkola} \& {Aarseth}}{{Mikkola} \&
  {Aarseth}}{1993}]{1m}
{Mikkola} S.,  {Aarseth} S.~J.,  1993, \mn@doi [Celest. Mech. Dyn. Astron.]
  {10.1007/BF00695714}, \href
  {http://adsabs.harvard.edu/abs/1993CeMDA..57..439M} {57, 439}

\bibitem[\protect\citeauthoryear{{Mikkola} \& {Aarseth}}{{Mikkola} \&
  {Aarseth}}{1998}]{mikkola1998}
{Mikkola} S.,  {Aarseth} S.~J.,  1998, \mn@doi [NewA]
  {10.1016/S1384-1076(98)00018-9}, \href
  {http://adsabs.harvard.edu/abs/1998NewA....3..309M} {3, 309}

\bibitem[\protect\citeauthoryear{{Mouri} \& {Taniguchi}}{{Mouri} \&
  {Taniguchi}}{2002}]{3a}
{Mouri} H.,  {Taniguchi} Y.,  2002, \mn@doi [\apj] {10.1086/343851}, \href
  {http://adsabs.harvard.edu/abs/2002ApJ...580..844M} {580, 844}

\bibitem[\protect\citeauthoryear{Olczak, Kaczmarek, Harfst, Pfalzner  \&
  Portegies~Zwart}{Olczak et~al.}{2012}]{olczak2012}
Olczak C.,  Kaczmarek T.,  Harfst S.,  Pfalzner S.,   Portegies~Zwart S.,
  2012, ApJ, 756, 15

\bibitem[\protect\citeauthoryear{{Parker}, {Wright}, {Goodwin}  \&
  {Meyer}}{{Parker} et~al.}{2014}]{parker}
{Parker} R.~J.,  {Wright} N.~J.,  {Goodwin} S.~P.,   {Meyer} M.~R.,  2014,
  \mn@doi [\mnras] {10.1093/mnras/stt2231}, \href
  {http://adsabs.harvard.edu/abs/2014MNRAS.438..620P} {438, 620}

\bibitem[\protect\citeauthoryear{{Pelupessy}, {van Elteren}, {de Vries},
  {McMillan}, {Drost}  \& {Portegies Zwart}}{{Pelupessy}
  et~al.}{2013}]{Pelupessy2013}
{Pelupessy} F.~I.,  {van Elteren} A.,  {de Vries} N.,  {McMillan} S.~L.~W.,
  {Drost} N.,   {Portegies Zwart} S.~F.,  2013, VizieR Online Data Catalog,
  \href {http://adsabs.harvard.edu/abs/2013yCat..35570084P} {355}

\bibitem[\protect\citeauthoryear{{Perets} \& {Kouwenhoven}}{{Perets} \&
  {Kouwenhoven}}{2012}]{perets2012}
{Perets} H.~B.,  {Kouwenhoven} M.~B.~N.,  2012, \mn@doi [\apj]
  {10.1088/0004-637X/750/1/83}, \href
  {http://adsabs.harvard.edu/abs/2012ApJ...750...83P} {750, 83}

\bibitem[\protect\citeauthoryear{{Plummer}}{{Plummer}}{1911}]{2a}
{Plummer} H.~C.,  1911, \mn@doi [\mnras] {10.1093/mnras/71.5.460}, \href
  {http://adsabs.harvard.edu/abs/1911MNRAS..71..460P} {71, 460}

\bibitem[\protect\citeauthoryear{{Portegies Zwart}}{{Portegies
  Zwart}}{2011}]{Portegies-Zwart2011}
{Portegies Zwart} S.,  2011, {AMUSE: Astrophysical Multipurpose Software
  Environment}, Astrophysics Source Code Library (\mn@eprint {ascl} {1107.007})

\bibitem[\protect\citeauthoryear{{Portegies Zwart}}{{Portegies
  Zwart}}{2016}]{11b}
{Portegies Zwart} S.~F.,  2016, \mn@doi [\mnras] {10.1093/mnras/stv2831}, \href
  {http://adsabs.harvard.edu/abs/2016MNRAS.457..313P} {457, 313}

\bibitem[\protect\citeauthoryear{{Portegies Zwart} \& {McMillan}}{{Portegies
  Zwart} \& {McMillan}}{2018}]{spzbook}
{Portegies Zwart} S.,  {McMillan} S.,  2018, {Astrophysical Recipes; The art of
  AMUSE}.
AAS, \mn@doi{10.1088/978-0-7503-1320-9}

\bibitem[\protect\citeauthoryear{{Portegies Zwart}, {Pelupessy}, {van Elteren},
  {Wijnen}  \& {Lugaro}}{{Portegies Zwart} et~al.}{2018}]{1e}
{Portegies Zwart} S.,  {Pelupessy} I.,  {van Elteren} A.,  {Wijnen} T.~P.~G.,
  {Lugaro} M.,  2018, \mn@doi [\aap] {10.1051/0004-6361/201732060}, \href
  {http://adsabs.harvard.edu/abs/2018A%26A...616A..85P} {616, A85}

\bibitem[\protect\citeauthoryear{{Portell de Mora}, {Garc{\'{\i}}a-Berro},
  {Estepa}, {Casta{\~n}eda}  \& {Clotet}}{{Portell de Mora}
  et~al.}{2011}]{hdf5}
{Portell de Mora} J.,  {Garc{\'{\i}}a-Berro} E.,  {Estepa} C.,  {Casta{\~n}eda}
  J.,   {Clotet} M.,  2011, in High-Performance Computing in Remote Sensing. p.
  818305, \mn@doi{10.1117/12.898203}

\bibitem[\protect\citeauthoryear{{Pu} \& {Lai}}{{Pu} \& {Lai}}{2018}]{puu}
{Pu} B.,  {Lai} D.,  2018, \mn@doi [\mnras] {10.1093/mnras/sty1098}, \href
  {http://adsabs.harvard.edu/abs/2018MNRAS.478..197P} {478, 197}

\bibitem[\protect\citeauthoryear{{Rein} \& {Liu}}{{Rein} \& {Liu}}{2012}]{1o}
{Rein} H.,  {Liu} S.-F.,  2012, \mn@doi [\aap] {10.1051/0004-6361/201118085},
  \href {http://adsabs.harvard.edu/abs/2012A%26A...537A.128R} {537, A128}

\bibitem[\protect\citeauthoryear{{Roy} \& {Haddow}}{{Roy} \&
  {Haddow}}{2003}]{k2}
{Roy} A.,  {Haddow} M.,  2003, Celest. Mech. Dyn. Astron., \href
  {http://adsabs.harvard.edu/abs/2003CeMDA..87..411R} {87, 411}

\bibitem[\protect\citeauthoryear{{Sabbi} et~al.,}{{Sabbi}
  et~al.}{2012}]{sabbi2012}
{Sabbi} E.,  et~al., 2012, \mn@doi [\apjl] {10.1088/2041-8205/754/2/L37}, \href
  {http://adsabs.harvard.edu/abs/2012ApJ...754L..37S} {754, L37}

\bibitem[\protect\citeauthoryear{Shara, Hurley  \& Mardling}{Shara
  et~al.}{2016}]{shara2016}
Shara M.,  Hurley J.,   Mardling R.,  2016, ApJ, 816, 8

\bibitem[\protect\citeauthoryear{{Shatsky} \& {Tokovinin}}{{Shatsky} \&
  {Tokovinin}}{2002}]{shatsky2002}
{Shatsky} N.,  {Tokovinin} A.,  2002, \mn@doi [\aap]
  {10.1051/0004-6361:20011542}, \href
  {http://adsabs.harvard.edu/abs/2002A%26A...382...92S} {382, 92}

\bibitem[\protect\citeauthoryear{{Spitzer}}{{Spitzer}}{1987}]{spitzer}
{Spitzer} L.,  1987, {Dynamical evolution of globular clusters}.
Princeton University Press

\bibitem[\protect\citeauthoryear{{Spurzem}}{{Spurzem}}{1999}]{1l}
{Spurzem} R.,  1999, Journal of Computational and Applied Mathematics, \href
  {http://adsabs.harvard.edu/abs/1999JCoAM.109..407S} {109, 407}

\bibitem[\protect\citeauthoryear{{Spurzem} \& {Giersz}}{{Spurzem} \&
  {Giersz}}{1996}]{spurzemgiersz1996}
{Spurzem} R.,  {Giersz} M.,  1996, \mn@doi [\mnras] {10.1093/mnras/283.3.805},
  \href {http://adsabs.harvard.edu/abs/1996MNRAS.283..805S} {283}

\bibitem[\protect\citeauthoryear{{Spurzem} \& {Takahashi}}{{Spurzem} \&
  {Takahashi}}{1995}]{spurzem1995}
{Spurzem} R.,  {Takahashi} K.,  1995, \mn@doi [\mnras]
  {10.1093/mnras/272.4.772}, \href
  {http://adsabs.harvard.edu/abs/1995MNRAS.272..772S} {272, 772}

\bibitem[\protect\citeauthoryear{{Spurzem}, {Giersz}, {Heggie}  \&
  {Lin}}{{Spurzem} et~al.}{2009}]{1g}
{Spurzem} R.,  {Giersz} M.,  {Heggie} D.~C.,   {Lin} D.~N.~C.,  2009, \mn@doi
  [\apj] {10.1088/0004-637X/697/1/458}, \href
  {http://adsabs.harvard.edu/abs/2009ApJ...697..458S} {697, 458}

\bibitem[\protect\citeauthoryear{{Tapamo}}{{Tapamo}}{2009}]{mpi}
{Tapamo} H.,  2009, in {Gracia} J.,  {de Colle} F.,   {Downes} T.,  eds,
  Lecture Notes in Physics, Berlin Springer Verlag Vol. 791, Jets From Young
  Stars V. p.~3, \mn@doi{10.1007/978-3-642-03370-4_1}

\bibitem[\protect\citeauthoryear{{Thies}, {Kroupa}  \& {Theis}}{{Thies}
  et~al.}{2005}]{thies2005}
{Thies} I.,  {Kroupa} P.,   {Theis} C.,  2005, \mn@doi [\mnras]
  {10.1111/j.1365-2966.2005.09644.x}, \href
  {http://adsabs.harvard.edu/abs/2005MNRAS.364..961T} {364, 961}

\bibitem[\protect\citeauthoryear{Thompson et~al.,}{Thompson
  et~al.}{2018}]{Thompson2018}
Thompson S.,  et~al., 2018, ApJ, 235, 49

\bibitem[\protect\citeauthoryear{Vinncke \& Pfalzner}{Vinncke \&
  Pfalzner}{2018}]{kirsten2018}
Vinncke K.,  Pfalzner S.,  2018, ApJ, 868, 14

\bibitem[\protect\citeauthoryear{{Wang}, {Kouwenhoven}, {Zheng}, {Church}  \&
  {Davies}}{{Wang} et~al.}{2015a}]{wangkouwenhoven2015}
{Wang} L.,  {Kouwenhoven} M.~B.~N.,  {Zheng} X.,  {Church} R.~P.,   {Davies}
  M.~B.,  2015a, \mn@doi [\mnras] {10.1093/mnras/stv542}, \href
  {http://adsabs.harvard.edu/abs/2015MNRAS.449.3543W} {449, 3543}

\bibitem[\protect\citeauthoryear{Wang, Spurzem, Aarseth, Nitadori, Berczik,
  Kouwenhoven  \& Naab}{Wang et~al.}{2015b}]{1p}
Wang L.,  Spurzem R.,  Aarseth S.,  Nitadori K.,  Berczik P.,  Kouwenhoven M.
  B.~N.,   Naab T.,  2015b, \mn@doi [MNRAS] {10.1093/mnras/stv817}, 450, 4070

\bibitem[\protect\citeauthoryear{{Wang} et~al.,}{{Wang}
  et~al.}{2016}]{wang2016}
{Wang} L.,  et~al., 2016, \mn@doi [\mnras] {10.1093/mnras/stw274}, \href
  {http://adsabs.harvard.edu/abs/2016MNRAS.458.1450W} {458, 1450}

\bibitem[\protect\citeauthoryear{{Yoshinaga}, {Kokubo}  \&
  {Makino}}{{Yoshinaga} et~al.}{1999}]{yoshinaga1999}
{Yoshinaga} K.,  {Kokubo} E.,   {Makino} J.,  1999, \mn@doi [\icarus]
  {10.1006/icar.1999.6098}, \href
  {http://adsabs.harvard.edu/abs/1999Icar..139..328Y} {139, 328}

\bibitem[\protect\citeauthoryear{{Zheng}, {Kouwenhoven}  \& {Wang}}{{Zheng}
  et~al.}{2015}]{11z}
{Zheng} X.,  {Kouwenhoven} M.~B.~N.,   {Wang} L.,  2015, \mn@doi [\mnras]
  {10.1093/mnras/stv1832}, \href
  {http://adsabs.harvard.edu/abs/2015MNRAS.453.2759Z} {453, 2759}

\bibitem[\protect\citeauthoryear{{de Grijs}}{{de Grijs}}{2009}]{11a}
{de Grijs} R.,  2009, \mn@doi [\apss] {10.1007/s10509-009-0100-0}, \href
  {http://adsabs.harvard.edu/abs/2009Ap%26SS.324..283D} {324, 283}

\bibitem[\protect\citeauthoryear{{de Grijs}, {Goodwin}, {Kouwenhoven}  \&
  {Kroupa}}{{de Grijs} et~al.}{2008}]{grijs2008}
{de Grijs} R.,  {Goodwin} S.~P.,  {Kouwenhoven} M.~B.~N.,   {Kroupa} P.,  2008,
  \mn@doi [\aap] {10.1051/0004-6361:200810251}, \href
  {http://adsabs.harvard.edu/abs/2008A%26A...492..685D} {492, 685}

\bibitem[\protect\citeauthoryear{{van Elteren}, {Portegies Zwart}, {Pelupessy},
  {Cai}  \& {McMillan}}{{van Elteren} et~al.}{2019}]{vanelteren2019}
{van Elteren} A.,  {Portegies Zwart} S.,  {Pelupessy} I.,  {Cai} M.~X.,
  {McMillan} S.~L.~W.,  2019, \mn@doi [\aap] {10.1051/0004-6361/201834641},
  \href {https://ui.adsabs.harvard.edu/abs/2019A&A...624A.120V} {624, A120}

\makeatother
\end{thebibliography}

\appendix

\section{Orbital element variations during and after close encounters} \label{section:appendix}

\begin{figure*}
\begin{tabular}{cc}
 \includegraphics[width=0.5\textwidth,height=!]{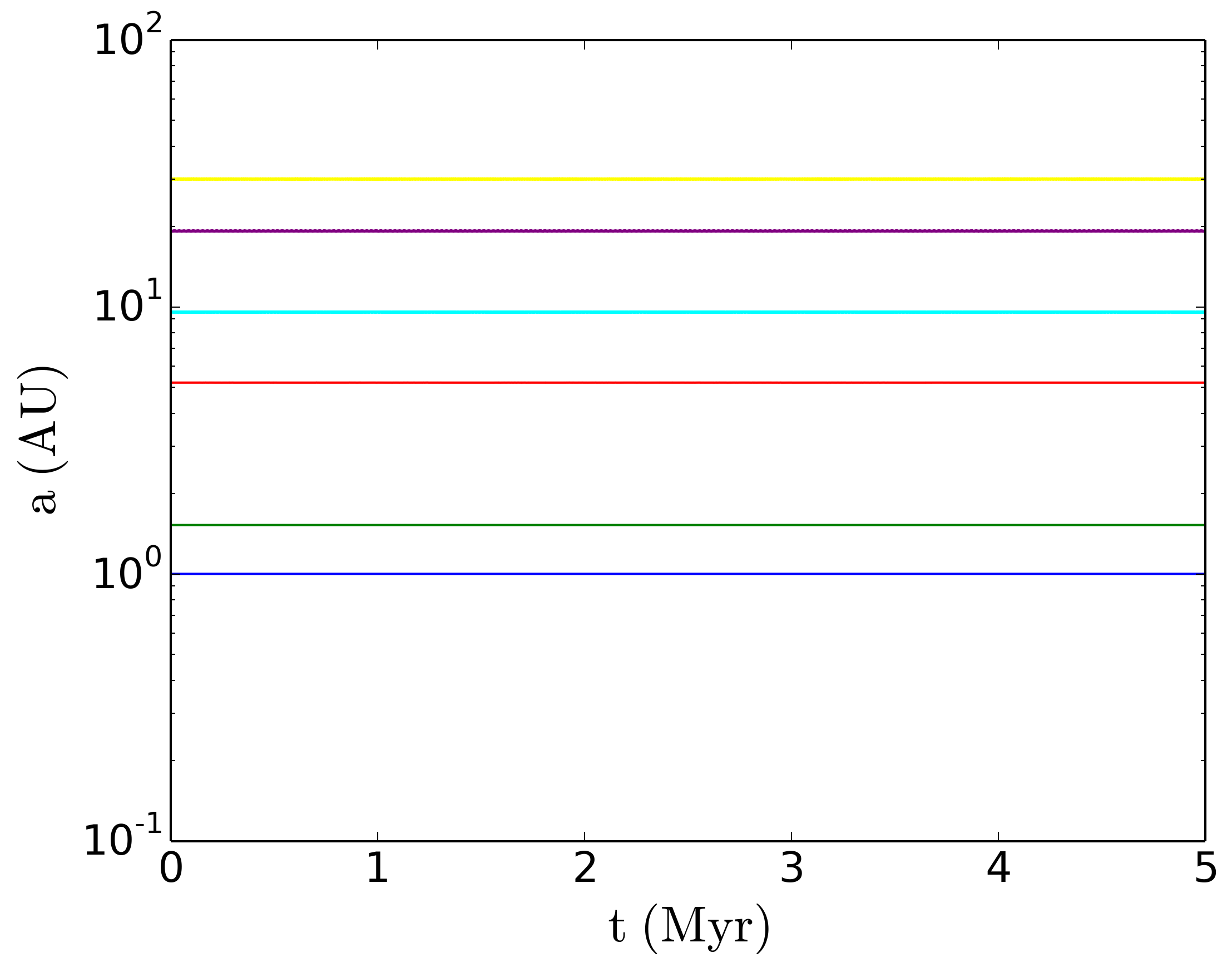} &
 \includegraphics[width=0.5\textwidth,height=!]{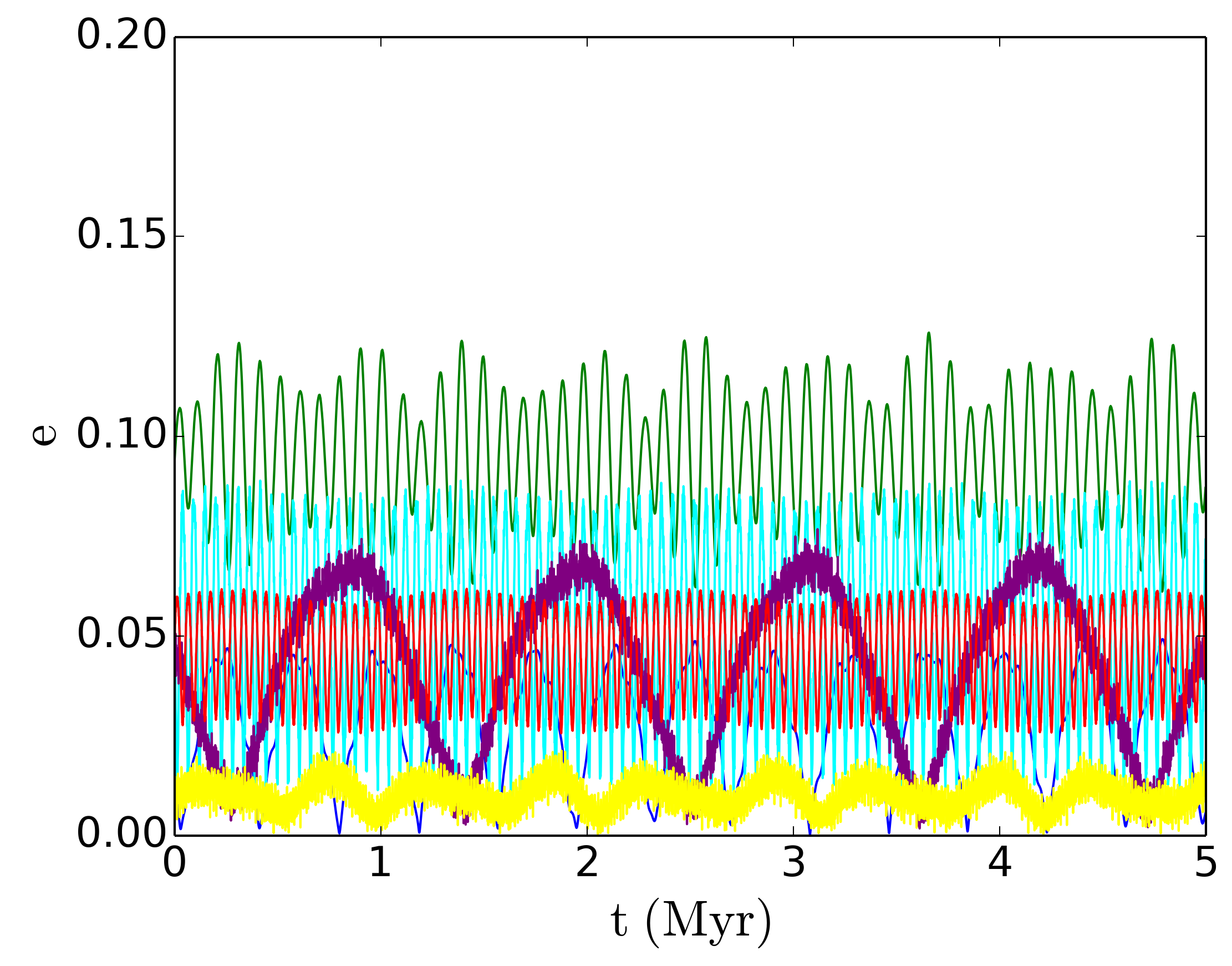} \\
\end{tabular}
\caption{The first 5~Myr of planetary system P194 in star cluster model C041E4 (cf. Figure~\ref{figure:comandchan}; top panels). None of the planets in this system experiences a significant perturbation from neighbouring stars, while planet-planet interactions are responsible for the secular evolution of the system.  }
\label{figure:app1}
\end{figure*}

\begin{figure*}
\begin{tabular}{cc}
 \includegraphics[width=0.5\textwidth,height=!]{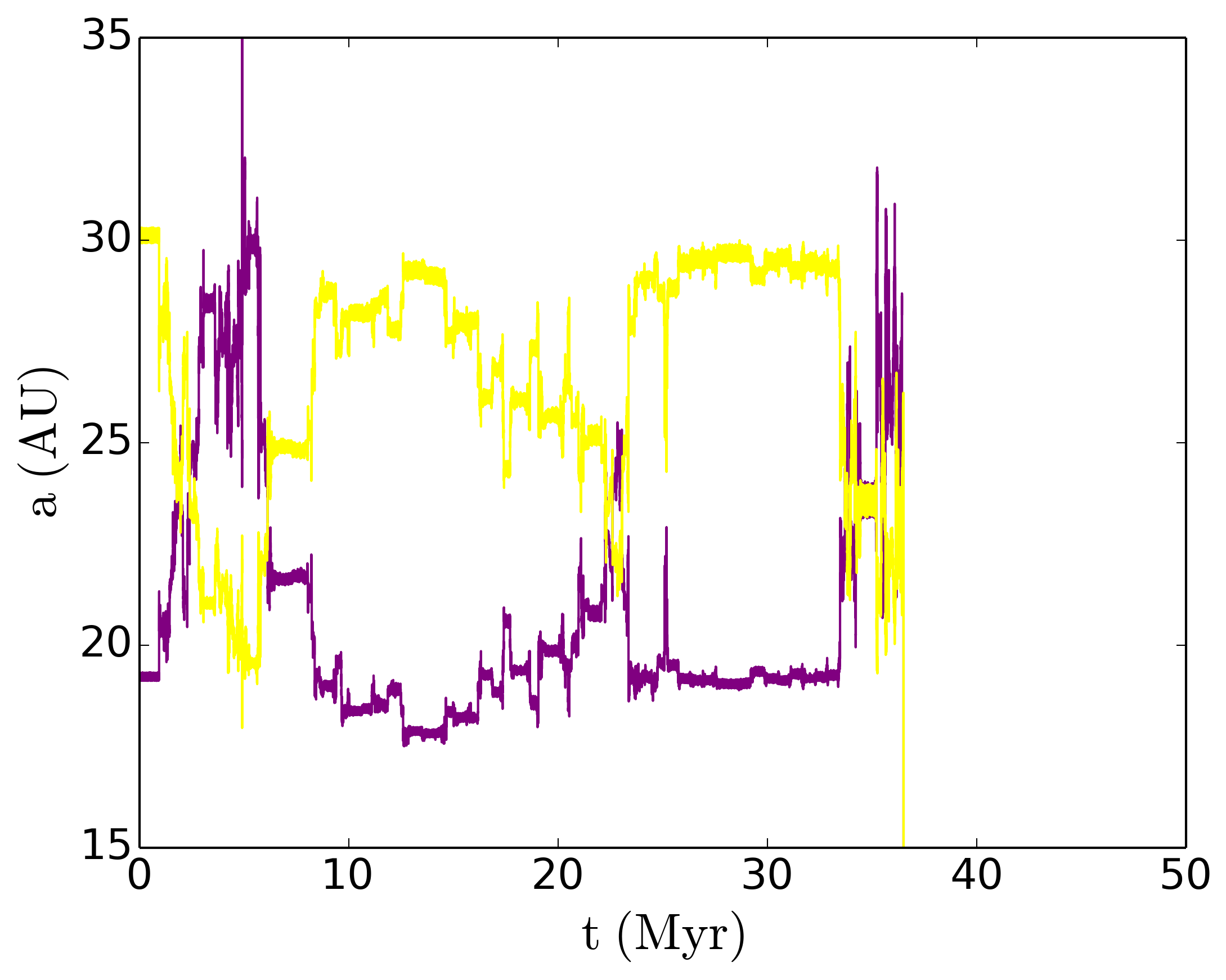} &
 \includegraphics[width=0.5\textwidth,height=!]{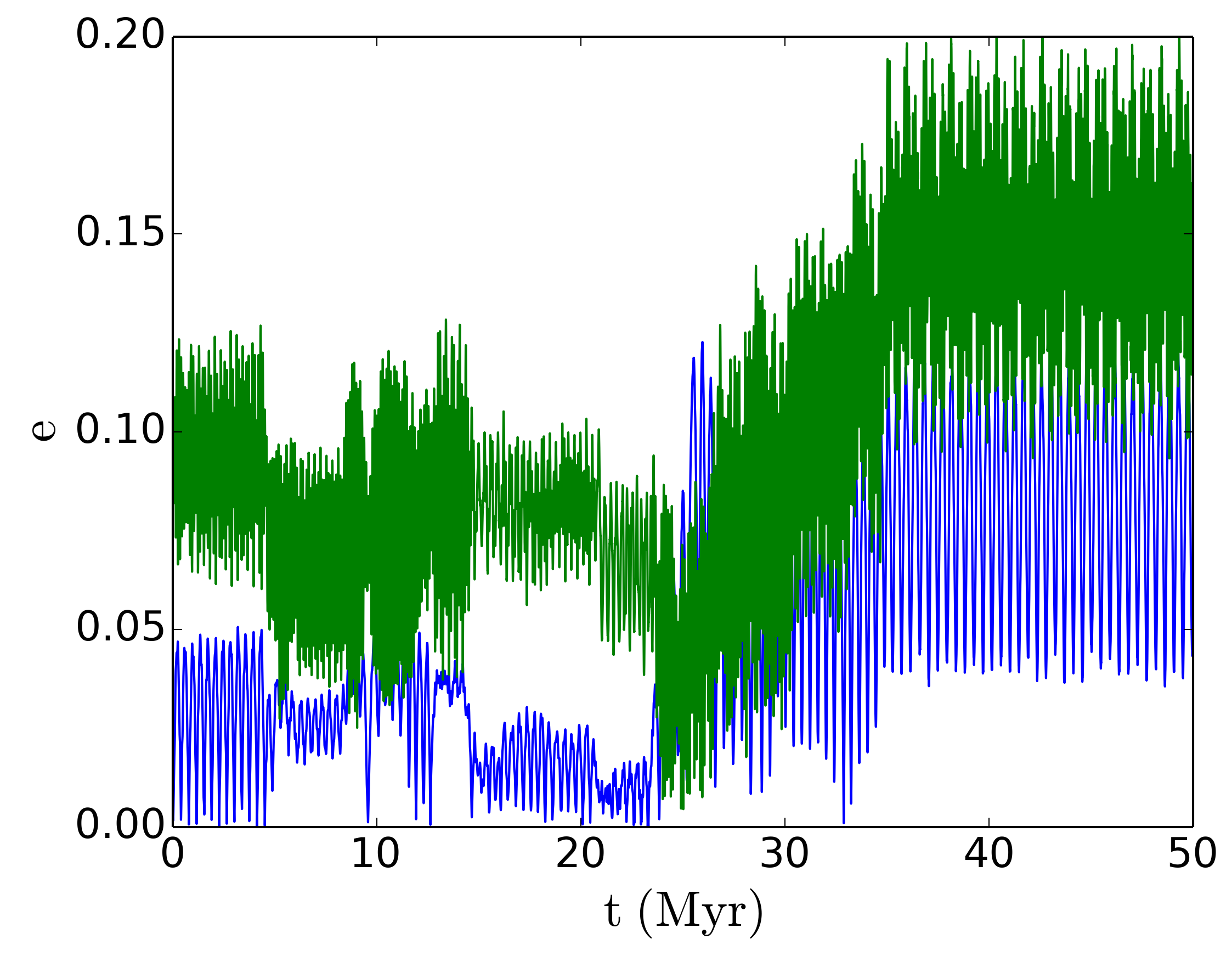} \\
 \includegraphics[width=0.5\textwidth,height=!]{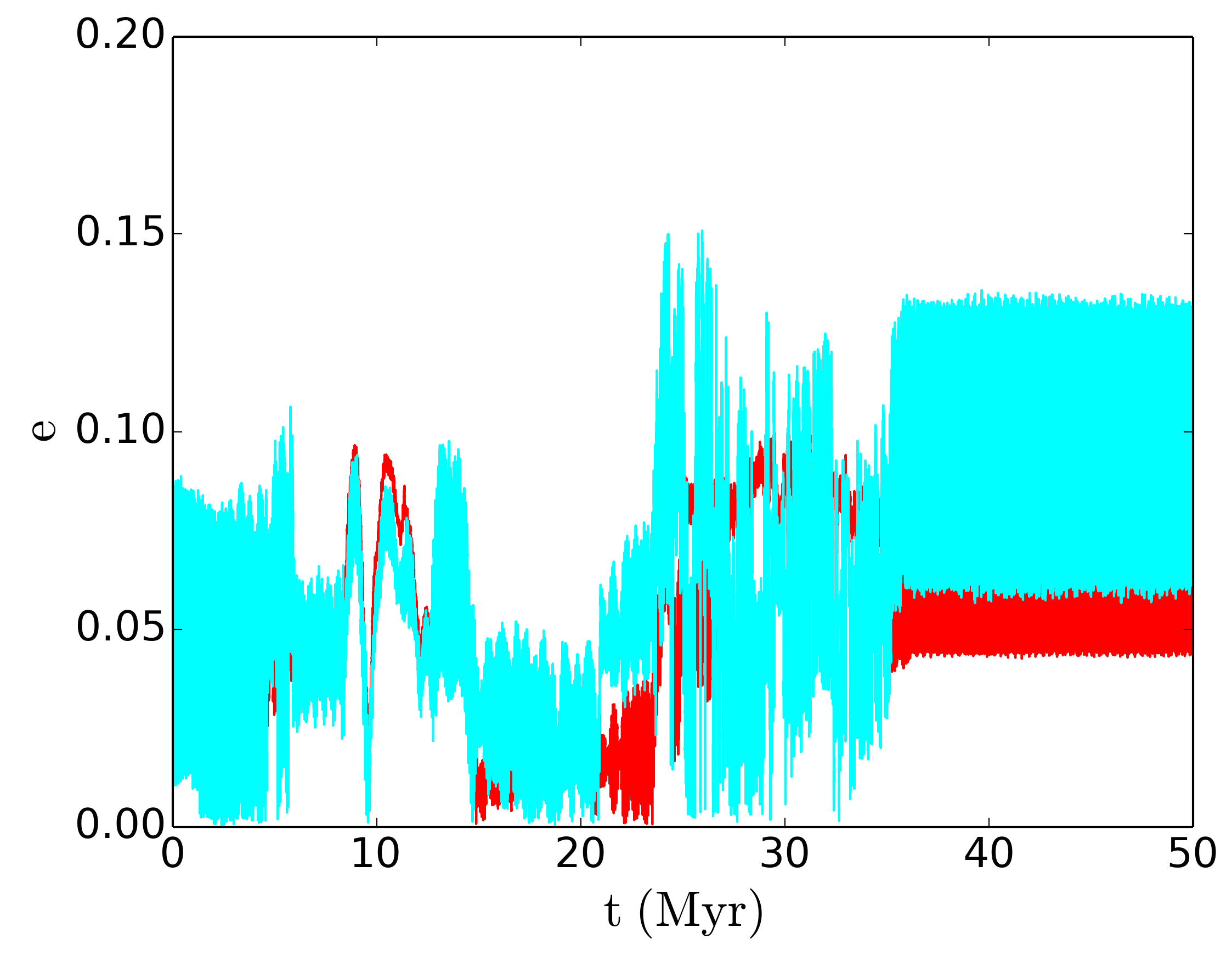} &
 \includegraphics[width=0.5\textwidth,height=!]{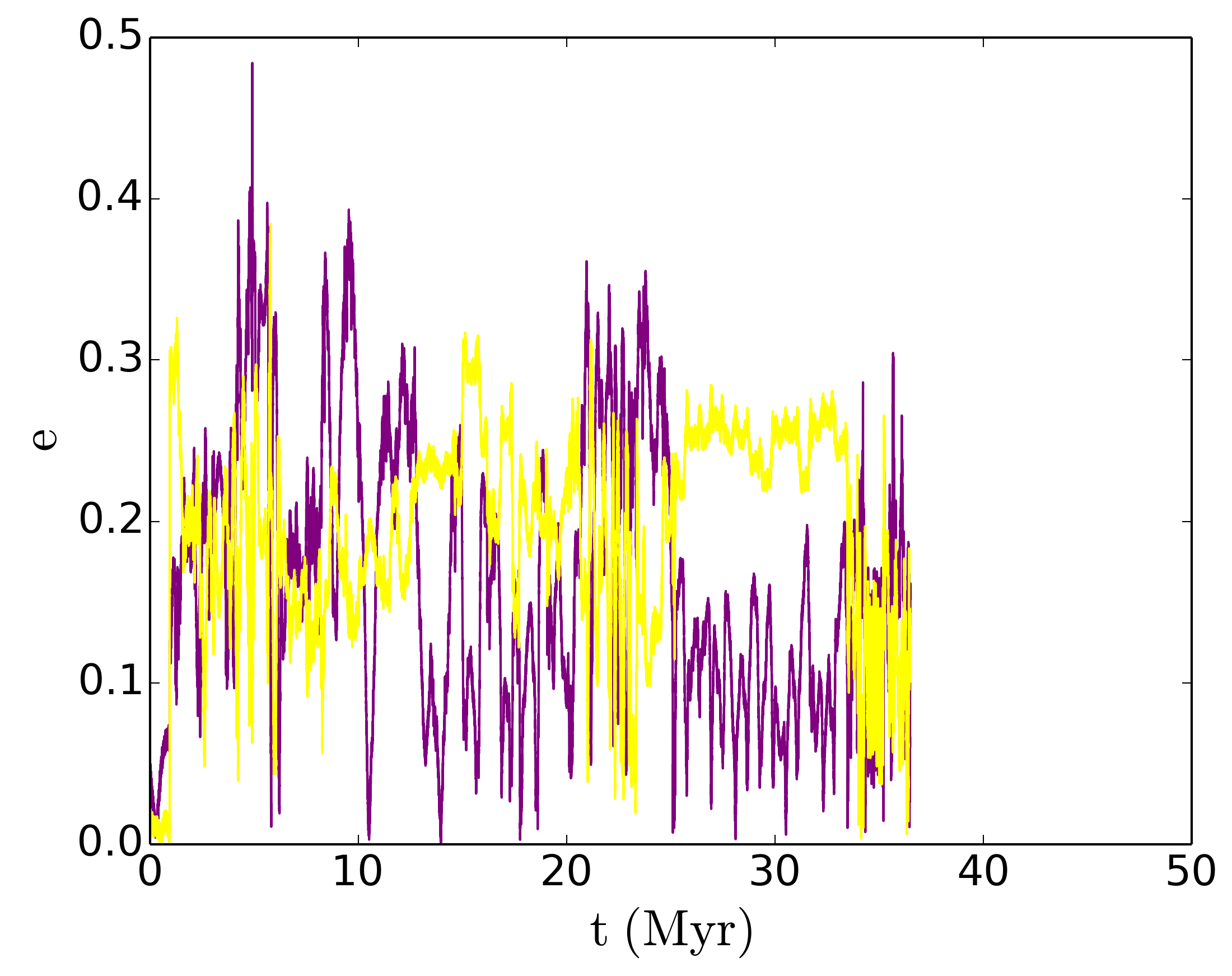} \\
\end{tabular}
\caption{Evolution of planetary system P161 in star cluster model C061E4 (cf. Figure~\ref{figure:comandchan}; bottom panels). The top-left panel shows the semi-major axis evolution of the Uranus and Neptune, while the remaining three panels show the eccentricity evolution of Earth and Mars (top-right), Jupiter and Saturn (bottom-left), and Uranus and Neptune (bottom-right). In this system, perturbations in the orbits of Uranus and Neptune due to interaction with a neighbouring star perturb increase the eccentricities of Jupiter and Saturn, which in turn perturb the terrestrial planets. Both Uranus and Neptune are expelled at $t\approx 38$~Myr, while all other planets remain bound.}
\label{figure:app2}
\end{figure*}

\begin{figure*}
\begin{tabular}{cc}
 \includegraphics[width=0.5\textwidth,height=!]{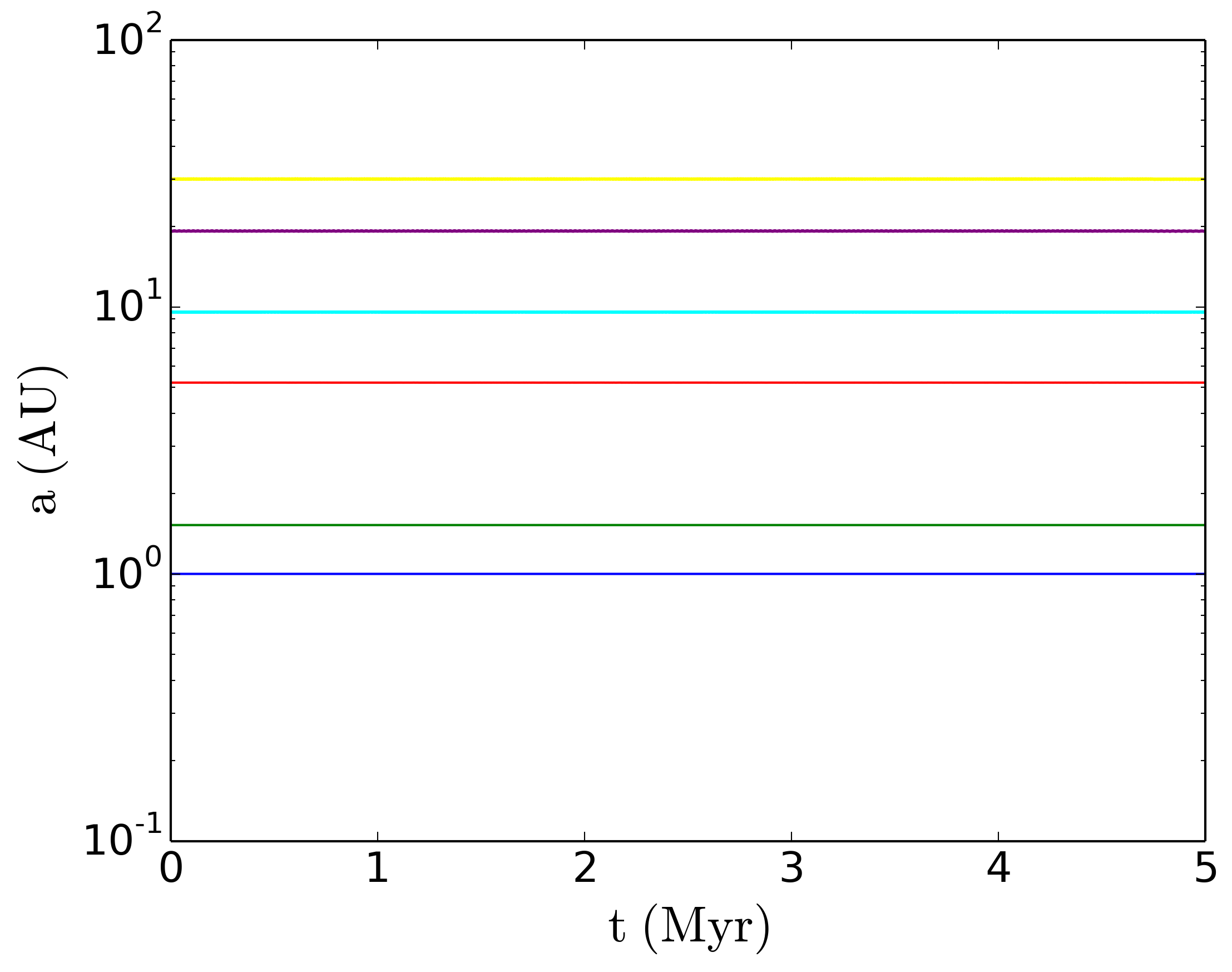} &
 \includegraphics[width=0.5\textwidth,height=!]{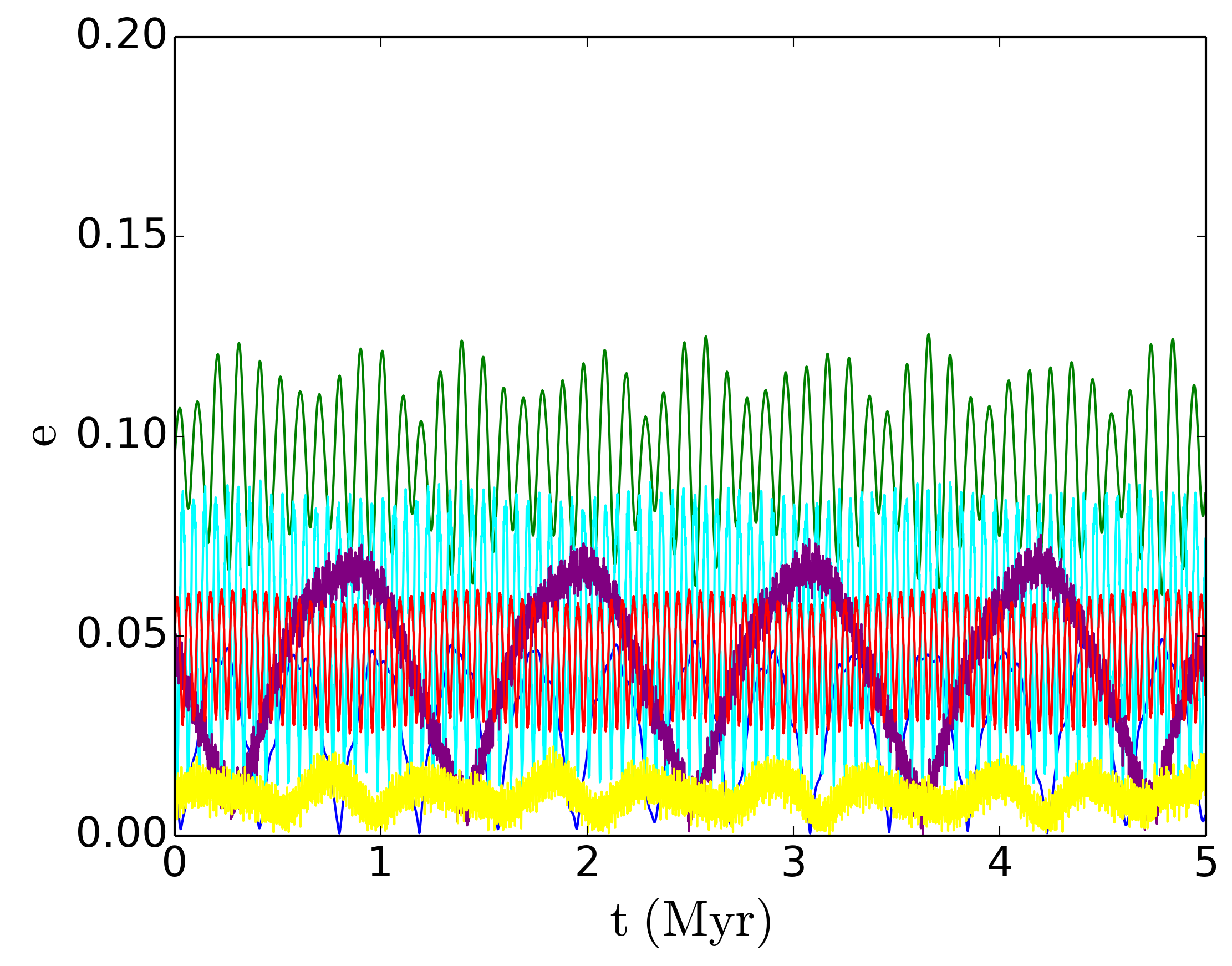} \\
\end{tabular}
\caption{First 5 Myr of planetary system P191 in star cluster model C051E4 (cf. Figure~\ref{figure:highandlowvel}; bottom panels). This planetary system escapes from the star cluster intact, with all planets in their original orbits (see also Figure~\ref{figure:app1}).}
\label{figure:app3}
\end{figure*}

\begin{figure*}
\begin{tabular}{cc}
 \includegraphics[width=0.5\textwidth,height=!]{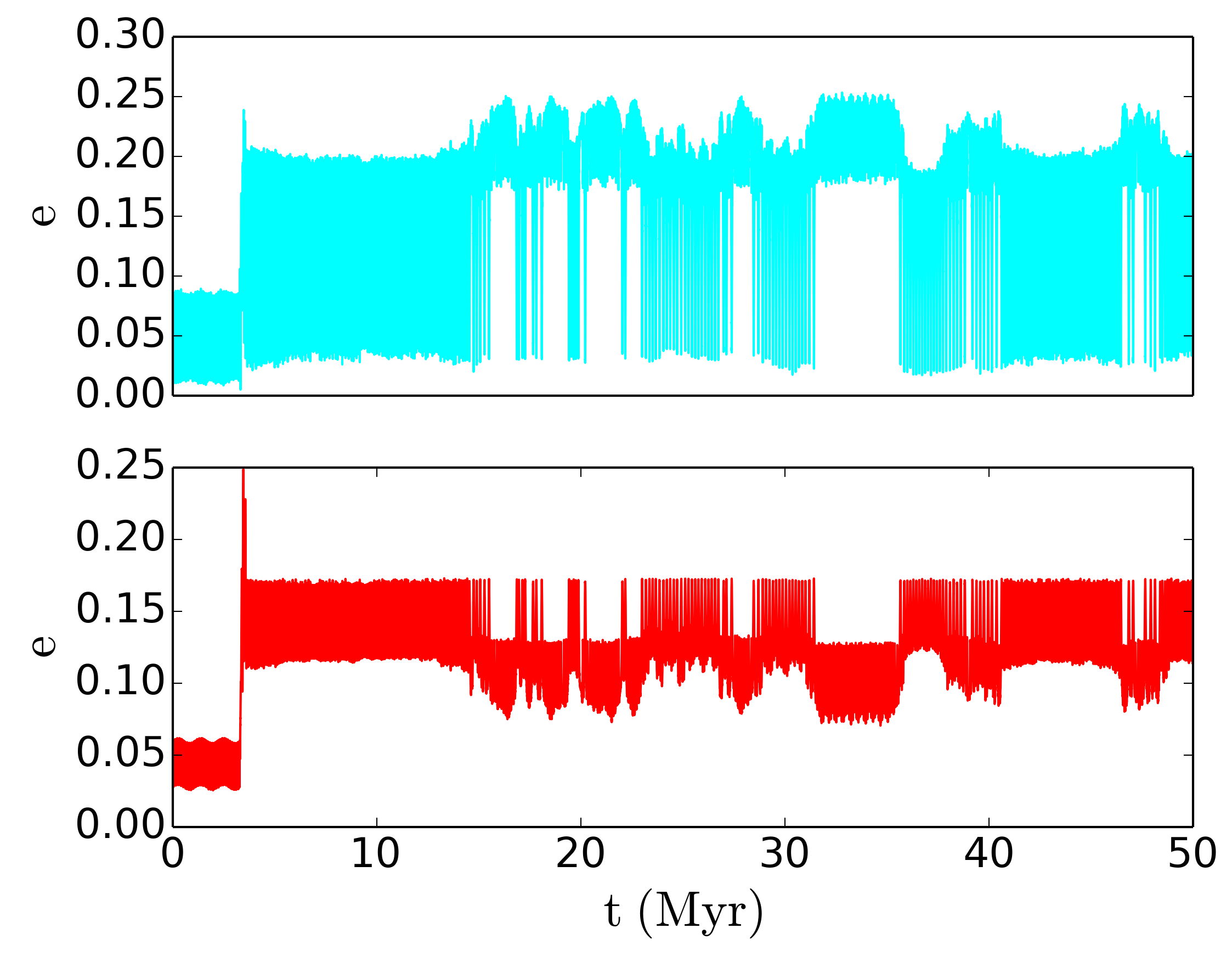} &
 \includegraphics[width=0.5\textwidth,height=!]{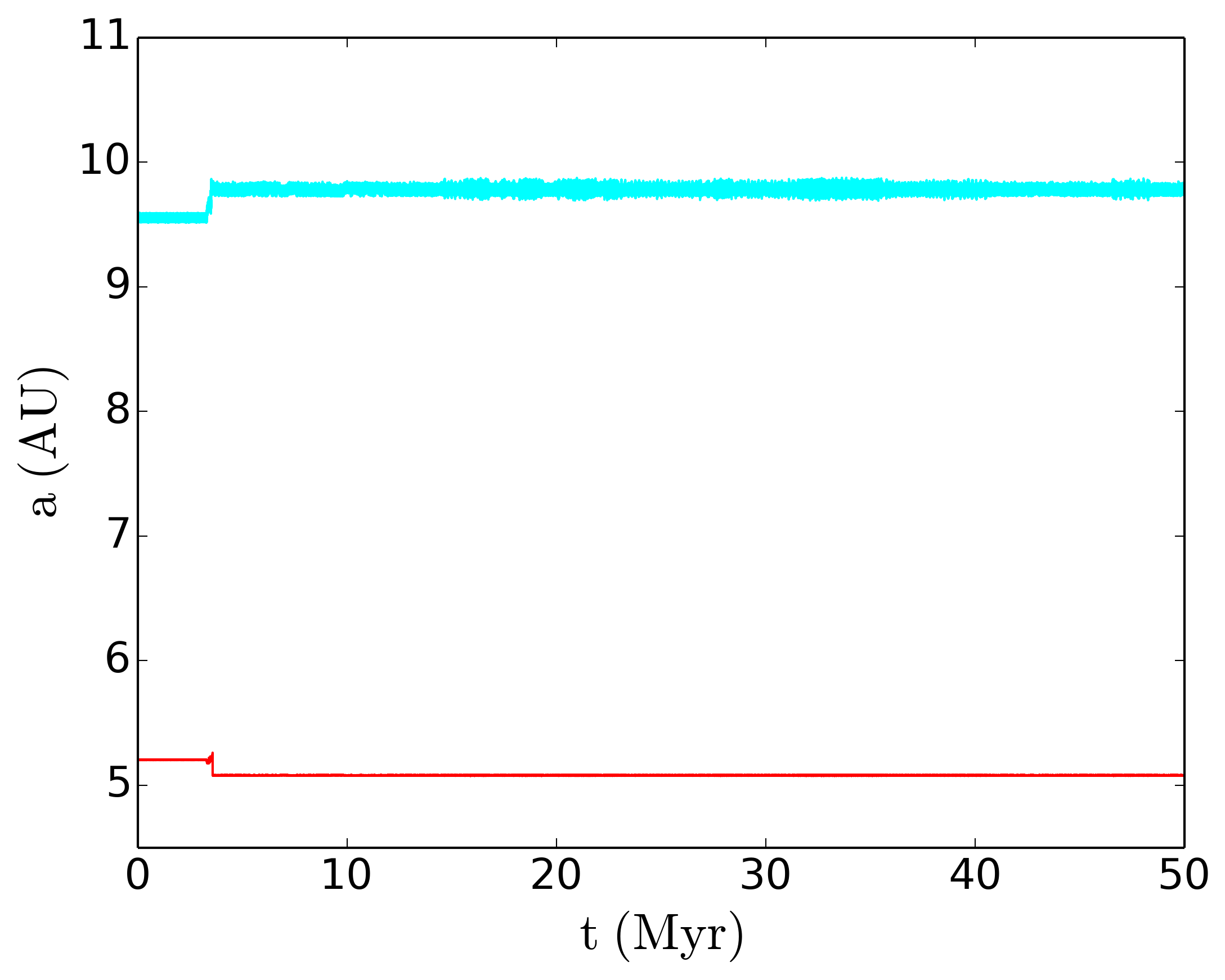} \\
\end{tabular}
\caption{Evolution of the semi-major axis and eccentricity Jupiter (red) and Saturn (blue) in planetary system P165 in star cluster model C061E4 (cf. Figure~\ref{figure:highandlowvel}; top panels). This system loses all its planets during a close encounter at $t\approx 4$~Myr, except Jupiter and Saturn. During this encounter, Uranus experiences a strong scattering event with Jupiter and Saturn, which obtain higher eccentricities and subsequently eject the terrestrial planets.  }
\label{figure:app4}
\end{figure*}

\begin{figure*}
\begin{tabular}{cc}
 \includegraphics[width=0.5\textwidth,height=!]{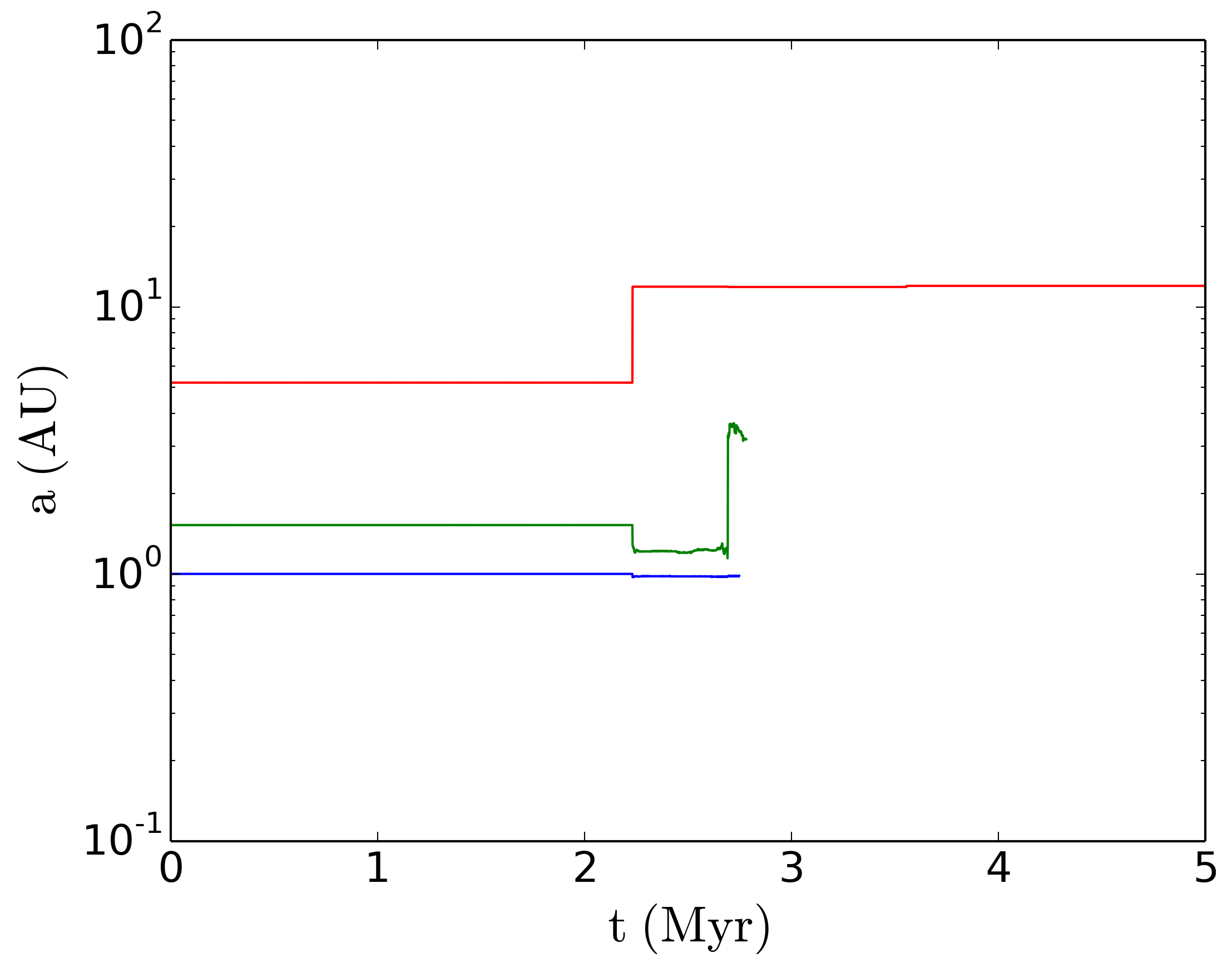} &
 \includegraphics[width=0.5\textwidth,height=!]{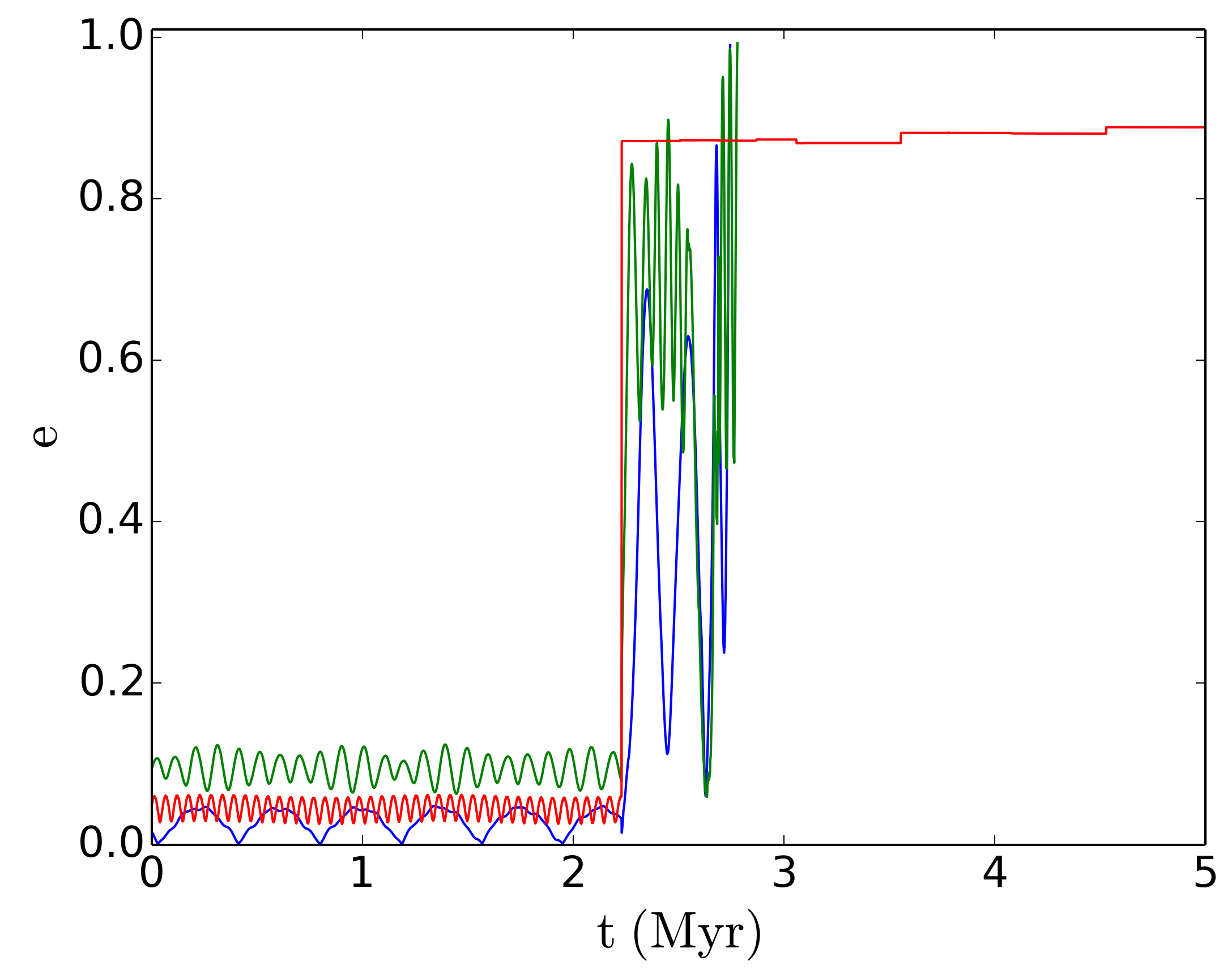} \\
 \end{tabular}
\caption{The first 5~Myr of evolution of planetary system P024 in star cluster model C051E4 (cf. Figure~\ref{figure:extraesc}; top panels). This figure demonstrates how a perturbation in Jupiter's orbit (in red) affects the rocky planets (Earth in blue; Mars in green). Although Jupiter itself remains bound, its change in eccentricity of Jupiter destabilises Mars and Earth, which are both ejected from the planetary system approximately a half a million years later. }
\label{figure:app5}
\end{figure*}

\begin{figure*}
\begin{tabular}{cc}
 \includegraphics[width=0.5\textwidth,height=!]{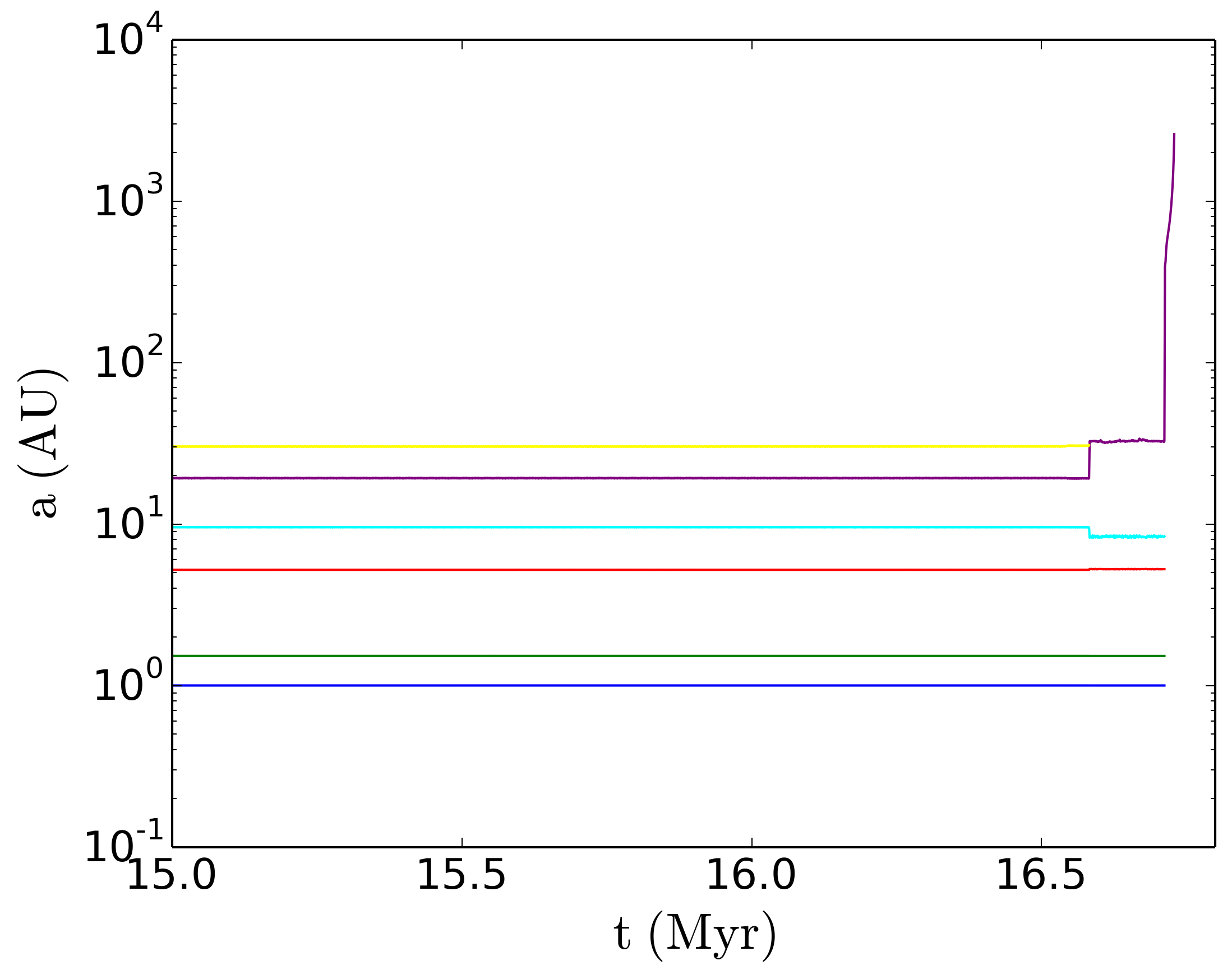} &
 \includegraphics[width=0.5\textwidth,height=!]{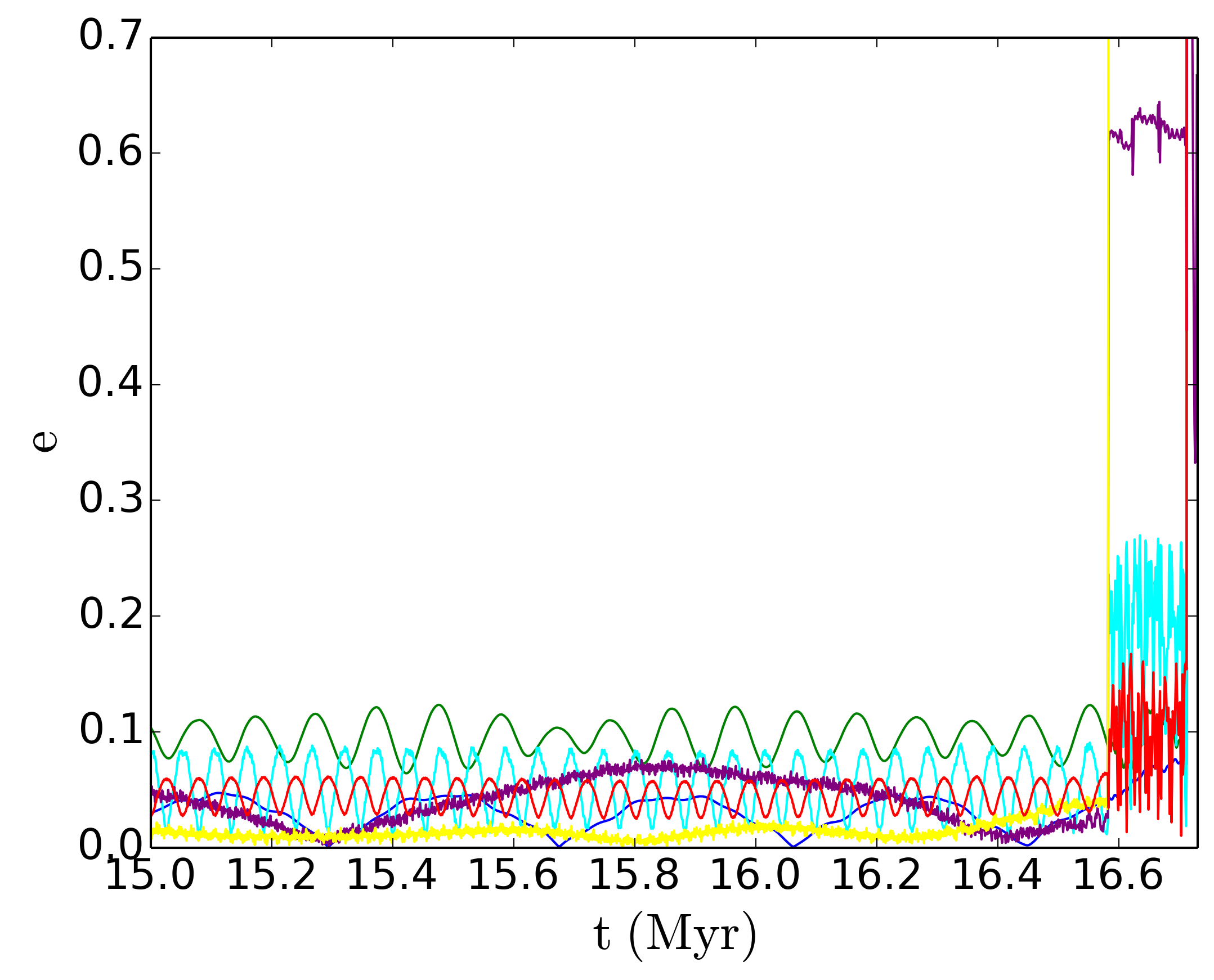} \\
 \end{tabular}
\caption{The  evolution of planetary system P187 in star cluster model C061E4 (cf. Figure~\ref{figure:extraesc}; bottom panels). This system interacts with two neighbour stars with that approach closer than 500~AU, within a time span of 0.2~Myr. The first encounter partially disrupts the outer planetary system, while the second encounter is responsible for the ejection of all remaining planets.}
\label{figure:app6}
\end{figure*}

This appendix illustrates in Figures~\ref{figure:app1}-\ref{figure:app6} the effect of stellar encounters and planetary interactions for the six planetary system described in Section~\ref{section:trends} (Figures~\ref{figure:comandchan}-\ref{figure:extraesc}).

\bsp	
\label{lastpage}
\end{document}